\documentclass[aps,prxquantum,reprint,groupedaddress,superscriptaddress,amsmath,amssymb]{revtex4-2}
\usepackage{graphicx}
\usepackage{dcolumn}
\usepackage{bbm}
\usepackage{braket}
\usepackage{algorithm}
\usepackage{hyperref}
\usepackage[caption=false,font=normalsize]{subfig}
\usepackage{url}
\usepackage[utf8]{inputenc}
\usepackage[T1]{fontenc}
\usepackage{array, tabularx, booktabs, lipsum, multirow}
\usepackage{siunitx, xcolor}
\usepackage[version=4]{mhchem}
\usepackage{newtxtext,newtxmath}
\usepackage{xcolor}
\usepackage{marvosym}
\hypersetup{colorlinks=true, citecolor=blue, urlcolor=blue, linkcolor=blue}
\usepackage[noend]{algpseudocode}
\usepackage{verbatim}
\usepackage{tikz,xcolor,hyperref}

\newcommand{\myparencite}[1]{\textcolor{blue}{(}\ref{#1}\textcolor{blue}{)}}

\definecolor{lime}{HTML}{A6CE39}
\DeclareRobustCommand{\orcidicon}{
	\begin{tikzpicture}
		\draw[lime, fill=lime] (0,0) 
		circle [radius=0.16] 
		node[white] {{\fontfamily{qag}\selectfont \tiny ID}};
		\draw[white, fill=white] (-0.0625,0.095) 
		circle [radius=0.007];
	\end{tikzpicture}
	\hspace{-2mm}
}
\foreach \x in {A, ..., Z}{%
	\expandafter\xdef\csname orcid\x\endcsname{\noexpand\href{https://orcid.org/\csname orcidauthor\x\endcsname}{\noexpand\orcidicon}}
}

\begin{document}

\title{Quantum machine learning for multiclass classification beyond kernel methods}

\author{Chao Ding\orcidA{}}
\affiliation{College of Electrical and Information Engineering, Hunan University, Changsha 410082, China}
\affiliation{Division of Physics and Applied Physics, School of Physical and Mathematical Sciences, Nanyang Technological University, Singapore, 637371, Singapore}
\affiliation{National Engineering Research Center of Robot Visual Perception and Control Technology, Hunan University, Changsha 410082, China}
\author{Shi Wang\orcidB{}}
\email[]{shi{\_}wang@hnu.edu.cn}
\author{Yaonan Wang\orcidC{}}
\affiliation{College of Electrical and Information Engineering, Hunan University, Changsha 410082, China}
\affiliation{National Engineering Research Center of Robot Visual Perception and Control Technology, Hunan University, Changsha 410082, China}
\author{Weibo Gao\orcidD{}}
\email[]{wbgao@ntu.edu.sg}
\affiliation{Division of Physics and Applied Physics, School of Physical and Mathematical Sciences, Nanyang Technological University, Singapore, 637371, Singapore}
\affiliation{Centre for Quantum Technologies, National University of Singapore, Singapore 117543, Singapore}
\affiliation{The Photonics Institute and Centre for Disruptive Photonic Technologies, Nanyang Technological University, Singapore 637371, Singapore}

\date{\today}

\begin{abstract}
Quantum machine learning is considered one of the current research fields with great potential. In recent years, Havl\'\i\v{c}ek $et \ al.$ [\href{https://doi.org/10.1038/s41586-019-0980-2}{Nature \textbf{567}, 209-212 (2019)}] have proposed a quantum machine learning algorithm with quantum-enhanced feature spaces, which effectively addressed a binary classification problem on a superconducting processor and offered a potential pathway to achieving quantum advantage. However, a straightforward binary classification algorithm falls short in solving multiclass classification problems. In this paper, we propose a quantum algorithm that rigorously demonstrates that quantum kernel methods enhance the efficiency of multiclass classification in real-world applications, providing quantum advantage. To demonstrate quantum advantage, we design six distinct quantum kernels within the quantum algorithm to map input data into quantum state spaces and estimate the corresponding quantum kernel matrices. The results from quantum simulations reveal that the quantum algorithm outperforms its classical counterpart in handling six real-world multiclass classification problems. Furthermore, we leverage a variety of performance metrics to comprehensively evaluate the classification and generalization performance of the quantum algorithm. The results demonstrate that the quantum algorithm achieves superior classification and better generalization performance relative to classical counterparts.

\end{abstract}

\maketitle

\section{introduction}\label{introduction}
Quantum machine learning stands as a remarkably promising avenue of research, poised to tackle extraordinarily complex computational challenges \cite{biamonte2017quantum,huangPowerDataQuantum2021,sajjan2022quantum}. It seeks to utilize the unique properties of entanglement and superposition among qubits to explore an exponentially large quantum state space \cite{schuld2019quantum}. As the number of qubits increases, the demands on quantum hardware and technology also rise, complicating the design of quantum algorithms. In the current era of noisy intermediate-scale quantum (NISQ) technology \cite{preskill2018quantum,mitarai2018quantum,zhu2019training,perez2020data,perez2021one,gonzalez2022error,Noisy2022Bharti,dutta2022single,chen2023complexity,MiaoNN2024,dutta2024trainability}, several strategies have been proposed to design quantum algorithms with limited hardware resources. Most of these strategies involve combining parameterized quantum circuits with classical methods to address complex computational tasks. Therefore, the combination of quantum and classical methods stands out as one of the most promising avenues towards achieving quantum advantage.

Supervised learning aims to learn the correlation between input feature vectors and output class labels \cite{JIANG2020675}. Intriguingly, this learning mechanism can tackle many real-world problems, from identifying diseases in medical diagnostics \cite{leachman2017final,zhang2019pathologist,myszczynska2020applications,cao2023large} to predicting molecular properties
in drug discovery \cite{feinberg2018potentialnet,ekins2019exploiting,patel2020machine,walters2020applications,chen2023artificial}. Quantum machine learning \cite{wan2017quantum,schuld2021effect,cerezo2021variational,caro2022generalization}, a groundbreaking learning mechanism, leverages parameterized quantum circuits to explore the intricate relationship between inputs and outputs. It relies on unique quantum properties to enhance learning efficiency and holds the potential to outperform traditional supervised learning in addressing real-world problems. Despite its proven advantages in certain specific problems \cite{rebentrost2014quantum, liu2018quantum, xia2018quantum, cong2019quantum, du2020expressive, liu2021rigorous}, the next frontier involves proving that quantum machine learning offers benefits across a wide range of real-world challenges.

A support vector machine (SVM) is a well-known supervised learning algorithm specifically designed to discover an optimal hyperplane that separates feature vectors into two different classes in a feature space \cite{hsu2002comparison,noble2006support,ding2023machine}. It typically utilizes kernel methods \cite{campbell2002kernel, sanchez2003advanced, pillonetto2014kernel} to implicitly map feature vectors to a higher-dimensional space, effectively handling classification problems that are not linearly separable. Some previous quantum implementations of support vector machines \cite{schuld2019quantum, havlivcek2019supervised, wu2021application}, which explore the connection between parameterized quantum circuits and kernel methods, have made strides on straightforward binary classification problems. At their core, these algorithms first perform a nonlinear mapping of feature vectors to quantum states through specialized parameterized quantum circuits. Following this mapping, they evaluate the overlap of pairwise quantum states to construct a kernel matrix. This matrix is then fed into a classical optimizer, which determines the optimal hyperplane by executing a convex quadratic program, effectively dividing the vectors into two classes. However, many classification problems encountered in real-world scenarios commonly involve multiple classes, posing unique challenges.

In this paper, we present quantum-enhanced multiclass SVMs that leverage quantum state space \cite{schuld2019quantum} as their feature space to address real-world multiclass classification tasks and demonstrate quantum advantage. First, we ensure that quantum kernels, devised through specific parameterized quantum circuits, meet the required standards. In addition, quantum kernel matrices must satisfy the condition of being positive semidefinite matrices. Building on this foundation, we utilize the unique capabilities of high-dimensional data representation provided by instantaneous quantum polynomial circuits \cite{havlivcek2019supervised} to develop three types of quantum kernels: full, linear, and circular. Also, we leverage parameterized quantum circuits featuring trainable single-qubit rotation layers to develop three types of quantum kernels: Pauli-X, Pauli-Y, and Pauli-Z. To compare the performance of quantum algorithms with various quantum kernels, we introduce six distinct real-world datasets, each with unique feature dimensions and class labels. The results demonstrate that the optimal quantum kernel is significantly contingent upon the distribution and structure of the real-world dataset. Furthermore, the results indicate that the quantum algorithm with the optimal quantum kernel outperforms its classical counterparts in solving multiclass classification tasks. Finally, we employ a comprehensive set of performance metrics to evaluate the quantum algorithm. The results from quantum simulations demonstrate that the quantum algorithm performs exceptionally well in classifying six real-world datasets.

The paper is organized as follows: Sec.~\ref{preliminary} presents classical multiclass SVMs and quantum kernel estimation. Sec.~\ref{kernel} elaborates on six distinct quantum kernels. Sec.~\ref{hybrid} proposes a quantum machine learning algorithm termed quantum-enhanced multiclass SVMs. Sec.~\ref{experiments} provides a comprehensive performance analysis of the quantum algorithm. Sec.~\ref{limitation} introduces the effects of exponential concentration and hardware noise on the quantum algorithm. The paper concludes with Sec.~\ref{conclusion}, encapsulating our findings.

\vspace{-1.5ex}
\section{Preliminaries}\label{preliminary}
\vspace{-1.5ex}
\subsection{Notations}
The relevant notations in this paper are listed as follows. The dot product is denoted by $\langle\vec{a}, \vec{b}\rangle=\vec{a} \cdot \vec{b}$. Consider two quantum states, $|\tau\rangle$ and $|\phi\rangle$, with the inner product denoted as $\langle\tau | \phi\rangle$, the outer product as $| \tau\rangle\langle\phi |$, and the overlap defined by $|\langle\tau | \phi\rangle|^2$. 
In addition, defining
\begin{equation}
	\delta_{i, p}= \begin{cases} \ 1, &\text{if}\ y_i =p \\ \ 0, &\text{otherwise,}\end{cases}
\end{equation}
where both $i$ and $p$ are indexes. Finally, three canonical rotations around the axes, corresponding to the Pauli matrices ($\sigma_x, \sigma_y, \sigma_z$), are denoted as:
\begin{align}
	&R_x(2\alpha) = e^{-i {\alpha} \sigma_x}=\sigma_0 \cos \alpha-i \sigma_x \sin \alpha, \label{rx} \\ 
	&R_y(2\alpha) = e^{-i {\alpha} \sigma_y}=\sigma_0 \cos \alpha-i \sigma_y \sin \alpha, \label{ry}\\
	&R_z(2\alpha) = e^{-i {\alpha} \sigma_z}=\sigma_0 \cos \alpha-i \sigma_z \sin \alpha,  \label{rotation_z}
\end{align}
where 
\begin{equation}
	\begin{array}{ll}
		\sigma_0=\left[\begin{array}{lr}
			1 & \ \ 0 \\
			0 & \ \ 1
		\end{array}\right], & \sigma_x=\left[\begin{array}{lr}
			0 & \ \ 1 \\
			1 & \ \ 0
		\end{array}\right], \\\\
		\sigma_y=\left[\begin{array}{lr}
			0 & -i \\
			i & 0 
		\end{array}\right], & \sigma_z =\left[\begin{array}{lr}
			1 & 0 \\
			0 & -1
		\end{array}\right].
	\end{array}
\end{equation}
The above summarizes the notations used in this paper.

\begin{figure}[t]
	\centering
	\includegraphics[width=\linewidth]{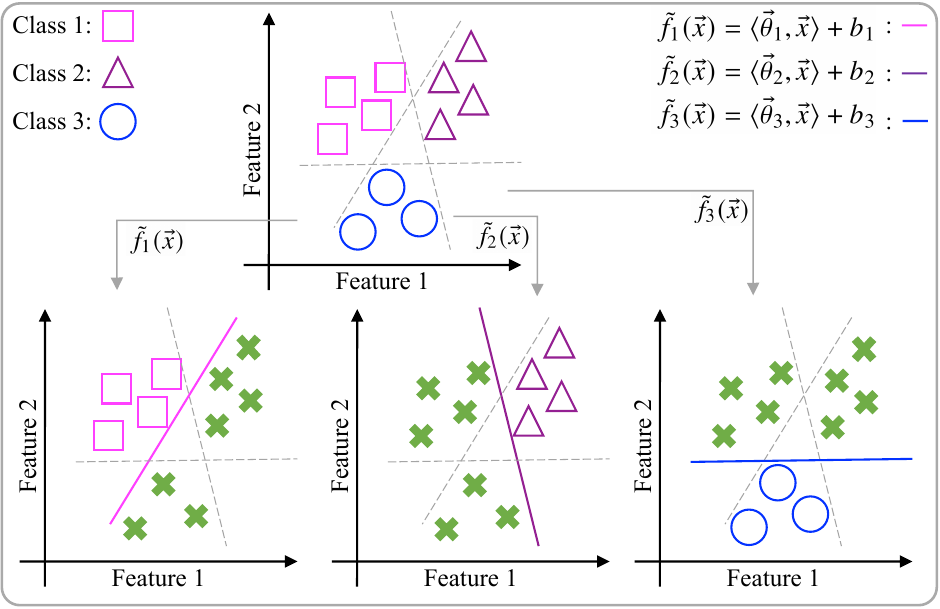}
	\caption{Illustration of multiclass classification employing multiclass SVMs through a one-versus-all approach.}
	\label{figure_01}
\end{figure}

\begin{figure*}[t]
	\centering
	\subfloat[]{\includegraphics[width=1.600in]{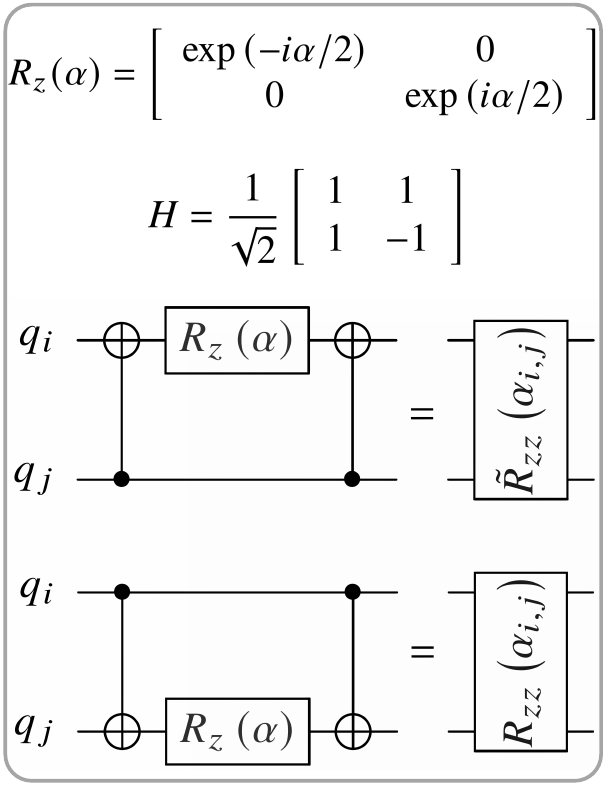}
		\label{figure_2a}}
	\hfill
	\subfloat[]{\includegraphics[width=5.393in]{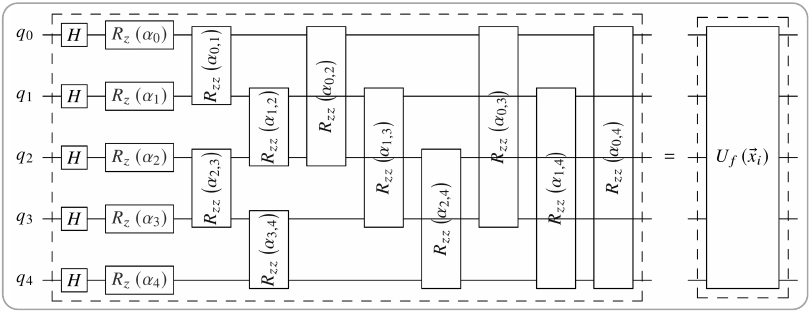}
		\label{figure_2b}}
	\hfill
	\subfloat[]{\includegraphics[width=3.215in]{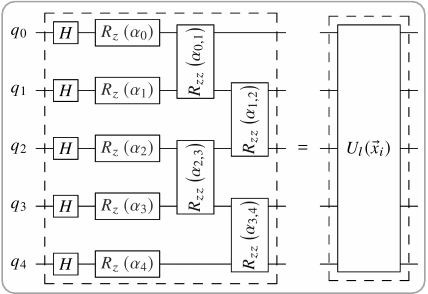}
		\label{figure_2c}}
	\hfill
	\subfloat[]{\includegraphics[width=3.778in]{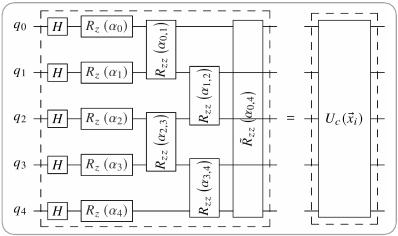}
		\label{figure_2d}}
	\hfill	
	\caption{Quantum circuit example with five qubits in the IQP circuit. (a) The matrix representation of single-qubit gates and the implementation of two-qubit gates. The two-qubit gates are constructed through $\text{CNOT}$ and $\text{RZ}$ gates. (b) The IQP circuit with a full entanglement layer. For a full entanglement layer with $\mathrm{N}$ qubits, where each qubit $q_{i}$ is entangled with every other qubit $q_{j}$, there are a total of $\mathrm{N}(\mathrm{N}-1)/2$ two-qubit gates and $\mathrm{N}(\mathrm{N}+1)/2$ training parameters. (c) The IQP circuit with a linear entanglement layer. For a linear entanglement layer with $\mathrm{N}$ qubits, where the qubit $q_i$ is entangled with the qubit $q_{i+1}$ and the index $i$ belongs to the set $\{0,1,\dots,\mathrm{N}-2\}$, there are a total of $\mathrm{N}-1$ two-qubit gates and $2\mathrm{N}-1$ training parameters. (d) The IQP circuit with a circular entanglement layer. For a circular entanglement layer with $\mathrm{N}$ qubits, the qubits are entangled in the same way as in the linear entanglement layer, but with an additional entanglement between the qubit $q_0$ and the qubit $q_{\mathrm{N}-1}$. There are a total of $\mathrm{N}$ two-qubit gates and $2\mathrm{N}$ training parameters.}
	\label{fig_2}
\end{figure*}

\begin{figure*}[t]
	\centering
	\subfloat[]{\includegraphics[width=2.2in]{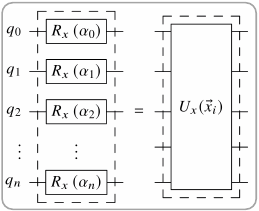}
		\label{figure_3a}}
	\hfill
	\subfloat[]{\includegraphics[width=2.2in]{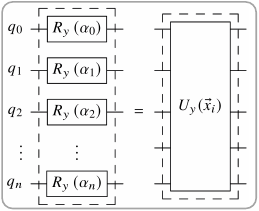}
		\label{figure_3b}}
	\hfill
	\subfloat[]{\includegraphics[width=2.51in]{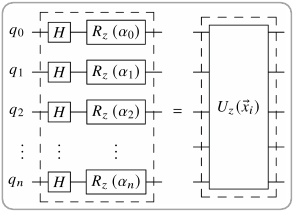}
		\label{figure_3c}}
	\caption{Quantum circuit example with a trainable single-qubit rotation layer. Each qubit corresponds to each feature of the input classical state $\vec{x}_i$. (a) The trainable single-qubit rotation layer with Pauli-X rotations. A Pauli-X rotation transforms $|0\rangle$ into the quantum state $|\phi_\textit{x}\left(x_t\right)\rangle=\cos (x_t/2)|0\rangle-i\sin (x_t/2)|1\rangle$. This rotation results in a superposition state of $|0\rangle$ and $|1\rangle$, significantly changing the probability amplitude distribution and introducing a phase difference between two basis states. (b) The trainable single-qubit rotation layer with Pauli-Y rotations \cite{schuld2021supervised}. A Pauli-Y rotation transforms $|0\rangle$ into the quantum state $|\phi_\textit{y}\left(x_t\right)\rangle=\cos (x_t/2)|0\rangle+\sin (x_t/2)|1\rangle$. Similar to the Pauli-X rotation, this rotation converts $|0\rangle$ into a superposition state of $|0\rangle$ and $|1\rangle$, altering the probability amplitude distribution of the two basis states and introducing a phase difference between them. (c) The trainable single-qubit rotation layer with Pauli-Z rotations. A Hadamard gate and Pauli-Z rotation transform $|0\rangle$ into the quantum state $|\phi_\textit{z}\left(x_t\right)\rangle=\sqrt{2}/2\left(\cos (x_t/2) - i\sin (x_t/2)\right) |0\rangle+\sqrt{2}/2\left(\cos (x_t/2) + i\sin (x_t/2)\right)|1\rangle$. As with the Pauli-X and Pauli-Y rotations, this operation generates a quantum state where the phase and probability amplitude distribution are both affected.}
	\label{fig_3}
\end{figure*}
\subsection{Multiclass SVMs}
One approach to address multiclass classification is by decomposing it into multiple binary classification tasks \cite{chang2011libsvm,garcia2011empirical,liu2015joint}. Multiclass SVMs typically adopt a one-versus-all approach \cite{hong2008probabilistic,xu2011extended,kumar2011reduced,sharan2015noise,faris2020medical}, illustrated in Fig.~\ref{figure_01}, wherein three binary classifiers are employed to address the classification task involving three classes. Suppose that there are a set of data points $\{(\vec{x}_i,y_i): \vec{x}_i \in \mathbb{R}^\mathrm{N}, y_i \in \mathbb{Y} \}_{i=1,\dots,m}$, where $\mathbb{Y} = \{1,\dots,l \}$, $y_i$ is the class label, and $\vec{x}_i$ is the feature vector. The task of the $s$th binary classifier is to differentiate data points into two classes: the $s$th class, which is assigned a positive label, and the combined class of all remaining classes, which receive a negative label. For the $s$th binary classifier, there exists a normal vector $\vec{\theta}_{s} \in \mathbb{R}^\mathrm{N}$ and a bias term $b_{s} \in \mathbb{R}$, defining the decision function $\tilde{f}_{s}(\vec{x})= \langle\vec{\theta}_{s}, \vec{x}\rangle + b_{s}$, where $\vec{x}$ is the feature vector of unknown data points. In pursuit of achieving optimal classification performance, it is imperative to address the optimization problem:
\begin{equation}\label{e1}
	\min _{\vec{\theta}, b, \vec{\xi}} L (\vec{\theta}, b, \vec{\xi}) = \enspace \frac{1}{2} \langle\vec{\theta}_{s}, \vec{\theta}_{s}\rangle + C \sum_{i=1}^{m} \xi_{i}^{s},
\end{equation}
subject to the constraints: 
\begin{equation}
\!\!\!	y_i[\langle\vec{\theta}_{s}, \vec{x}_i \rangle +b_{s}] \geqslant 1-\xi_{i}^{s}, y_i =  \pm 1, \xi_{i}^{s} \geqslant 0, \forall i=1, \dots, m.
\end{equation}
The dual formulation to this optimization problem is 
\begin{equation}\label{e2}
	\max _{\vec{\alpha}} L(\vec{\alpha})=\sum_{i=1}^m \alpha_i^{s}-\frac{1}{2} \sum_{i,j} \alpha_i^{s} \alpha_j^{s} y_i y_j \langle\vec{x}_i, \vec{x}_j\rangle, 
\end{equation}
subject to the constraints: 
\begin{equation}
	C \geqslant \alpha_i^{s} \geqslant 0, \sum_{i=1}^{m} y_i \alpha_i^{s}=0.
\end{equation}
The Lagrange multipliers $\alpha_i^{s}$ here are used to determine support vectors, and the index set of the support vectors is denoted by $\Omega$. Hence, the decision function of the $s$th binary classifier is redefined as 
\begin{equation}\label{e3}
	\tilde{f}_{s}(\vec{x})= \sum_{i \in {\Omega} } \alpha_i^{s} y_i \left\langle\vec{x}_i, \vec{x}\right\rangle +b_{s}.
\end{equation}
As there are $l$ binary classifiers, the corresponding $l$ decision functions are expressed as follows:
\begin{eqnarray}\label{e4}
	\tilde{f}_{1}(\vec{x})= \langle\vec{\theta}_{1}, \vec{x}\rangle +b_{1}, \ldots, \tilde{f}_{l}(\vec{x})= \langle\vec{\theta}_{l}, \vec{x}\rangle +b_{l}.
\end{eqnarray}
This yields the resulting decision function:
\begin{equation}\label{e5}
	\tilde{f}(\vec{x}) = \arg \max_{s=1,\ldots,l} \left[\langle\vec{\theta}_{s}, \vec{x}\rangle+ b_{s}\right].
\end{equation} 
Therefore, we determine the class with the highest value by calculating values for each decision function.

\subsection{Quantum kernel estimation}
Suppose that a feature vector $\vec{x}$ is mapped to a quantum state $|\phi(\vec{x})\rangle = \mathcal{S}(\vec{x})|0\rangle^{\otimes \mathrm{N}}$, the density matrix of $|\phi(\vec{x})\rangle$ is represented as $\rho(\vec{x})=|\phi(\vec{x})\rangle\langle\phi(\vec{x})| \in \mathbb{C}^{2^{\mathrm{N}}\times2^{\mathrm{N}}}$. For $\ket{\phi(\vec{x}_i)}$ and $|\phi(\vec{x})\rangle$, we define the quantum kernel \cite{wang2021towards,schuld2021supervised,hubregtsen2022training,jerbi2023quantum,paine2023quantum} using their overlap, i.e., $\kappa(\vec{x}_i,\vec{x})= |\langle \phi(\vec{x}_i) | \phi(\vec{x})\rangle|^2$.
Given a set of data points $\{(\vec{x}_i,y_i): \vec{x}_i \in \mathbb{R}^\mathrm{N}, y_i \in \mathbb{Y} \}_{i=0,\dots,\mathrm{N-1}}$, we can generalize the quantum kernel $\kappa(\vec{x}_i,\vec{x}_j)$ to a quantum kernel matrix $K \in \mathbb{R}^{\mathrm{N} \times \mathrm{N}}$, with the individual elements denoted by
\begin{equation}\label{e6}
	K_{ij} = |\langle \phi(\vec{x}_i) | \phi(\vec{x}_j) \rangle|^2.
\end{equation}
We can simulate and compute each element $K_{ij}$ of the quantum kernel matrix $K$ on an NISQ device. Furthermore, in accordance with Mercer condition \cite{ghojogh2021reproducing}, it is imperative to establish that the quantum kernel matrix $K$ satisfies the criteria of being a positive semidefinite matrix. Here, we readily demonstrate that the quantum kernel $\kappa(\vec{x}_i,\vec{x}_j)$ is an admissible kernel based on the straightforward inference from $K_{ij} = K_{ji}$ and $K_{ij} \geqslant 0$.

\begin{table*}[!htp]
	\centering
	\caption{\label{tab:table1}Examples of kernel function representations for input feature vectors $\vec{x}_i$ and $\vec{x}_j$, including classical kernels and quantum kernels.}
	\begin{ruledtabular}
		\begin{tabular}{lllc}
			Type& Name&Kernel function& Hyperparameters \\
			\hline
			\rule{0pt}{10pt}Classical \cite{campbell2002kernel}&Linear kernel (LK) & $\kappa_(\vec{x}_{i},\vec{x}_{j}) = \langle\vec{x}_{i},\vec{x}_{j}\rangle$ & $\textrm{None}$ \\
			&Polynomial kernel (PK) & $\kappa(\vec{x}_{i},\vec{x}_{j}) = \left(\varkappa\langle\vec{x}_{i},\vec{x}_{j}\rangle \right)^{\mathrm{d}}$& $\varkappa>0, \mathrm{d} \geqslant 1$\\
			&Sigmoid kernel (SK)  & $\kappa(\vec{x}_{i},\vec{x}_{j}) = \tanh\left(\varkappa \langle \vec{x}_{i}, \vec{x}_{j}\rangle + c\right)$& $\varkappa>0, c<0$\\
			&Gaussian kernel (GK) & $\kappa(\vec{x}_{i},\vec{x}_{j}) = \textrm{exp}(-\varkappa\cdot\Vert\vec{x}_{i}-\vec{x}_{j}\Vert^{2})$& $\varkappa>0$\\
			Quantum	&Full quantum kernel (FQK)  & $\kappa(\vec{x}_{i},\vec{x}_{j}) = \operatorname{Tr}[\left(|0\rangle\langle 0|\right)^{\otimes \mathrm{N}}(U_\textit{f}^{\dag}\left(\vec{x}_j\right)U_\textit{{f}}\left(\vec{x}_i\right)\left(|0\rangle\langle 0|\right)^{\otimes \mathrm{N}} U_\textit{f}^{\dag}\left(\vec{x}_i\right) U_\textit{f}\left(\vec{x}_j\right))]$& $\textrm{None}$ \\
			&Linear quantum kernel (LQK)  & $\kappa(\vec{x}_{i},\vec{x}_{j}) = \operatorname{Tr}[\left(|0\rangle\langle 0|\right)^{\otimes \mathrm{N}}(U_\textit{l}^{\dag}\left(\vec{x}_j\right)U_\textit{{l}}\left(\vec{x}_i\right)\left(|0\rangle\langle 0|\right)^{\otimes \mathrm{N}} U_\textit{l}^{\dag}\left(\vec{x}_i\right) U_\textit{l}\left(\vec{x}_j\right))]$& $\textrm{None}$ \\
			&Circular quantum kernel (CQK)  & $\kappa(\vec{x}_{i},\vec{x}_{j}) = \operatorname{Tr}[\left(|0\rangle\langle 0|\right)^{\otimes \mathrm{N}}(U_\textit{c}^{\dag}\left(\vec{x}_j\right)U_\textit{{c}}\left(\vec{x}_i\right)\left(|0\rangle\langle 0|\right)^{\otimes \mathrm{N}} U_\textit{c}^{\dag}\left(\vec{x}_i\right) U_\textit{c}\left(\vec{x}_j\right))]$& $\textrm{None}$ \\
			&Pauli-X quantum kernel (XQK)  & $\kappa(\vec{x}_{i},\vec{x}_{j}) = \operatorname{Tr}[\left(|0\rangle\langle 0|\right)^{\otimes \mathrm{N}}(U_\textit{x}^{\dag}\left(\vec{x}_j\right)U_\textit{{x}}\left(\vec{x}_i\right)\left(|0\rangle\langle 0|\right)^{\otimes \mathrm{N}} U_\textit{x}^{\dag}\left(\vec{x}_i\right) U_\textit{x}\left(\vec{x}_j\right))]$& $\textrm{None}$ \\
			&Pauli-Y quantum kernel (YQK)  & $\kappa(\vec{x}_{i},\vec{x}_{j}) = \operatorname{Tr}[\left(|0\rangle\langle 0|\right)^{\otimes \mathrm{N}}(U_\textit{y}^{\dag}\left(\vec{x}_j\right)U_\textit{{y}}\left(\vec{x}_i\right)\left(|0\rangle\langle 0|\right)^{\otimes \mathrm{N}} U_\textit{y}^{\dag}\left(\vec{x}_i\right) U_\textit{y}\left(\vec{x}_j\right))]$& $\textrm{None}$ \\
			&Pauli-Z quantum kernel (ZQK)  & $\kappa(\vec{x}_{i},\vec{x}_{j}) = \operatorname{Tr}[\left(|0\rangle\langle 0|\right)^{\otimes \mathrm{N}}(U_\textit{z}^{\dag}\left(\vec{x}_j\right)U_\textit{{z}}\left(\vec{x}_i\right)\left(|0\rangle\langle 0|\right)^{\otimes \mathrm{N}} U_\textit{z}^{\dag}\left(\vec{x}_i\right) U_\textit{z}\left(\vec{x}_j\right))]$& $\textrm{None}$ \\	
		\end{tabular}
	\end{ruledtabular}
\end{table*}

\vspace{-2ex}
\section{Quantum kernels with learnable rotations}\label{kernel}
In this section, we develop six quantum kernels for multiclass classification, each featuring learnable rotations and built from distinct parameterized quantum circuits.
\vspace{-1ex}
\subsection{Full, linear, and circular quantum kernels}
In this part, we utilize an instantaneous quantum polynomial (IQP) circuit \cite{havlivcek2019supervised} to efficiently implement quantum kernels. The IQP circuit includes a Hadamard layer, a single-qubit rotation layer, and a two-qubit entanglement layer, with its performance influenced by the quantum circuit structure. To explore this, we examine three distinct quantum circuit structures in the IQP circuit: full, linear, and circular entanglement layers, illustrated in Figs.~\ref{figure_2b}, \ref{figure_2c}, and \ref{figure_2d}, respectively. Furthermore, three unique quantum circuit structures yield three distinct quantum kernels: full, linear, and circular.

Quantum kernels can map the classical input state $\vec{x}_i = (x_0, x_1, \ldots, x_{\mathrm{N}-1})^T \in \mathbb{R}^{\mathrm{N}}$ to a quantum state space. For the full quantum kernel, $\vec{x}_i$ is mapped to $|\phi_{\mathrm{\textit{f}}}(\vec{x}_i)\rangle = U_{\mathrm{\textit{f}}}(\vec{x}_i)|0\rangle^{\otimes \mathrm{N}}$, where $|0\rangle^{\otimes \mathrm{N}}$ is the initial state. As illustrated in Fig.~\ref{figure_2b}, implementing a Hadamard gate on each qubit yields the following superposition state:
\begin{align}\label{superposition}
	\left|\phi_1\right\rangle =\left(\frac{1}{\sqrt{2}}\right)^{{\mathrm{N}}} \sum_{q_0=0}^1 \sum_{q_1=0}^1 \ldots \sum_{q_{\mathrm{N}-1}=0}^1 \left|q_0, q_1, \ldots, q_{\mathrm{N}-1}\right\rangle,
\end{align}
where $q_i$ describes the state of the $i$th qubit, which can be either $0$ or $1$. Due to $q=q_{\mathrm{N}-1} 2^0 + q_{\mathrm{N}-2} 2^1 + \cdots + q_0 2^{\mathrm{N}-1}$, Eq.~\myparencite{superposition} simplifies to $|\phi_1\rangle = ({1}/{\sqrt{2}})^{\mathrm{N}} \sum_{q=0}^{2^{\mathrm{N}}-1}|q\rangle$. According to Eq.~\myparencite{rotation_z}, we have $R_z\left(\alpha_i\right)|0\rangle=|0\rangle$ and $R_z\left(\alpha_i\right)|1\rangle=e^{i \alpha_i}|1\rangle$. Applying the Pauli-Z rotation gate to each qubit transforms $\left|\phi_1\right\rangle$ to 
\begin{equation}
	\left|\phi_2\right\rangle = \left(\frac{1}{\sqrt{2}}\right)^{\mathrm{N}} \sum_{q=0}^{2^{\mathrm{N}}-1} \exp{\left(i\sum_{i=0}^{\mathrm{N}-1}\alpha_i q_i  \right)}     |q\rangle.
\end{equation}
From Fig.~\ref{figure_2a}, the two-qubit gate is defined as follows:
\begin{equation}\label{rzz}
	R_{zz}(\alpha) = \tilde{R}_{zz}(\alpha) = \begin{bmatrix}
		e^{-i \alpha/2} & 0 & 0 & 0 \\
		0 & e^{i \alpha/2} & 0 & 0 \\
		0 & 0 & e^{i \alpha/2} & 0 \\
		0 & 0 & 0 & e^{-i \alpha/2}
	\end{bmatrix}
\end{equation}
with $\alpha = \alpha_{i,j} = \alpha_i \alpha_j$. Each pair of qubits receives a phase adjustment according to its state. Based on Eq.~\myparencite{rzz}, we have $R_{zz}(\alpha)|00\rangle = e^{-i\alpha/2}|00\rangle$, $R_{zz}(\alpha)|01\rangle = e^{i\alpha/2} |01\rangle$, $R_{zz}(\alpha)|11\rangle = e^{-i\alpha/2}|11\rangle$, and $R_{zz}(\alpha)|10\rangle = e^{i\alpha/2} |10\rangle$. Therefore, $\left|\phi_2\right\rangle$ is transformed to 
\begin{equation}
	\left|\phi_3\right\rangle = \left(\frac{1}{\sqrt{2}}\right)^{\mathrm{N}} \sum_{q=0}^{2^{\mathrm{N}}-1} \exp{\left(i\sum_{i=0}^{\mathrm{N}-1}\alpha_i q_i  \right)} \left(\prod_{(i, j) \in I} \digamma_{(i,j)}  \right) |q\rangle
\end{equation}
with $\digamma_{(i,j) } =  \exp ({i \frac{-\alpha_{i}\alpha_{j}}{2} \left(-1\right)^{q_i \oplus q_j}})$ and $I=\{(i, j) \mid 0 \leqslant i<j \leqslant \mathrm{N}-1\}$. The feature $x_i$ corresponds to $\alpha_{i}$, and $x_ix_j$ corresponds to $\alpha_{i}\alpha_{j}$. The final quantum state is formulated as
\begin{equation}
\!\!\!	|\phi_{\mathrm{\textit{f}}}(\vec{x}_i)\rangle \!=\! \left(\frac{1}{\sqrt{2}}\right)^{\mathrm{N}} \sum_{q=0}^{2^{\mathrm{N}}-1} \exp{\left(i\sum_{i=0}^{\mathrm{N}-1}x_i q_i  \right)} \left(\prod_{(i, j) \in I} \digamma^{\prime}_{(i,j)}  \right) |q\rangle
\end{equation}
with $\digamma^{\prime}_{(i,j) } =  \exp ({i \frac{-x_{i}x_{j}}{2} \left(-1\right)^{q_i \oplus q_j}})$. For the linear quantum kernel, $\vec{x}_i$ is mapped to $|\phi_{\textit{l}}(\vec{x}_i)\rangle = U_{\textit{l}}(\vec{x}_i)|0\rangle^{\otimes \mathrm{N}}$. Similar to the full quantum kernel, the resulting quantum state is denoted as 
\begin{equation}
	\!\!\!	|\phi_\textit{l}(\vec{x}_i)\rangle \!=\! \left(\frac{1}{\sqrt{2}}\right)^{\mathrm{N}} \sum_{q=0}^{2^{\mathrm{N}}-1} \exp{\left(i\sum_{i=0}^{\mathrm{N}-1}x_i q_i  \right)} \left(\prod_{(i, j) \in \bar{I}} \! \digamma^{\prime}_{(i,j)}  \right) |q\rangle
\end{equation}
with $\bar{I}=\{(i, j) \mid 0 \leqslant i<j \leqslant \mathrm{N}-1, j = i+1\}$. For the circular quantum kernel, $\vec{x}_i$ is mapped to $|\phi_{\textit{c}}(\vec{x}_i)\rangle = U_{\textit{c}}(\vec{x}_i)|0\rangle^{\otimes \mathrm{N}}$. Similar to the full and linear quantum kernels, the obtained quantum state is represented as
\begin{equation}
	\!\!\!	|\phi_\textit{c}(\vec{x}_i)\rangle \!=\! \left(\frac{1}{\sqrt{2}}\right)^{\mathrm{N}} \sum_{q=0}^{2^{\mathrm{N}}-1} \exp{\left(i\sum_{i=0}^{\mathrm{N}-1}x_i q_i  \right)} \left(\prod_{(i, j) \in \hat{I}} \digamma^{\prime}_{(i,j)} \right) |q\rangle
\end{equation}
with $\hat{I}=\{(i, j) \mid 0 \leqslant i \leqslant \mathrm{N}-1, j = (i+1) \mod \mathrm{N}\}$.

\subsection{Pauli-X, Pauli-Y, and Pauli-Z quantum kernels}
In this part, we utilize a parameterized quantum circuit featuring a trainable single-qubit rotation layer to implement quantum kernels. In terms of the single-qubit rotation layer, we employ three distinct quantum circuit structures: Pauli-X, Y, and Z rotations, illustrated in Figs.~\ref{figure_3a}, \ref{figure_3b}, and \ref{figure_3c}, respectively. Similarly, these quantum circuit structures give rise to three unique quantum kernels: Pauli-X, Pauli-Y, and Pauli-Z. For the Pauli-X quantum kernel, $\vec{x}_i$ is mapped to $|\phi_x(\vec{x}_i)\rangle = U_\textit{x}(\vec{x}_i)|0\rangle^{\otimes \mathrm{N}}$, where $U_\textit{x}(\vec{x}_i)=  R_x\left(x_0\right)\otimes \cdots \otimes R_x\left(x_{\mathrm{N}-1}\right)$. Therefore, the quantum state $|\phi_\textit{x}(\vec{x}_i)\rangle$ is reformulated as
\begin{equation}\label{state_x}
 \!\!\!	|\phi_\textit{x}(\vec{x}_i)\rangle \!=\! \sum_{q_0=0}^1 \dots \! \sum_{q_{\mathrm{N}-1}=0}^1 \! \prod_{i=0}^{\mathrm{N}-1}(\cos \frac{x_i}{2})^{1-q_i}(-i \sin \frac{x_i}{2})^{q_i} \left|q\right\rangle.
\end{equation}
For the Pauli-Y quantum kernel, $\vec{x}_i$ is mapped to $|\phi_\textit{y}(\vec{x}_i)\rangle = U_\textit{y}(\vec{x}_i)|0\rangle^{\otimes \mathrm{N}}$, where $U_\textit{y}(\vec{x}_i)= R_y\left(x_0\right) \otimes \cdots \otimes R_y\left(x_{\mathrm{N}-1}\right)$. Similar to Eq.~\myparencite{state_x}, this results in the following quantum state:
\begin{equation}\label{state_y}
\!\!	|\phi_\textit{y}(\vec{x}_i)\rangle \!=\! \sum_{q_0=0}^1 \! \ldots \! \sum_{q_{\mathrm{N}-1}=0}^1 \prod_{i=0}^{\mathrm{N}-1}(\cos \frac{x_i}{2})^{1-q_i}(\sin \frac{x_i}{2})^{q_i} \left|q\right\rangle.
\end{equation}
For the Pauli-Z quantum kernel, we observe that Pauli-Z rotations affect only the phase of the quantum state. Therefore, we introduce a Hadamard layer before the Pauli-Z rotation layer to optimize this quantum feature mapping. As illustrated in Fig.~\ref{figure_3c}, $\vec{x}_i$ is mapped to $|\phi_\textit{z}(\vec{x}_i)\rangle = U_\textit{z}(\vec{x}_i)|0\rangle^{\otimes \mathrm{N}}$, where $U_\textit{z}(\vec{x}_i)= R_z\left(x_0\right)H \otimes \cdots \otimes R_z\left(x_{\mathrm{N}-1}\right)H$. The Pauli-Z rotation gate is represented as $R_z(x_i)=e^{-i x_i/2}|0\rangle\langle 0|+e^{i x_i/2}| 1\rangle\langle 1|$. Therefore, the quantum state $|\phi_z(\vec{x}_i)\rangle$ has the form
\begin{equation}\label{state_z}
	\begin{split}
		|\phi_\textit{z}(\vec{x}_i)\rangle = \left(\frac{1}{\sqrt{2}}\right)^{\mathrm{N}} \sum_{q=0}^{2^{\mathrm{N}}-1} \exp{\left(i\sum_{i=0}^{\mathrm{N}-1}x_i q_i  \right)}\left|q\right\rangle.
	\end{split}
\end{equation}
\begin{figure}[t]
	\centering
	\includegraphics[width=\linewidth]{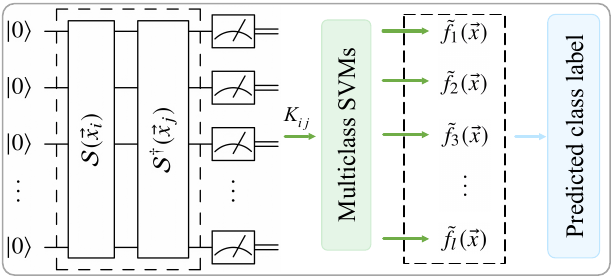}
	\caption{Schematic diagram of quantum-enhanced multiclass SVMs designed by integrating quantum kernels with classical multiclass SVMs, where $K_{ij}=|\langle 0|^{\otimes \mathrm{N}} \mathcal{S}^{\dag}(\vec{x}_j) \mathcal{S}\left(\vec{x}_i\right)|0\rangle^{\otimes \mathrm{N}}|^2$. The quantum feature mapping $\mathcal{S}(\vec{x}_i)$ is restricted to elements of the set $\{ U_\textit{f}{\left(\vec{x}_i\right)}, U_\textit{l}{\left(\vec{x}_i\right)}, U_\textit{c}{\left(\vec{x}_i\right)}, U_\textit{x}{\left(\vec{x}_i\right)}, U_\textit{y}{\left(\vec{x}_i\right)}, U_\textit{z}{\left(\vec{x}_i\right)}\}$.}
	\label{figure_04}
\end{figure}
\section{Quantum-Enhanced Multiclass SVMs}\label{hybrid}
In this section, we propose quantum-enhanced multiclass SVMs. The pseudocode for the quantum algorithm can be found in Appendix~\ref{appendixe}. The quantum algorithm includes two main stages, as illustrated in Fig.~\ref{figure_04}. At the first stage, it calculates the quantum kernel matrix on an NISQ device to evaluate the similarity between quantum states. This includes performing multiple measurements on a parameterized quantum circuit to derive the elements of the quantum kernel matrix. Note that the quantum kernels we used for multiclass classification are detailed in Table~\ref{tab:table1}. In the following stage, the quantum kernel matrix is utilized to formulate and solve a quadratic programming problem for multiclass classification. This includes generalizing the quadratic programming problem presented in Eq.~\myparencite{e1} to the following form \cite{hsu2002comparison}:

\begin{equation}\label{e9}
	\min_{\vec{\theta}, b, \vec{\xi}} \tilde{L}(\vec{\theta}, b, \vec{\xi}) = \frac{1}{2} \sum_{s=1}^{l}\langle\vec{\theta}_{s}, \vec{\theta}_{s}\rangle+C \sum_{i=1}^{m} \sum_{s \neq y_i}^{l} \xi_i^s,
\end{equation}
subject to the constraints: 
\begin{gather}
	\xi_i^s \geqslant 0, \  \langle\vec{\theta}_{y_i}, \vec{x}_i \rangle+b_{y_i} \geqslant \langle\vec{\theta}_s,\vec{x}_i\rangle+b_s+2-\xi_i^s, 	\\  \nonumber
	\forall s \neq y_i, \  \forall i=1,\dots, m,
\end{gather}
where $\xi_i^s$ and $C$ are slack variables and penalty parameters, respectively. To address the quadratic programming problem in Eq.~\myparencite{e9}, we formulate the following generalized Lagrangian function: 
\begin{eqnarray}\label{e10}
	\begin{split}
		\tilde{L} = &-\sum_{i,s}{\alpha_i^s\left[\langle\vec{\theta}_{y_i}-\vec{\theta}_s, \vec{x}_i\rangle + b_{y_i} - b_s- 2 + \xi_i^s\right]} \\
		&-\sum_{i,s} \mu_i^s \xi_i^s +\frac{1}{2} \sum_{s=1}^{l}\langle\vec{\theta}_{s}, \vec{\theta}_{s}\rangle +C \sum_{i,s} \xi_i^s,
	\end{split}
\end{eqnarray}
subject to the constraints: 
\begin{align}
	\alpha_i^{s}, \mu_i^{s}, \xi_i^s \geqslant 0, \ \forall s \neq y_i, \ \forall i=1,\dots, m,
\end{align}
where $\alpha_i^{s}$ and $\mu_i^{s}$ are Lagrange multipliers. Notably, the variables $\alpha_i^{y_i}=0$, $\mu_i^{y_i}=0$, $\xi_i^{y_i} = 2$, and $\forall i=1,\dots, m$. Then, the optimal parameter set is attainable by performing min-max optimization $\arg \max_{\vec{\alpha}, \vec{\mu}} \enspace  \arg \min_{\vec{\theta}, b, \vec{\xi}} \enspace \tilde{L}$. The partial derivatives of Eq.~\myparencite{e10} regarding the primal variables ($\vec{\theta}_{s^{\prime}}$, $b_{s^{\prime}}$, $\xi^{s^{\prime}}_i$) are expressed as follows:
\begin{gather}
	\frac{\partial {\tilde{L}}}{\partial \vec{\theta}_{s^{\prime}}}= -\sum_{i,s}\alpha_i^s \delta_{i,{s^{\prime}}} \vec{x}_i + \sum_{i=1}^{m} \alpha_i^{s^{\prime}} \vec{x}_i+\vec{\theta}_{s^{\prime}}, \label{e11.1}\\
	\frac{\partial {\tilde{L}}}{\partial b_{s^{\prime}}}= -\sum_{i,s} \alpha_i^s \delta_{i,{s^{\prime}}} +\sum_{i=1}^{m} \alpha_i^{s^{\prime}}, \label{e11.2} \\
	\frac{\partial {\tilde{L}}}{\partial {\xi}^{s^{\prime}}_i}= -\alpha_i^{s^{\prime}}-\mu_i^{s^{\prime}} +C.  \label{e11.3} 
\end{gather}
Derived from the Karush-Kuhn-Tucker (KKT) conditions \cite{ghojogh2021kkt}, we have
\begin{equation}
	\frac{\partial {\tilde{L}}}{\partial \vec{\theta}_{s^{\prime}}} = 0, \  \frac{\partial {\tilde{L}}}{\partial b_{s^{\prime}}} = 0, \  \frac{\partial {\tilde{L}}}{\partial {\xi}^{s^{\prime}}_i} = 0.
\end{equation}
Therefore, the optimal conditions are as follows:
\begin{gather}
	\vec{\theta}_{s^{\prime}}=\sum_{i=1}^m \left(\delta_{i,{s^{\prime}}} \sum_{s=1}^{l}\alpha_i^s - \alpha_i^{s^{\prime}} \right) \vec{x}_i, \label{e12.1}\\
	\sum_{i=1}^{m} \alpha_i^{s^{\prime}} = \sum_{i,s} \alpha_i^s \delta_{i,{s^{\prime}}}, \label{e12.2} \\
	C = \alpha_i^{s^{\prime}}+\mu_i^{s^{\prime}}, \quad C \geqslant \alpha_i^{{s^{\prime}}} \geqslant 0. \label{e12.3}
\end{gather}
Introducing the formulas
\begin{equation}
	\!\!\!\!\!	\left(\sum_{i,s}{\alpha_{i}^s}\right) b_{y_i} \!=\! \sum_{s=1}^{l}{b_s} \left(\sum_{i=1}^{m}{\delta_{i, s}\sum_{s=1}^{l}{\alpha_{i}^s}}\right) \!=\! \sum_{s=1}^{l}{b_s}\left(\sum_{i,s} \alpha_i^s \delta_{i,s}\right)
\end{equation}
and
\begin{equation}
	\sum_{i,s}{\alpha_{i}^s} b_{s} = \sum_{s=1}^{l}{b_s} \left(\sum_{i=1}^{m}{\alpha_i^s}\right).
\end{equation}
According to Eq.~\myparencite{e12.2}, we get
\begin{equation}\label{cancel_b}
	\left(\sum_{i,s}{\alpha_{i}^s}\right) b_{y_i} = \sum_{i,s}{\alpha_{i}^s} b_{s}.
\end{equation}
Substituting Eq.~\myparencite{e12.1} into Eq.~\myparencite{e10} results in the dual formulation, expressed as
\begin{align}\label{e13}
\!\! \max_{\vec{\alpha}} \tilde{L}(\vec{\alpha}) =&2\sum_{s,i}\alpha_i^s +\frac{1}{2} \sum_{s,i,j} \left[\left(\delta_{i,s} \sum_{s=1}^{l}\alpha_i^s\right)\left(\delta_{j,s} \sum_{s=1}^{l}\alpha_j^s\right) \right. \\ \nonumber
	&\left. - \alpha_j^s \left(\delta_{i,s} \sum_{s=1}^{l}\alpha_i^s\right)+ \alpha_i^s \left(\delta_{j,s} \sum_{s=1}^{l}\alpha_j^s\right) \right. \\ \nonumber
	&\left. - 2 \alpha_i^s \left(\delta_{j,y_i} \sum_{s=1}^{l}\alpha_j^s\right) + 2 \alpha_i^s \alpha_j^{y_i} -\alpha_i^s \alpha_j^s \right] \langle\vec{x}_i, \vec{x}_j\rangle,
\end{align}
subject to the constraints: 
\begin{gather}
	\alpha_i^{y_i}=0,\  \sum_{i=1}^{m} \alpha_i^{s^{\prime}} = \sum_{i,s} \alpha_i^s \delta_{i,{s^{\prime}}},\  C \geqslant \alpha_i^{{s^{\prime}}} \geqslant 0, \\ \nonumber
	\forall s \neq y_i, \ \forall i=1,\dots, m, \ \forall {s^{\prime}}=1,\dots, l.
\end{gather}
The solution to Eq.~\myparencite{e13} is typically simpler than that of Eq.~\myparencite{e9}. According to the formula
\begin{equation}
	\sum_{s,i,j} \alpha_j^s \left(\delta_{i,s} \sum_{s=1}^{l}\alpha_i^s\right) = \sum_{s,i,j} \alpha_i^s \left(\delta_{j,s} \sum_{s=1}^{l}\alpha_j^s\right),
\end{equation}
the dual formulation can be reformulated as
\begin{align}\label{e14}
\!\! \max_{\vec{\alpha}} \tilde{L}(\vec{\alpha}) = &2 \sum_{s,i}\alpha_i^s \!+\! \frac{1}{2} \sum_{s,i,j}\left[\left(\delta_{i,s} \sum_{s=1}^{l}\alpha_i^s\right)\left(\delta_{j,s} \sum_{s=1}^{l}\alpha_j^s\right) \right. \\ \nonumber
	&\left.- 2 \alpha_i^s \left(\delta_{j,y_i} \sum_{s=1}^{l}\alpha_j^s\right) + 2\alpha_i^s \alpha_j^{y_i} - \alpha_i^s\alpha_j^s \right]\langle\vec{x}_i, \vec{x}_j\rangle.
\end{align}
In addition, considering the following formula
\begin{equation}
	\sum_{s=1}^{l} \delta_{i,s} \delta_{j,s} = \delta_{i,y_j} = \delta_{j,y_i} = \begin{cases} \ 1, &\text{if}\ y_i =y_j \\ \ 0, &\text{otherwise,}\end{cases}
\end{equation}
Eq.~\myparencite{e14} can be simplified to
\begin{align}\label{e15}
	\max_{\vec{\alpha}} \tilde{L}(\vec{\alpha}) = &2 \sum_{s,i}\alpha_i^s +\frac{1}{2} \sum_{s,i,j}\left[-\delta_{j,y_i} \left(\sum_{s=1}^{l}\alpha_i^s\right) \left( \sum_{s=1}^{l}\alpha_j^s\right) \right. \\ \nonumber
	&\left. + 2\alpha_i^s \alpha_j^{y_i} - \alpha_i^s\alpha_j^s \right.\Bigg] \langle\vec{x}_i, \vec{x}_j\rangle.
\end{align}
The solution to Eq.~\myparencite{e15} can be obtained through the sequential minimal optimization (SMO) algorithm \cite{cao2006parallel,kuan2011vlsi,huang2015sequential}. Finally, this yields the decision function
\begin{align}\label{e16}
\!\!\!\!\tilde{f}(\vec{x}, \vec{\alpha})\!=\! \arg \max_{s^{\prime}} \left[\sum_{i: y_i=s^{\prime}}\left(\sum_{s=1}^{l}\alpha_i^s\right)\!-\!\! \sum_{i: y_i \neq s^{\prime}}\!\! \alpha_i^{s^{\prime}} \right]\langle\vec{x}_i, \vec{x}\rangle \!+ \! b_{s^{\prime}}. 
\end{align}
Typically, we can replace $\langle\vec{x}_i, \vec{x}_j\rangle$ in Eqs.~\myparencite{e13}, \myparencite{e14}, and \myparencite{e15} with a quantum kernel $\kappa(\vec{x}_i,\vec{x}_j)$. Therefore, the resulting decision function is given by
\begin{equation}
	\tilde{f}(\vec{x}, \vec{\alpha}) = \arg \max_{s^{\prime}} \sum_{i=1}^{m}\left(\delta_{i,s^{\prime}}\sum_{s=1}^{l}\alpha_i^{s} - \alpha_i^{s^{\prime}} \right)\kappa(\vec{x}_i,\vec{x}) + b_{s^{\prime}}.
\end{equation}

\begin{table}[!b]
	\centering
	\caption{\label{tab:table2}Real-world datasets used in experimental result analysis.}
	\begin{ruledtabular}
		\begin{tabular}{lccc}
			Dataset& $\#$ Instances & $\#$ Features & $\#$ Class \\ 
			\hline \rule{0pt}{10pt}Iris \cite{kelly2023uci_iris}& 150 & 4 & 3   \\
			Tae \cite{kelly2023uci_tae}& 151 & 5 & 3  \\
			Penguin \cite{penguins2020}& 344 & 5 & 3 \\
			Glass \cite{kelly2023uci_glass}& 214 & 9 & 6  \\
			Ecoli \cite{kelly2023uci_ecoli}& 336 & 7 & 8  \\
			Vowel \cite{kelly2023uci_vowel}& 528 & 10 & 11  \\
		\end{tabular}
	\end{ruledtabular}
\end{table}
\begin{table}[!b]
	\centering
	\caption{Dataset description following a $70$-$30$ training-testing split. $N_{\text{train}}$ and $N_{\text{test}}$ represent the sizes of training and testing sets, respectively.}
	\label{tab:table3}
	\begin{ruledtabular}
		\begin{tabular}{ccccccc}
			& \multicolumn{6}{c}{Dataset}  \\ 
			\cline{2-7} 
			& \rule{0pt}{10pt}Iris & Tae & Penguin & Glass & Ecoli & Vowel  \\ 
			\hline
			\rule{0pt}{10pt}$N_{\text{train}}$ & 105 & 105 & 233 & 149 & 235 & 369  \\ 
			$N_{\text{test}}$ & 45 & 46 & 100 & 65 & 101 & 159 \\ 
		\end{tabular}
	\end{ruledtabular}
\end{table}

\section{Experiments}\label{experiments}
This section begins with an exploration of real-world datasets and simulation platforms, followed by an evaluation of the proposed quantum-enhanced multiclass SVMs using quantum simulations.
\begin{figure}[t]
	\centering
	\includegraphics[width=\linewidth]{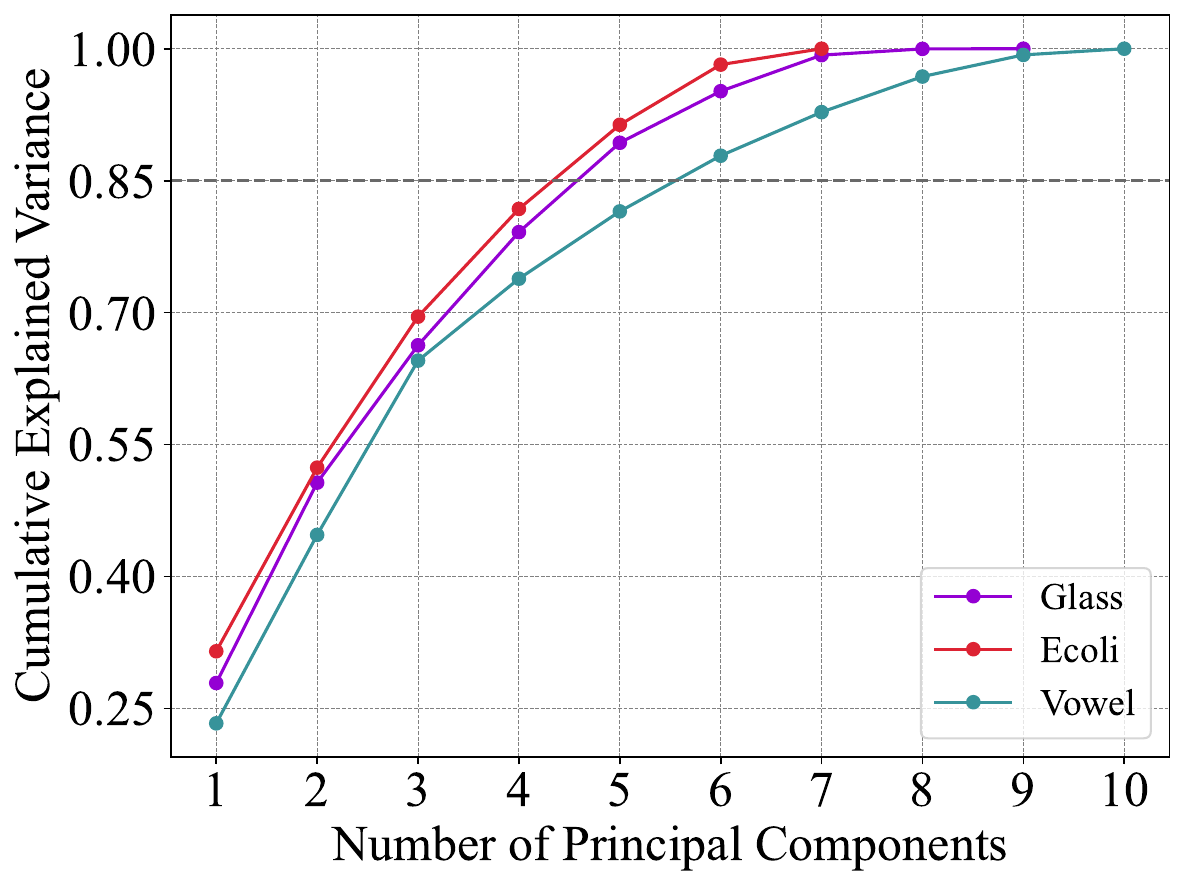}
	\caption{Cumulative explained variance analysis by principal components. The gray dashed line marks the threshold where the cumulative explained variance reaches 85\%.}
	\label{figure_05}
\end{figure}
\begin{figure}[b]
	\centering
	\includegraphics[width=\linewidth]{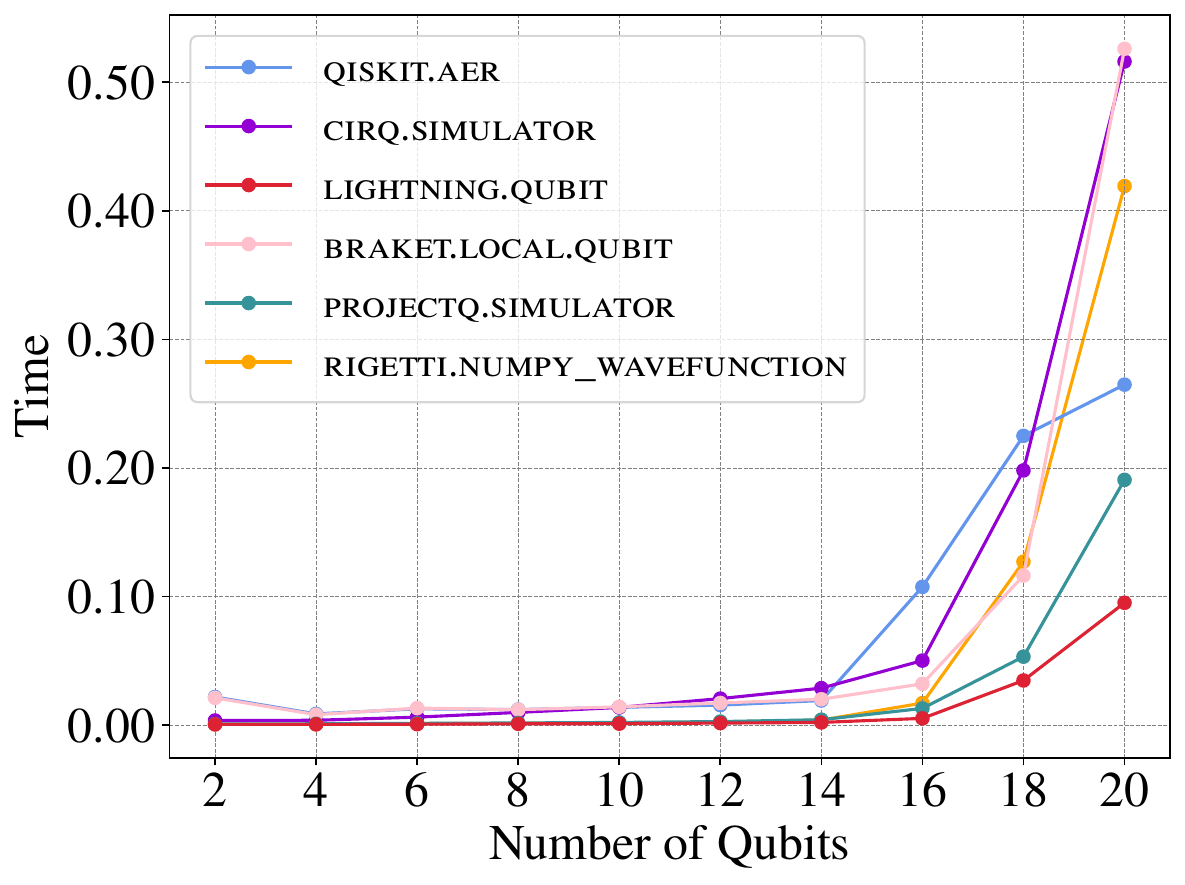}
	\caption{Runtime comparison of quantum simulators across different numbers of qubits, evaluated using expectation values of Pauli-Z operators in quantum kernel computations.}
	\label{figure_06}
\end{figure}
\begin{figure*}[t]
	\centering	
	\subfloat[]{\includegraphics[width=2.321in]{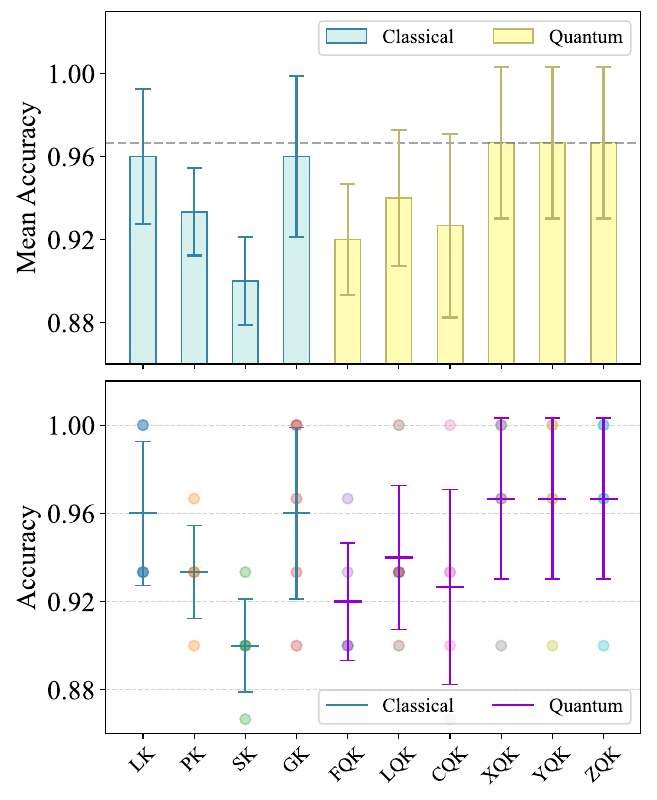}
		\label{figure_7a}}
	\hfill
	\subfloat[]{\includegraphics[width=2.321in]{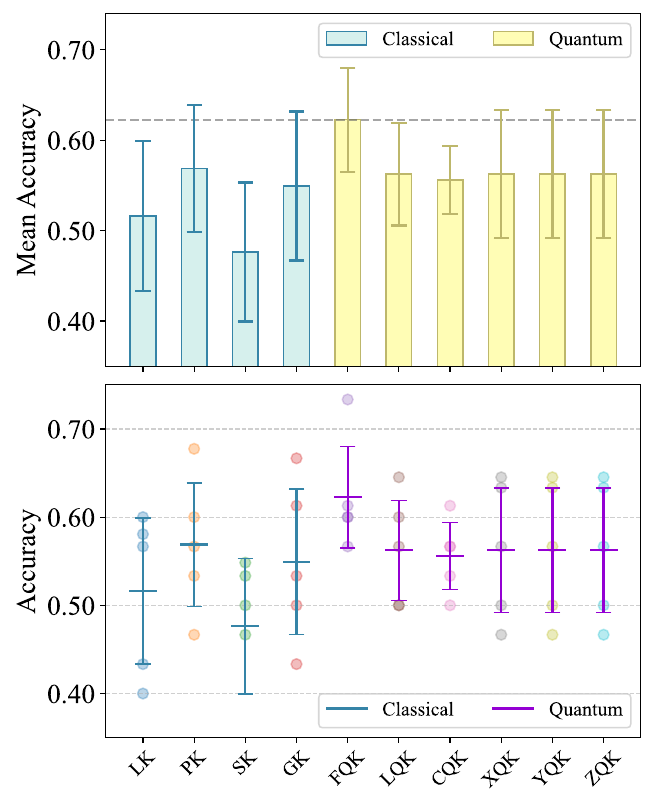}
		\label{figure_7b}}
	\hfill
	\subfloat[]{\includegraphics[width=2.321in]{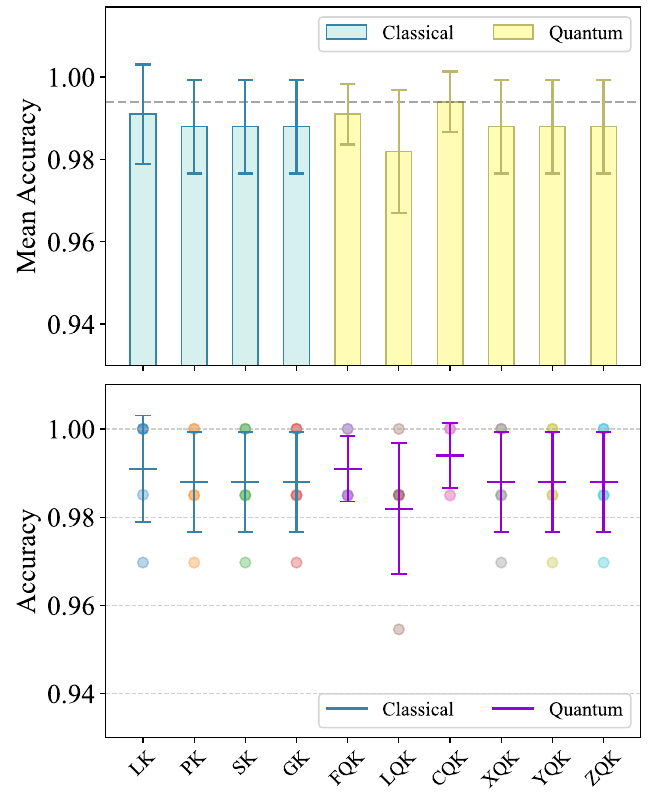}
		\label{figure_7c}}
	\hfill
	\subfloat[]{\includegraphics[width=2.321in]{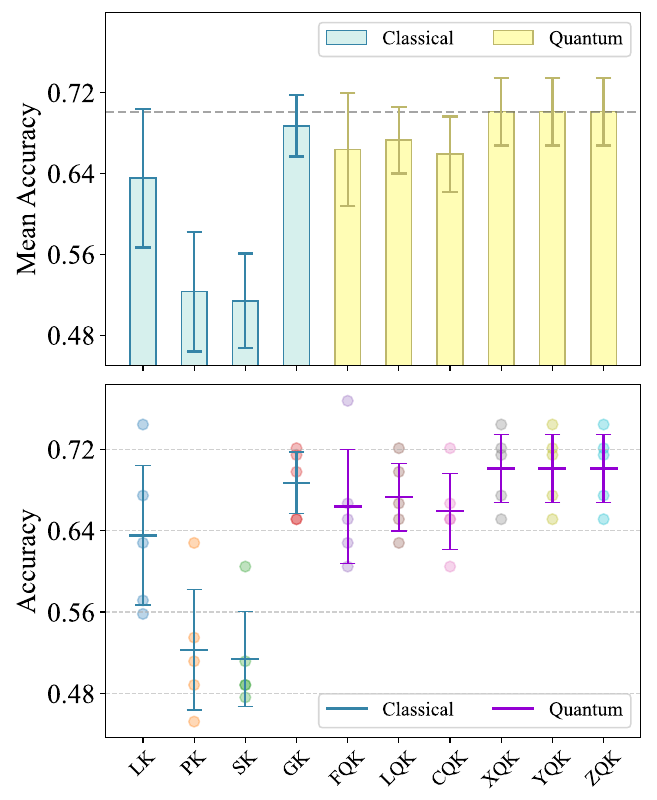}
		\label{figure_7d}}
	\hfill
	\subfloat[]{\includegraphics[width=2.321in]{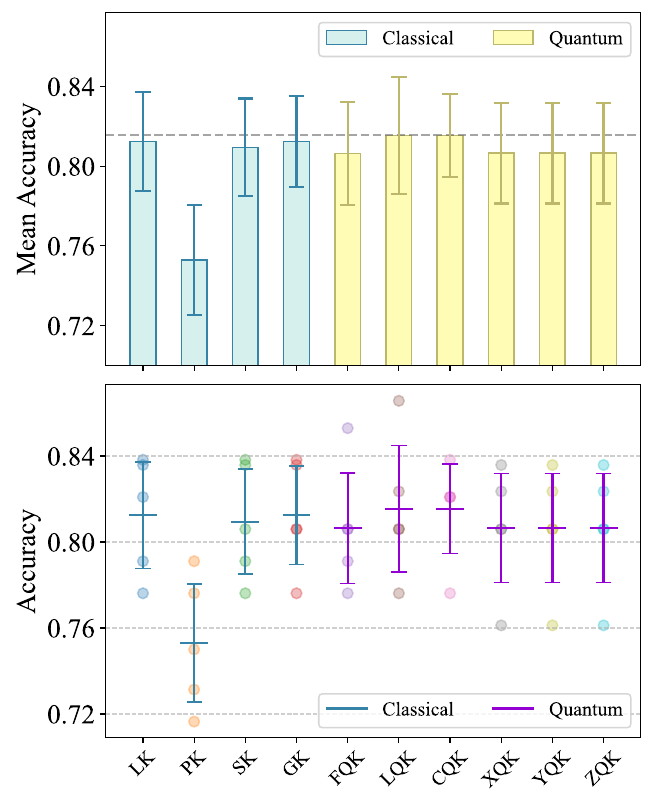}
		\label{figure_7e}}
	\hfill
	\subfloat[]{\includegraphics[width=2.321in]{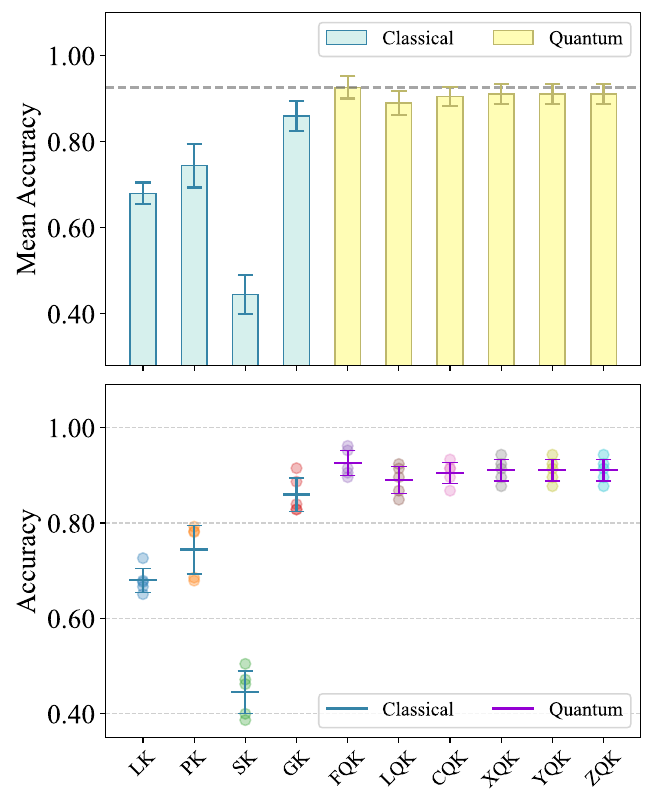}
		\label{figure_7f}}
	\caption{Performance comparison across datasets: (a) Iris, (b) Tae, (c) Penguin, (d) Glass, (e) Ecoli, and (f) Vowel. The bar chart above shows the mean accuracy comparison between classical and quantum kernels, and the scatter plot below illustrates the accuracy distribution for each cross-validation iteration.}
	\vspace{1.5ex}
	\label{figure_07}
\end{figure*}

\begin{table*}[t]
	\centering
	\caption{\label{tab:table4}Performance analysis of the proposed quantum-enhanced multiclass SVMs using optimal kernels across various real-world datasets.}
	\begin{ruledtabular}
		\begin{tabular}{llcccccccccc}
			& &
			\multicolumn{3}{c}{Macroaverage} &
			\multicolumn{3}{c}{Microaverage} &
			\multicolumn{3}{c}{Weighted average} &\\
			\cline{3-5}   \cline{6-8}  \cline{9-11}  
			\rule{0pt}{10pt}Dataset&Kernel function & Precision & Recall & F1 score & Precision & Recall & F1 score & Precision & Recall & F1 score & Accuracy \\  \hline
			\rule{0pt}{10pt}Iris 
			&XQK, YQK, ZQK  &
			1.0000 &
			1.0000 &
			1.0000 &
			1.0000 &
			1.0000 &
			1.0000 &
			1.0000 &
			1.0000 &
			1.0000 &
			1.0000\\
			Tae 
			&FQK &
			0.5926 &
			0.5875 &
			0.5827 &
			0.5870 &
			0.5870 &
			0.5870 &
			0.5919 &
			0.5870 &
			0.5822 &
			0.5870\\
			Penguin 
			&CQK &
			0.9932 &
			0.9815 &
			0.9870 &
			0.9900 &
			0.9900 &
			0.9900 &
			0.9902 &
			0.9900 &
			0.9899 &
			0.9900\\
			Glass& XQK, YQK, ZQK    & 0.7357    & 0.6562 & 0.6796   & 0.7692    & 0.7692 & 0.7692   & 0.7451    & 0.7692 & 0.7406   &0.7692 \\
			Ecoli & LQK & 0.7584    & 0.6925 & 0.7141   & 0.8515    & 0.8515 & 0.8515   & 0.8423  & 0.8515 & 0.8437 &0.8515 \\
			Vowel & FQK & 0.9299    & 0.9331 & 0.9267   & 0.9308    & 0.9308 & 0.9308   & 0.9355  & 0.9308 & 0.9278 &0.9308 \\
		\end{tabular}
	\end{ruledtabular}
\end{table*}

\begin{figure*}[t]
	\centering	
	\subfloat[]{\includegraphics[width=2.321in]{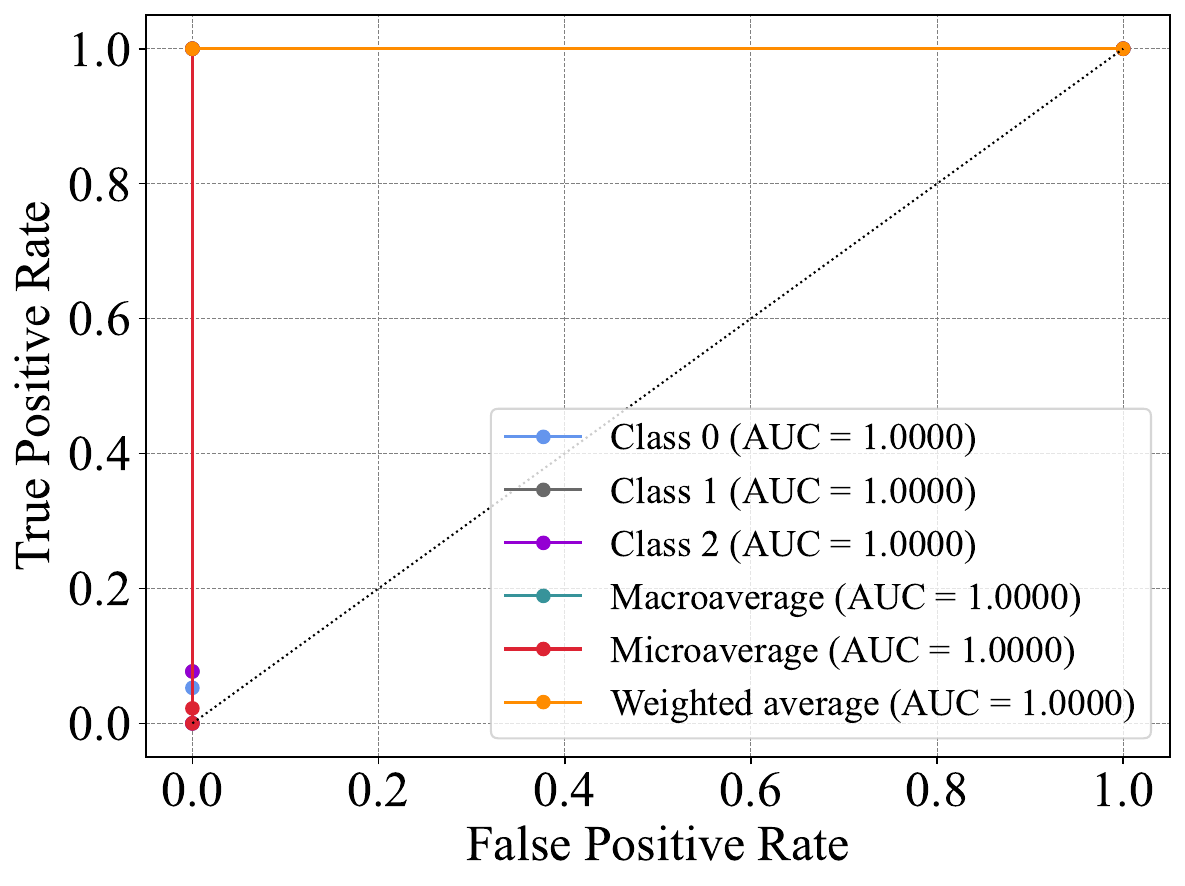}
		\label{figure_8a}}
	\hfill
	\subfloat[]{\includegraphics[width=2.321in]{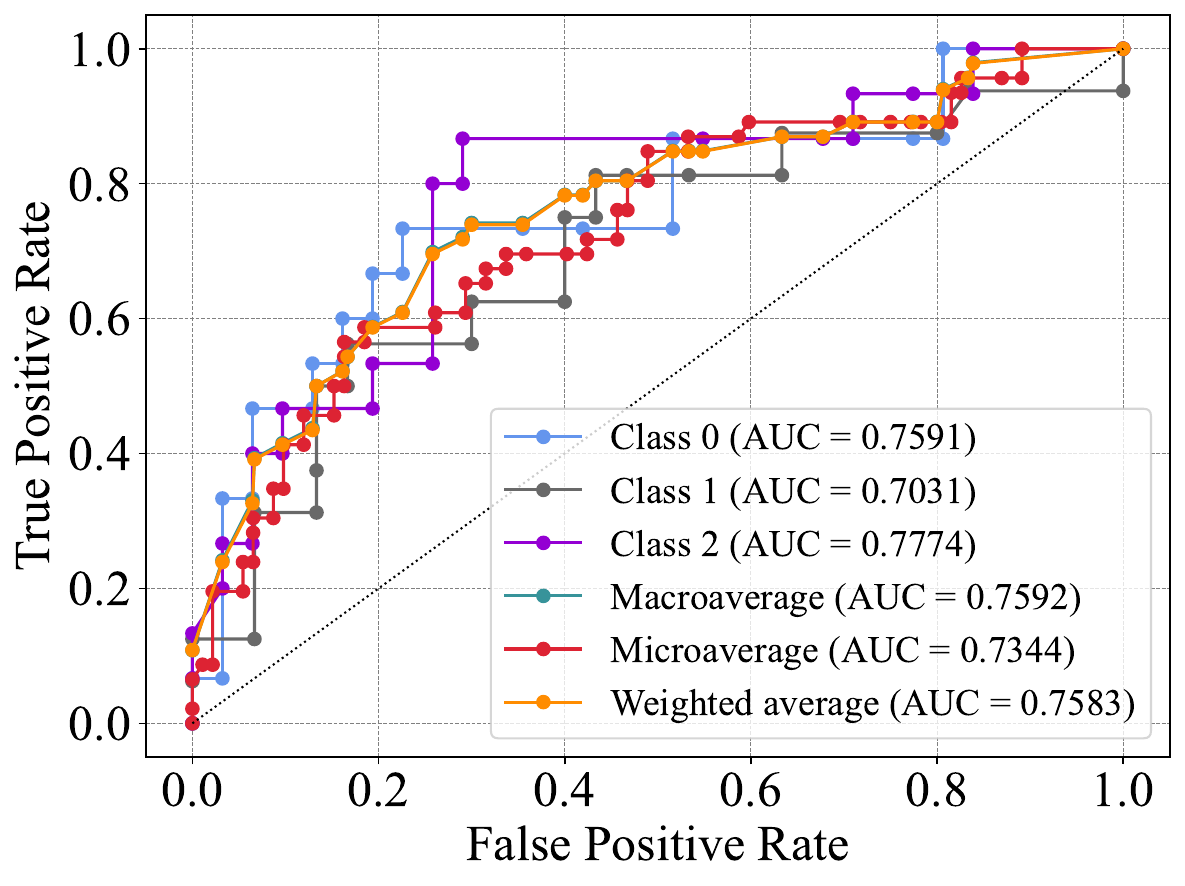}
		\label{figure_8b}}
	\hfill
	\subfloat[]{\includegraphics[width=2.321in]{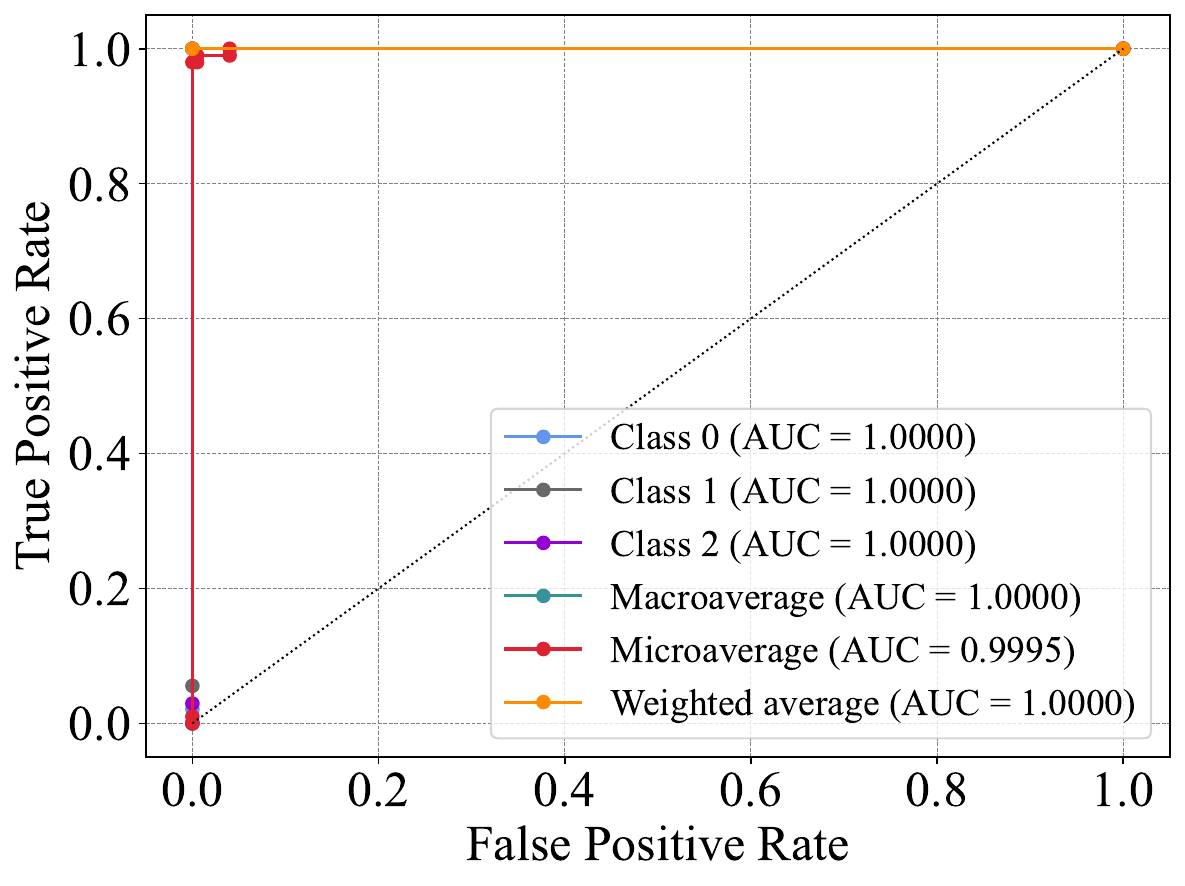}
		\label{figure_8c}}
	\hfill
	\subfloat[]{\includegraphics[width=2.321in]{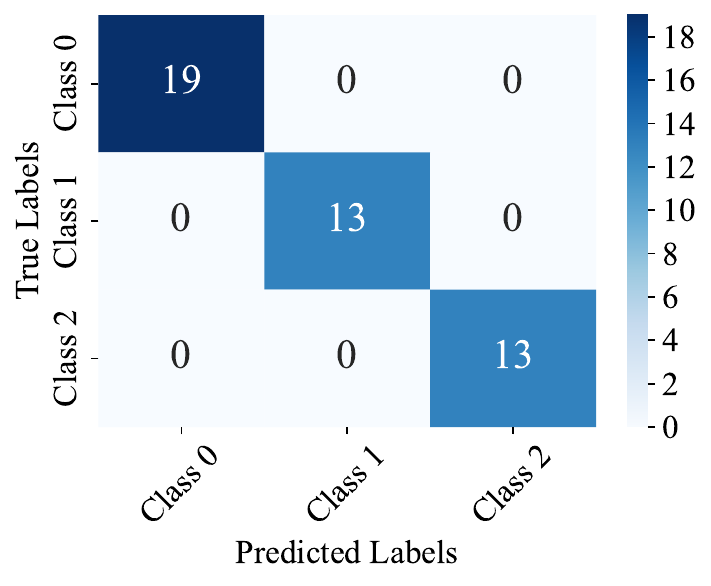}
		\label{figure_8d}}
	\hfill
	\subfloat[]{\includegraphics[width=2.321in]{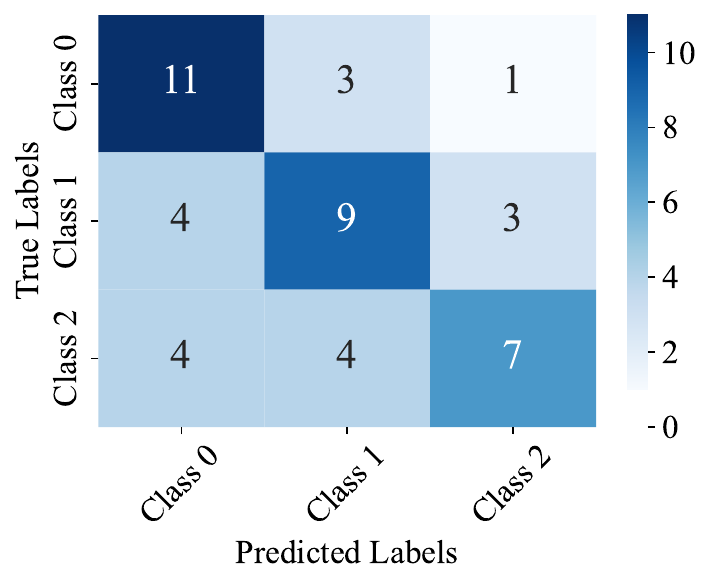}
		\label{figure_8e}}
	\hfill
	\subfloat[]{\includegraphics[width=2.321in]{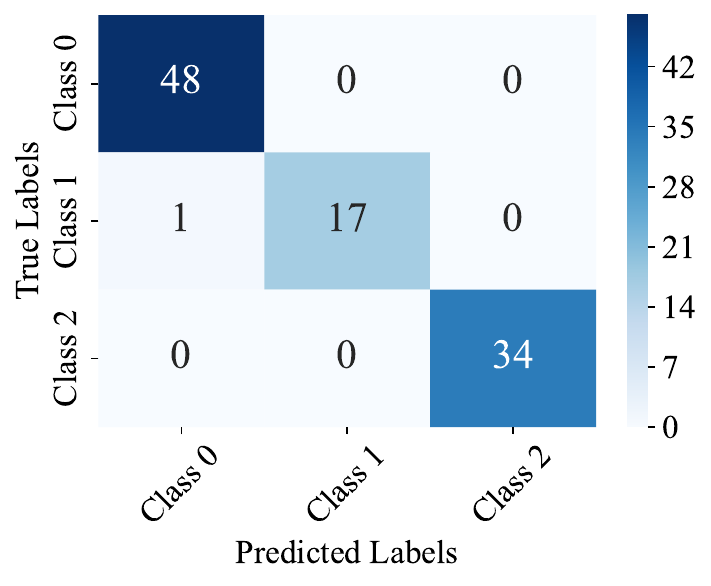}
		\label{figure_8f}}
	\caption{ROC curve and confusion matrix for each dataset: (a) and (d) for the Iris dataset, (b) and (e) for the Tae dataset, and (c) and (f) for the Penguin dataset. The ROC curve illustrates the trade-off between true positive rate (TPR) and false positive rate (FPR) for each class, and the confusion matrix shows classification accuracy, highlighting correct and incorrect predictions among classes.}
	\label{figure_08}
\end{figure*}

\subsection{Experimental datasets and simulation platforms}
In this part, we describe the real-world datasets and simulation platforms; details of the datasets are provided in Table~\ref{tab:table2}. Notably, the Penguin dataset includes $11$ instances with missing values, which have been removed prior to the experiment. Among the six real-world datasets, we categorize the three with fewer features (i.e., Iris, Tae, and Penguin) as low-dimensional datasets, and the three with more features (i.e., Glass, Ecoli, and Vowel) as high-dimensional datasets. We apply z-score normalization \cite{patro2015normalization} to standardize the six real-world datasets during a preprocessing stage. For high-dimensional datasets, we utilize the cumulative explained variance in principal component analysis (PCA) \cite{ringner2008principal,shlens2014tutorial,greenacre2022principal} to select principal components. Analysis of Fig.~\ref{figure_05} indicates that achieving 85\% cumulative explained variance requires $5$ principal components for the Glass and Ecoli datasets, and $6$ principal components for the Vowel dataset. Therefore, features are adjusted from $9$ to $5$, $7$ to $5$, and $10$ to $6$ for the Glass, Ecoli, and Vowel datasets, respectively. 

\newpage
To validate feasibility and efficiency, the proposed quantum algorithm is implemented with a quantum simulator and applied to the six multiclass real-world datasets. Here are some well-known quantum simulators from various providers: \textsc{qiskit.aer} \cite{aleksandrowicz2019qiskit}, \textsc{cirq.simulator} \cite{cirq_developers_2024_11398048}, \textsc{lightning.qubit} \cite{bergholm2018pennylane}, \textsc{braket.local.qubit} \cite{braket2022}, \textsc{projectq.simulator} \cite{steiger2018projectq}, and \textsc{rigetti.numpy\_wavefunction} \cite{pyquil2020}. As shown in Fig.~\ref{figure_06}, the \textsc{lightning.qubit} quantum simulator outperforms other quantum simulators in speed. Hence, all noise-free quantum simulations for training and testing in this paper are performed using the \textsc{lightning.qubit} quantum simulator. Moreover, noisy quantum simulations are conducted with the \textsc{default.mixed} quantum simulator.

\vspace{-1.5ex}
\subsection{Numerical result analysis}
\vspace{-1.5ex}
The numerical results of the proposed quantum algorithm are detailed in this part, which includes two stages. First, we analyze and compare the performance of various kernels across six different multiclass real-world datasets to choose the optimal kernel for each dataset. Second, we assess the performance of the quantum algorithm, using the optimal kernel for each dataset, based on a family of performance metrics. The performance metrics are detailed in Appendix~\ref{appendixa}.
\newpage
At the first stage, we employ a stratified k-fold cross-validation method \cite{bhagat2022implementation, prusty2022skcv, mahesh2023stratified} to evaluate the proposed quantum algorithm. For each dataset, we divide it into $5$ folds, ensuring that the class label proportions in each fold are consistent with those in the entire dataset. Based on the comparative analysis in Fig.~\ref{figure_07}, we can obtain the optimal kernel for each dataset. For the Iris and Glass datasets, the Pauli-X, Pauli-Y, and Pauli-Z quantum kernels perform equally well, with the proof of their equivalence provided in the Appendix~\ref{appendixb}. For the Tae and Vowel datasets, the full quantum kernel is optimal. The Penguin dataset benefits most from the circular quantum kernel, and the Ecoli dataset is best served by the linear quantum kernel. Therefore, the optimal quantum kernel is significantly contingent upon the distribution and structure of real-world datasets. In addition, the proposed quantum algorithm with the optimal quantum kernel exhibits superior classification performance and strong generalization ability compared to its classical counterparts across all six real-world datasets.\\
\begin{figure}[t]
	\centering
	\subfloat[]{\includegraphics[width=0.9\linewidth]{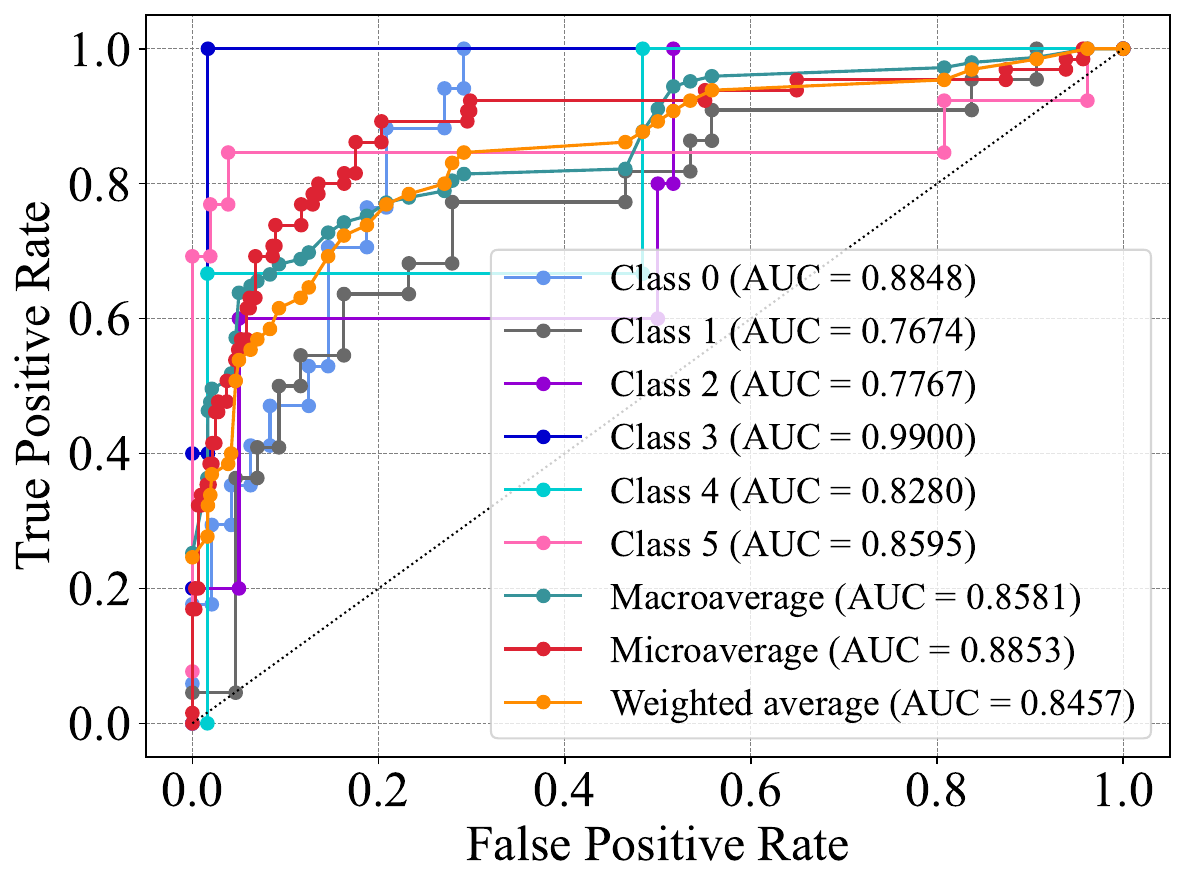}
		\label{figure_9a}}
	\hfill
	\subfloat[]{\includegraphics[width=0.8\linewidth]{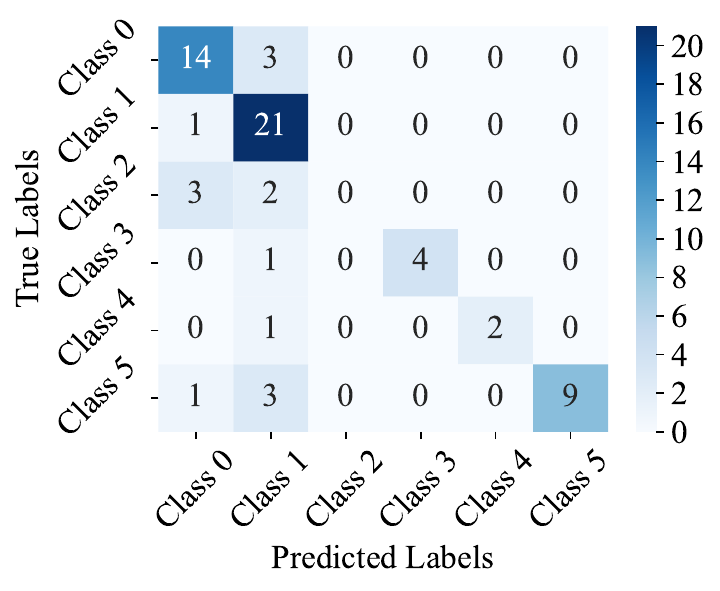}
		\label{figure_9b}}
	\caption{ROC curve (a) and confusion matrix (b) for the Glass dataset.}
	\label{figure_09}
\end{figure}
\begin{figure}[t]
	\centering
	\subfloat[]{\includegraphics[width=0.9\linewidth]{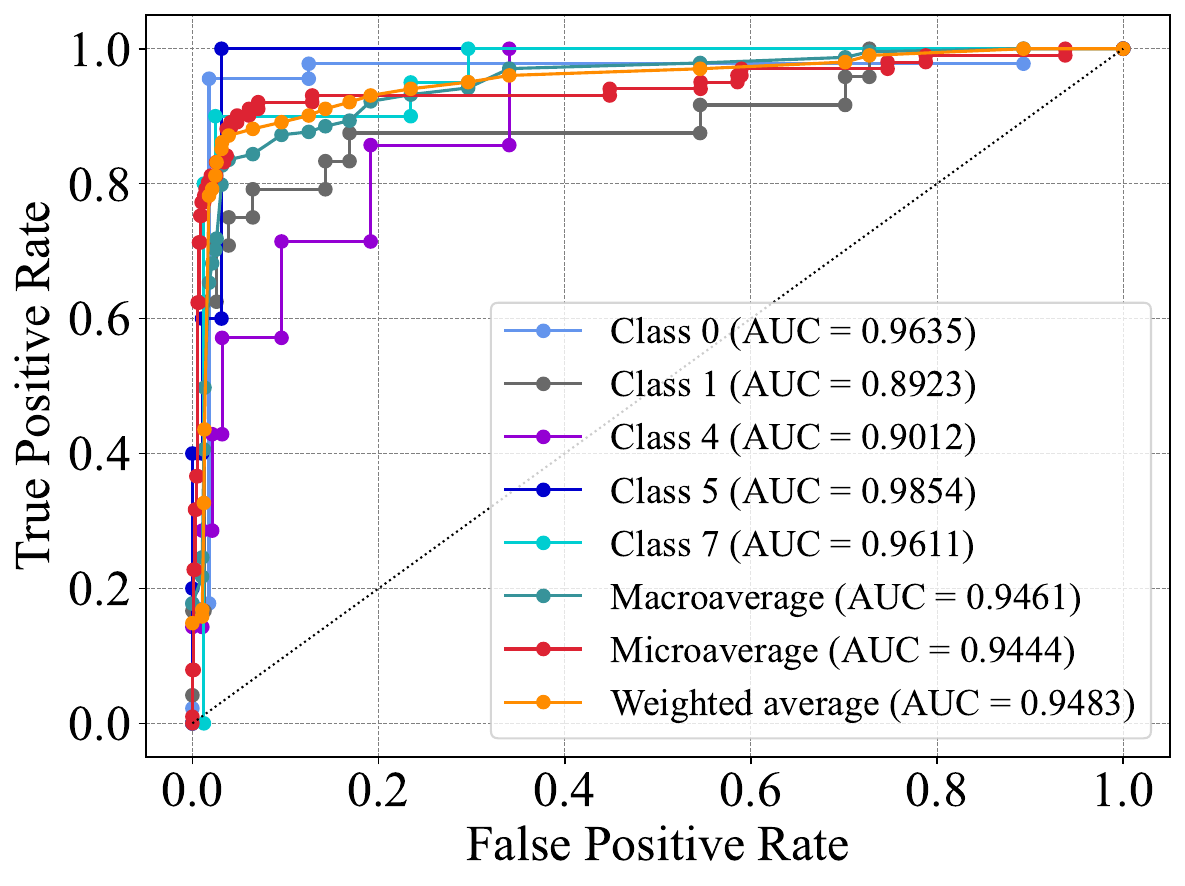}
		\label{figure_10a}}
	\hfill
	\subfloat[]{\includegraphics[width=0.8\linewidth]{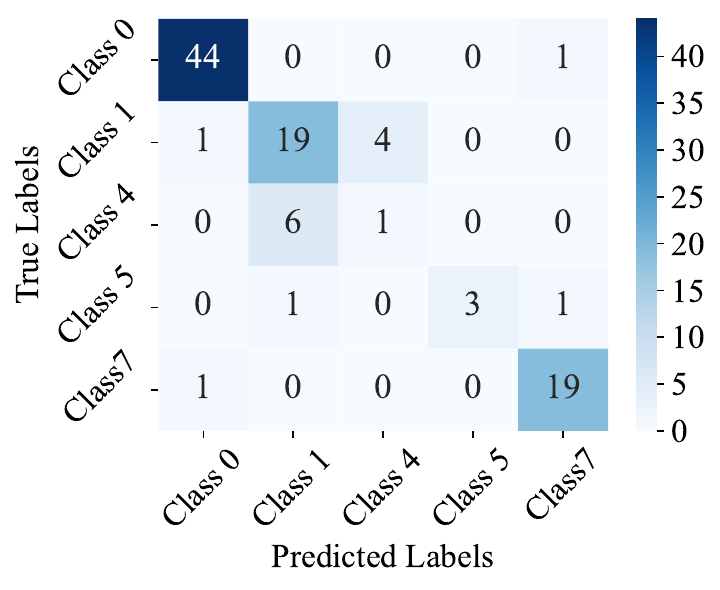}
		\label{figure_10b}}
	\caption{ROC curve (a) and confusion matrix (b) for the Ecoli dataset.}
	\label{figure_10}
\end{figure}
\begin{figure}[t]
	\centering
	\includegraphics[width=\linewidth]{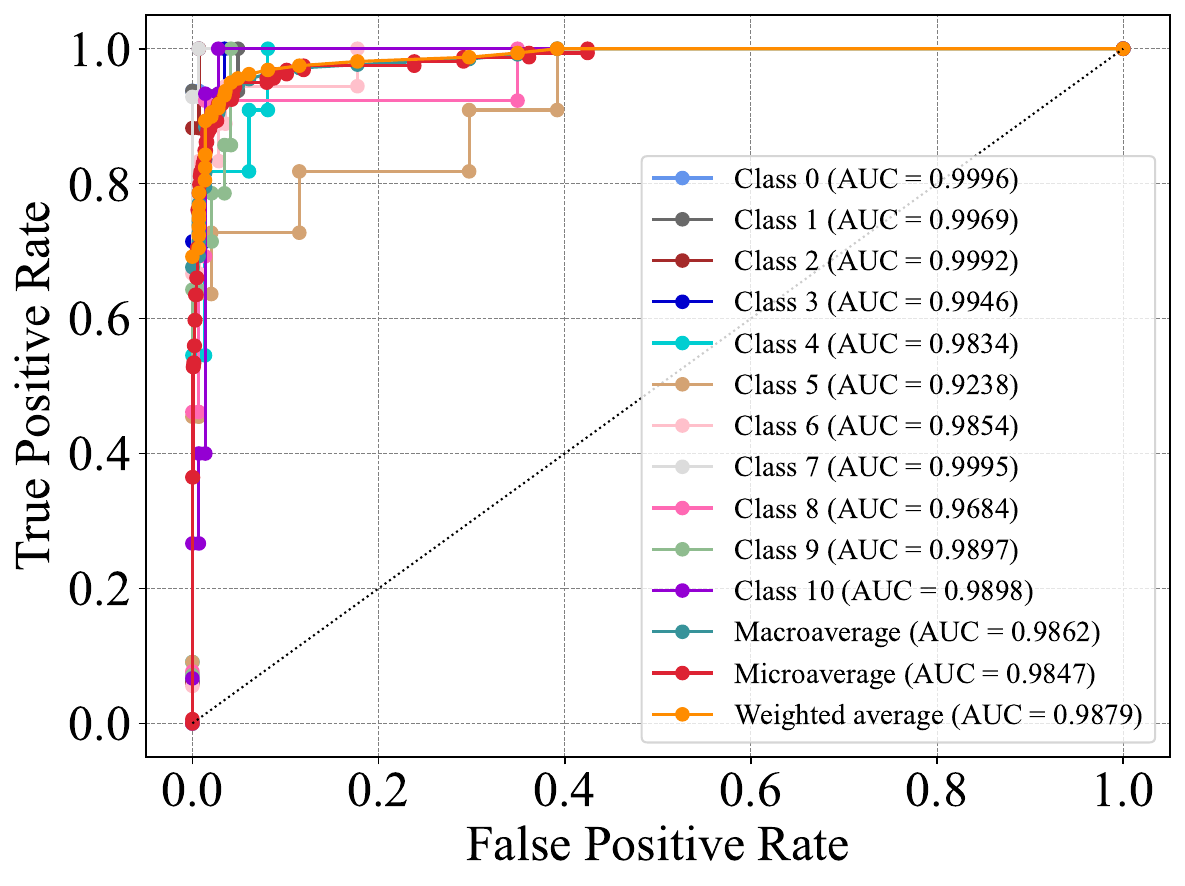}
	\caption{\label{figure_11}ROC curve for Vowel dataset. The AUCs all exceed $0.9000$.}
\end{figure}
\begin{figure}[t]
	\centering
	\includegraphics[width=\linewidth]{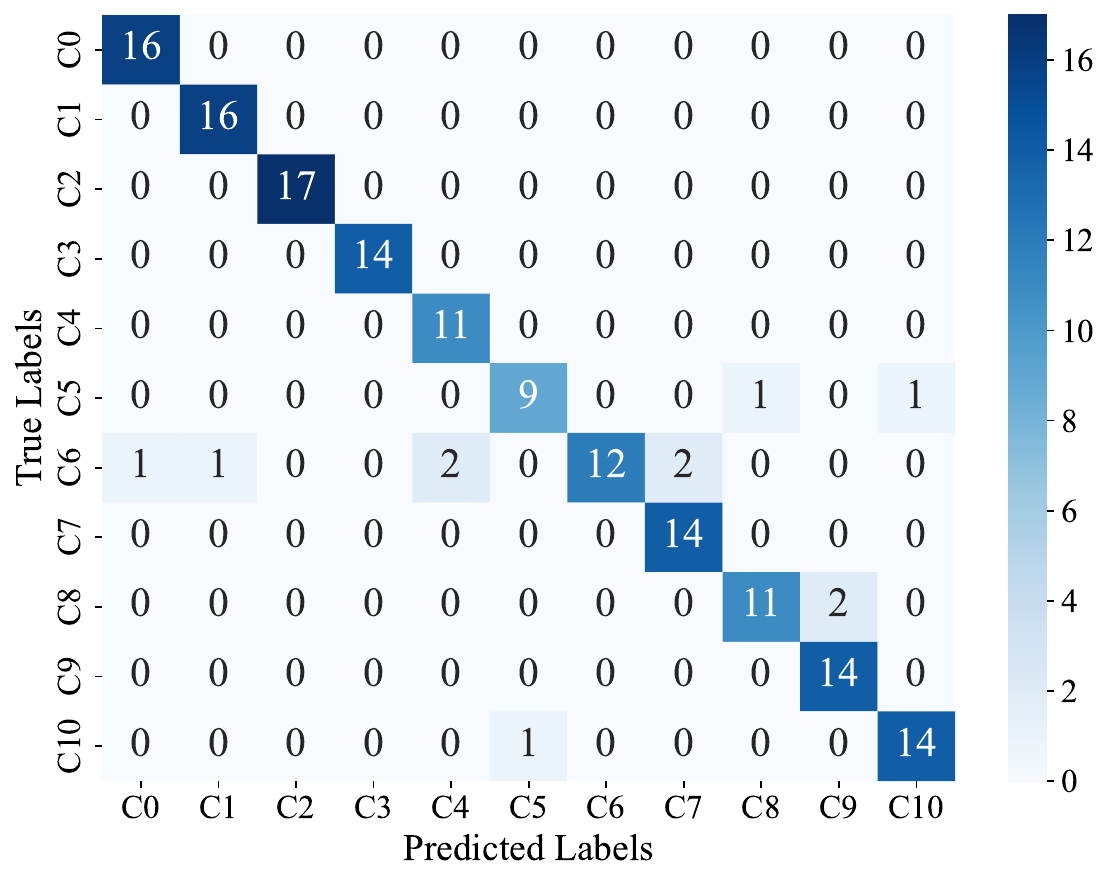}
	\caption{Confusion matrix for Vowel dataset. C0--C10, Class $0$--$10$.}
	\vspace{-1ex}
	\label{figure_12}
\end{figure}
\begin{figure}[t]
	\centering
	\includegraphics[width=\linewidth]{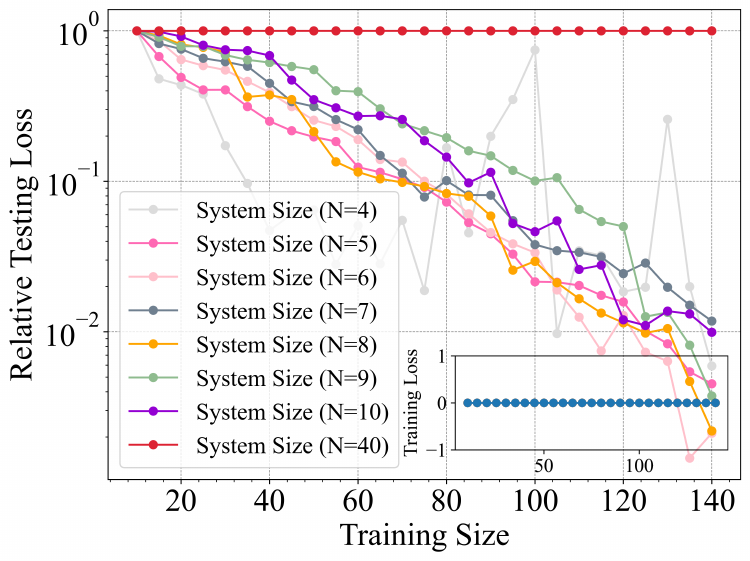}
	\caption{Effect of exponential concentration on generalization performance. In alignment with the parameters in Ref.~\cite{thanasilp2024exponential}, we adopt the following settings: $150$ data points, $1000$ measurement shots, and $20$ testing data points. The random seed is set to $15$. Each data point is randomly generated in the interval $[0,2 \pi)$. The main plot illustrates how the relative testing loss changes with increasing training size, while the inset shows the variation in absolute training loss as the training size increases.}
	\label{figure_13}
\end{figure}
In the following stage, we split each dataset into a 70:30 training and testing set, as shown in Table~\ref{tab:table3}. We then train and test the proposed quantum algorithm with optimal kernels. Analysis of Table~\ref{tab:table4} reveals that the accuracy values are consistent with the precision, recall, and F1 score values computed using microaverage methods. Moreover, they also match the recall values derived from weighted average methods. Therefore, these five performance metrics serve the same function in evaluating the quantum algorithm. Further analysis indicates that the quantum algorithm achieves satisfactory classification accuracy, along with robust precision, recall, and F1 score across macroaverage, microaverage, and weighted average evaluations. Specifically, the quantum algorithm achieves a perfect classification accuracy of $100\%$ for the Iris dataset and an impressive $99.00\%$ accuracy for the Penguin dataset. It also demonstrates strong performance on the Vowel dataset with an accuracy of $93.08\%$ and the Ecoli dataset with $85.15\%$. For the Glass dataset, it achieves an accuracy of $76.92\%$, while for the Tae dataset, it attains a satisfactory accuracy of $58.70\%$.

To evaluate the proposed quantum algorithm with optimal quantum kernels, we also employ receiver operating characteristic (ROC) curves and confusion matrices. Drawing from a comprehensive analysis of Figs.~\ref{figure_08}, \ref{figure_09}, \ref{figure_10}, \ref{figure_11}, and \ref{figure_12}, we evaluate the classification performance of quantum algorithms employing optimal quantum kernels across various classes in each dataset, as well as the overall performance using macroaverage, microaverage, and weighted average methods. For the Iris dataset, the area under the ROC curve (AUC) for each class, as well as the macroaverage, microaverage, and weighted average AUCs, are all $1.0000$, demonstrating that the quantum algorithm achieves perfect classification. For the Tae dataset, Class $2$ exhibits the strongest classification performance, while Class $1$ is relatively weaker; the macroaverage, microaverage, and weighted average AUCs confirm robust overall performance. For the Penguin dataset, the AUC for each class, as well as the macroaverage, microaverage, and weighted average AUCs, are either $1.0000$ or close to $1.0000$, indicating near-perfect classification performance. For the Glass dataset, Class $3$ shows the highest performance, while Class $1$ is the lowest; the macroaverage, microaverage, and weighted average AUCs reflect solid overall performance. For the Ecoli dataset, Class $5$ achieves the best classification performance, while Class $1$ is the worst; the macroaverage, microaverage, and weighted average AUCs demonstrate strong overall performance. Finally, for the Vowel dataset, Class $0$ exhibits the strongest classification performance, while Class $5$ is the weakest; the macroaverage, microaverage, and weighted average AUCs indicate near-optimal overall performance. Interestingly, as illustrated in Fig.~\ref{figure_10}, the quantum algorithm classifies only five classes in the Ecoli dataset, despite it containing eight classes. This is due to a class imbalance in the dataset, with Classes $2$, $3$, and $6$ having insufficient instances, resulting in their exclusion from prediction.

\begin{table*}[t]
	\centering
	\caption{\label{tab:table5}Examples of representing noisy quantum kernel functions for input feature vectors $\vec{x}_i$ and $\vec{x}_j$.}
	\begin{ruledtabular}
		\begin{tabular}{lcc}
			Name&Kernel function& Hyperparameters \\
			\hline
			\rule{0pt}{10pt}Noisy full quantum kernel (NFQK)  & $\kappa^{\prime}(\vec{x}_{i},\vec{x}_{j}) = \operatorname{Tr}[\left(|0\rangle\langle 0|\right)^{\otimes \mathrm{N}} (\mathcal{N}_{\bar{p}} \circ {U}^{\dag}_\textit{{f}}\left(\vec{x}_j\right) \circ \mathcal{N}_{\bar{p}} \circ {U}_\textit{{f}}\left(\vec{x}_i\right))\left(|0\rangle\langle 0|\right)^{\otimes \mathrm{N}}]$ & $\bar{p} \in [ 0,1]$ \\
			Noisy linear quantum kernel (NLQK)  & $\kappa^{\prime}(\vec{x}_{i},\vec{x}_{j}) = \operatorname{Tr}[\left(|0\rangle\langle 0|\right)^{\otimes \mathrm{N}} (\mathcal{N}_{\bar{p}} \circ {U}^{\dag}_\textit{{l}}\left(\vec{x}_j\right) \circ \mathcal{N}_{\bar{p}} \circ {U}_\textit{{l}}\left(\vec{x}_i\right))\left(|0\rangle\langle 0|\right)^{\otimes \mathrm{N}}]$ & $\bar{p} \in [ 0,1]$ \\
			Noisy circular quantum kernel (NCQK)  & $\kappa^{\prime}(\vec{x}_{i},\vec{x}_{j}) = \operatorname{Tr}[\left(|0\rangle\langle 0|\right)^{\otimes \mathrm{N}} (\mathcal{N}_{\bar{p}} \circ {U}^{\dag}_\textit{{c}}\left(\vec{x}_j\right) \circ \mathcal{N}_{\bar{p}} \circ {U}_\textit{{c}}\left(\vec{x}_i\right))\left(|0\rangle\langle 0|\right)^{\otimes \mathrm{N}}]$ & $\bar{p} \in [ 0,1]$\\
			Noisy Pauli-X quantum kernel (NXQK)  & $\kappa^{\prime}(\vec{x}_{i},\vec{x}_{j}) = \operatorname{Tr}[\left(|0\rangle\langle 0|\right)^{\otimes \mathrm{N}} (\mathcal{N}_{\bar{p}} \circ {U}^{\dag}_\textit{{x}}\left(\vec{x}_j\right) \circ \mathcal{N}_{\bar{p}} \circ {U}_\textit{{x}}\left(\vec{x}_i\right))\left(|0\rangle\langle 0|\right)^{\otimes \mathrm{N}}]$ & $\bar{p} \in [ 0,1]$ \\
			Noisy Pauli-Y quantum kernel (NYQK)  & $\kappa^{\prime}(\vec{x}_{i},\vec{x}_{j}) = \operatorname{Tr}[\left(|0\rangle\langle 0|\right)^{\otimes \mathrm{N}} (\mathcal{N}_{\bar{p}} \circ {U}^{\dag}_\textit{{y}}\left(\vec{x}_j\right) \circ \mathcal{N}_{\bar{p}} \circ {U}_\textit{{y}}\left(\vec{x}_i\right))\left(|0\rangle\langle 0|\right)^{\otimes \mathrm{N}}]$ & $\bar{p} \in [ 0,1]$ \\
			Noisy Pauli-Z quantum kernel (NZQK)  & $\kappa^{\prime}(\vec{x}_{i},\vec{x}_{j}) = \operatorname{Tr}[\left(|0\rangle\langle 0|\right)^{\otimes \mathrm{N}} (\mathcal{N}_{\bar{p}} \circ {U}^{\dag}_\textit{{z}}\left(\vec{x}_j\right) \circ \mathcal{N}_{\bar{p}} \circ {U}_\textit{{z}}\left(\vec{x}_i\right))\left(|0\rangle\langle 0|\right)^{\otimes \mathrm{N}}]$ & $\bar{p} \in [ 0,1]$ \\	
		\end{tabular}
	\end{ruledtabular}
\end{table*}

\section{In-depth performance analysis}\label{limitation}
Up to now, we only consider the ideal, noise-free scenarios, focusing on the evaluation of parameterized quantum circuits for quantum kernels under perfect scenarios. In fact, when implementing quantum kernels on quantum devices, we also need to consider the effects of exponential concentration and hardware noise on the quantum kernel. In this section, we first evaluate the effect of exponential concentration on the performance of the proposed quantum algorithm, followed by an assessment of the effect of hardware noise.

\begin{figure*}[t]
	\centering	
	\subfloat[]{\includegraphics[width=2.321in]{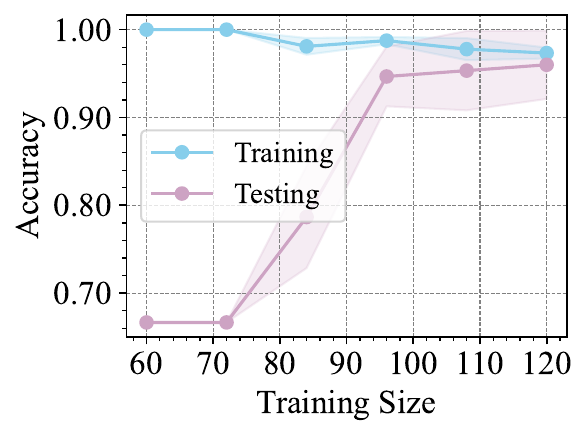}
		\label{figure_14a}}
	\hfill
	\subfloat[]{\includegraphics[width=2.321in]{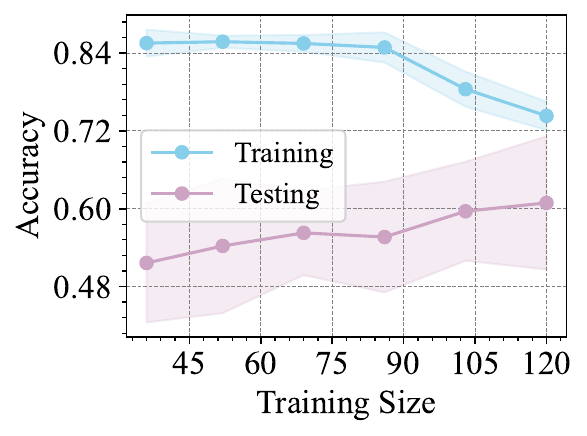}
		\label{figure_14b}}
	\hfill
	\subfloat[]{\includegraphics[width=2.321in]{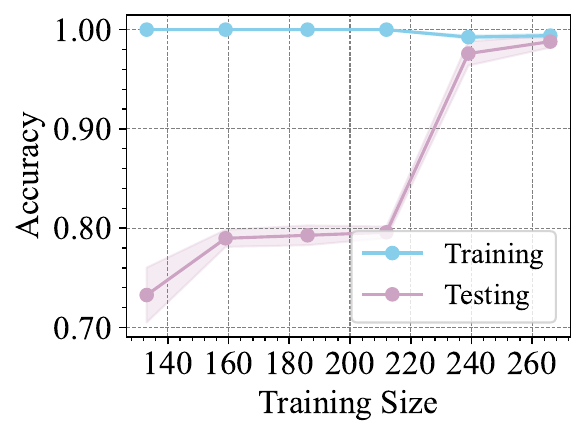}
		\label{figure_14c}}
	\hfill
	\subfloat[]{\includegraphics[width=2.321in]{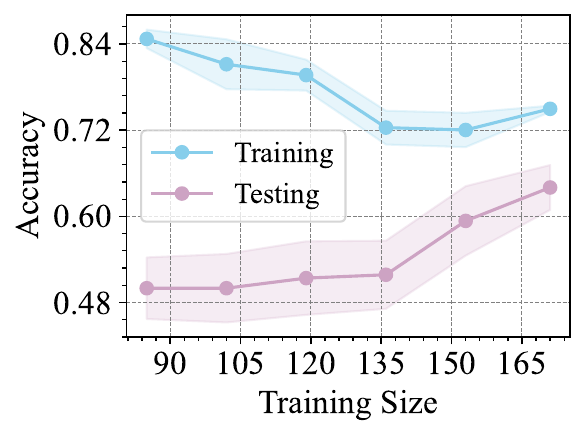}
		\label{figure_14d}}
	\hfill
	\subfloat[]{\includegraphics[width=2.321in]{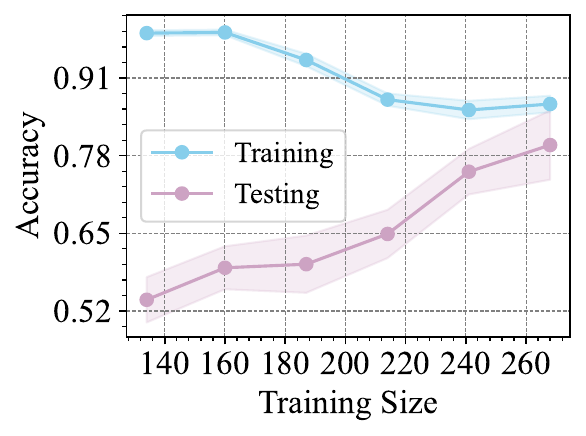}
		\label{figure_14e}}
	\hfill
	\subfloat[]{\includegraphics[width=2.321in]{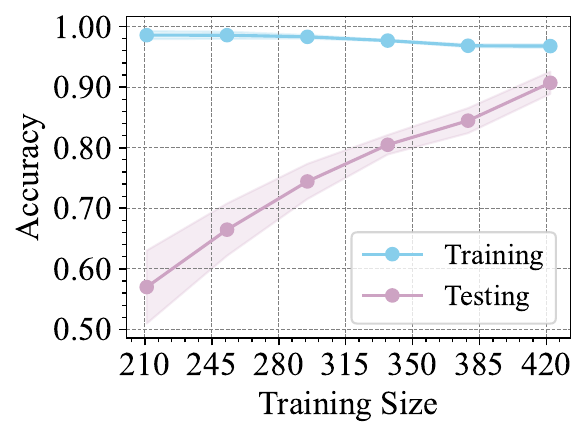}
		\label{figure_14f}}
	\caption{Learning curves for the proposed quantum algorithm utilizing optimal quantum kernels on the following datasets: (a) Iris, (b) Tae, (c) Penguin, (d) Glass, (e) Ecoli, and (f) Vowel. The learning curve of training illustrates how the training accuracy evolves as the training size increases, and the learning curve of testing depicts the variation in testing accuracy with the growing training size. The blue and purple shaded areas represent the variability in training and testing accuracy, respectively. The generalization performance of the quantum algorithm is assessed on six real-world datasets using a stratified $5$-fold cross-validation method.}
	\label{figure_14}
\end{figure*}

\subsection{Effect of exponential concentration}
Exponential concentration in quantum kernel methods captures the effect where, as the problem size grows, the kernel values between data points mapped by quantum feature mapping progressively approach a fixed constant \cite{thanasilp2024exponential}. Due to the inherent uncertainty of measurements on quantum devices, we must conduct repeated measurements to ensure an accurate estimation of the quantum kernel. Therefore, more measurement shots are essential to distinguish between different kernel values. However, under the constraints of a polynomial number of measurement shots, exponential concentration reduces the distinguishability of kernel values, preventing the proposed quantum algorithm from effectively capturing the patterns of the input data. That is, the trained quantum algorithm may make incorrect predictions on unseen inputs, resulting in poor generalization performance.\\
\begin{figure}[t]
	\vspace{1ex}
	\centering
	\includegraphics[width=\linewidth]{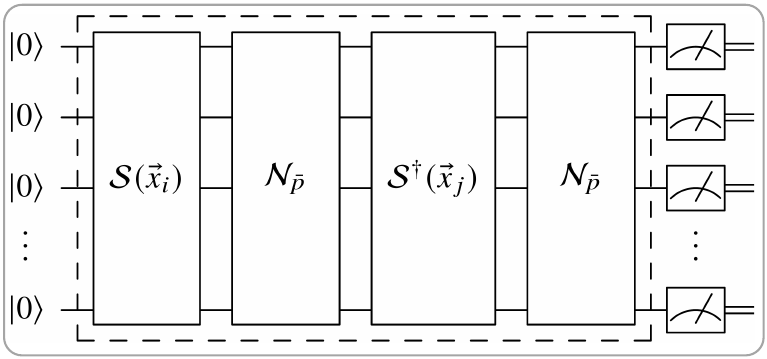}
	\caption{\label{figure_15}Parameterized quantum circuit to generate noisy quantum kernels for feature vectors $\vec{x}_i$ and $\vec{x}_j$. The implementation of the circuit consists of the following steps: first, applying the unitary operation $\mathcal{S}(\vec{x}_i)$, followed by a depolarizing noise channel $\mathcal{N}_{\bar{p}}$. Then, the inverse operation $\mathcal{S}^{\dag}(\vec{x}_j)$ is performed, followed again by a depolarizing noise channel. Finally, a projector $\Pi$ is employed to measure the resulting quantum state.}
\end{figure}
\begin{table*}[t]
	\centering
	\caption{\label{tab:table6}Performance analysis of the proposed quantum algorithm across various real-world datasets under depolarizing noise.}
	\begin{ruledtabular}
		\begin{tabular}{llcccccccccc}
			& &
			\multicolumn{3}{c}{Macroaverage} &
			\multicolumn{3}{c}{Microaverage} &
			\multicolumn{3}{c}{Weighted average} &\\
			\cline{3-5}   \cline{6-8}  \cline{9-11}  
			\rule{0pt}{10pt}Dataset&Kernel function & Precision & Recall & F1 score & Precision & Recall & F1 score & Precision & Recall & F1 score & Accuracy \\  \hline
			\rule{0pt}{10pt}Iris 
			&NXQK, NYQK, NZQK  &
			1.0000 &
			1.0000 &
			1.0000 &
			1.0000 &
			1.0000 &
			1.0000 &
			1.0000 &
			1.0000 &
			1.0000 &
			1.0000\\
			Tae 
			&NFQK &
			0.6380 &
			0.6319 &
			0.6295 &
			0.6304 &
			0.6304 &
			0.6304 &
			0.6392 &
			0.6304 &
			0.6293 &
			0.6304\\
			Penguin 
			&NCQK &
			0.9932 &
			0.9815 &
			0.9870 &
			0.9900 &
			0.9900 &
			0.9900 &
			0.9902 &
			0.9900 &
			0.9899 &
			0.9900\\
			Glass& NXQK, NYQK, NZQK    & 0.7289    & 0.5895 & 0.6218   & 0.7385    & 0.7385 & 0.7385   & 0.7312    & 0.7385 & 0.7064   &0.7385 \\
			Ecoli & NLQK & 0.7515    & 0.6841 & 0.7074   & 0.8416    & 0.8416 & 0.8416   & 0.8355  & 0.8416 & 0.8357 &0.8416 \\
			Vowel & NFQK & 0.8676    & 0.8676 & 0.8536   & 0.8616    & 0.8616 & 0.8616   & 0.8779  & 0.8616 & 0.8535 &0.8616 \\
		\end{tabular}
	\end{ruledtabular}
\end{table*}
\begin{figure*}[t]
	\centering	
	\subfloat[]{\includegraphics[width=2.321in]{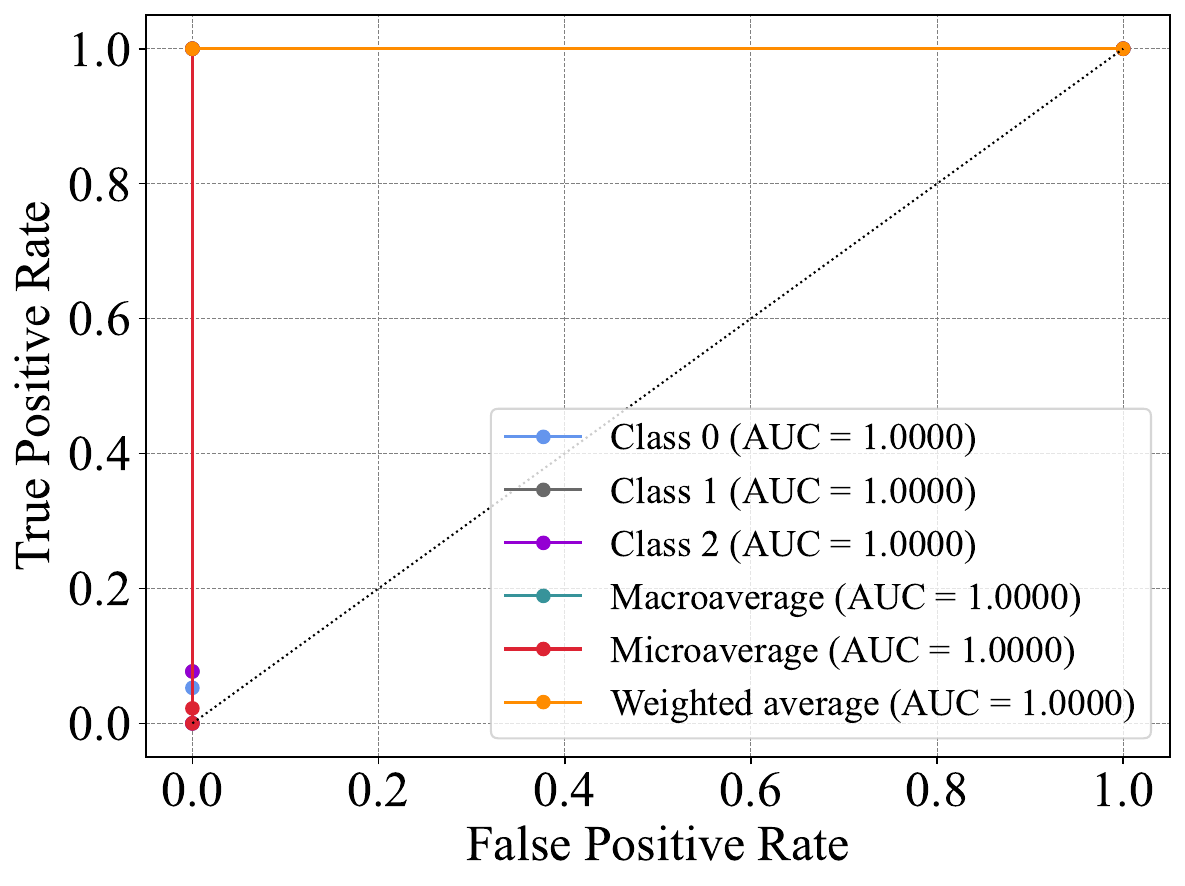}
		\label{figure_16a}}
	\hfill
	\subfloat[]{\includegraphics[width=2.321in]{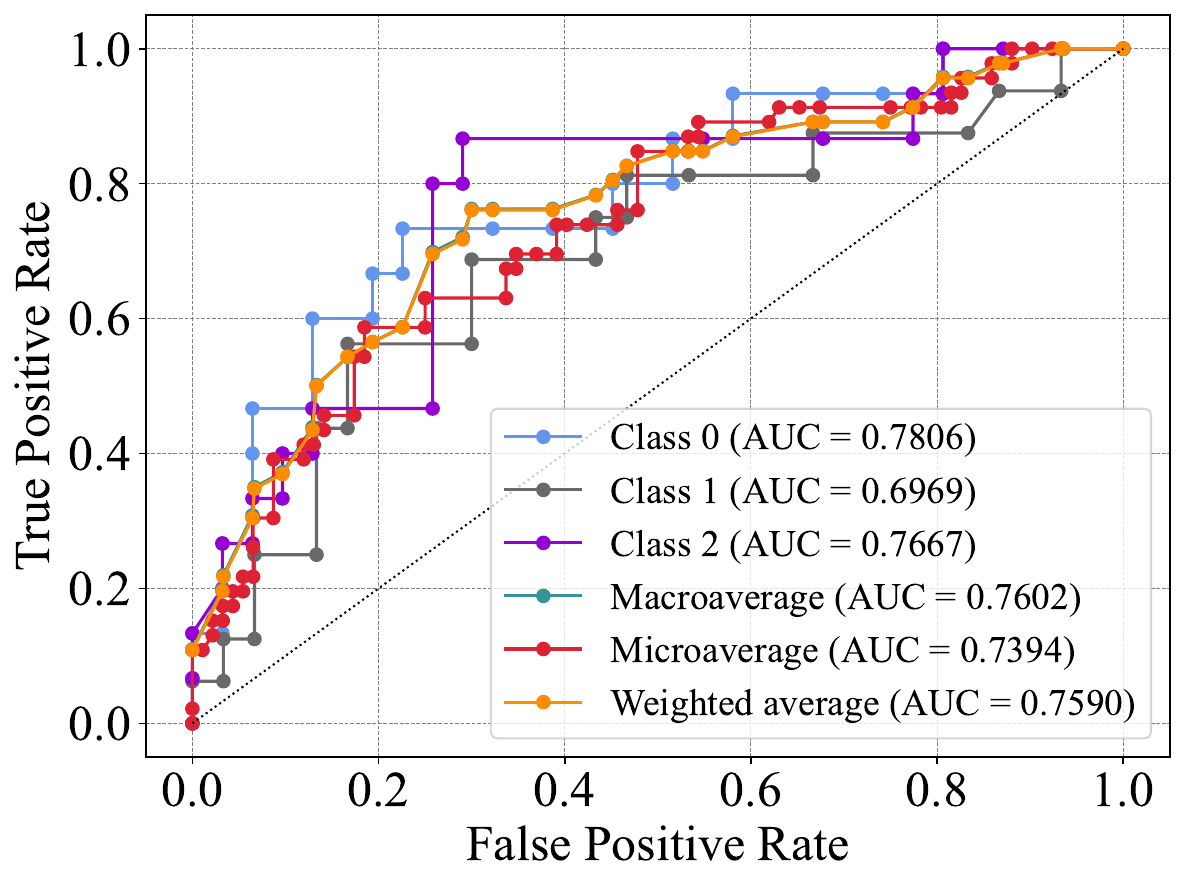}
		\label{figure_16b}}
	\hfill
	\subfloat[]{\includegraphics[width=2.321in]{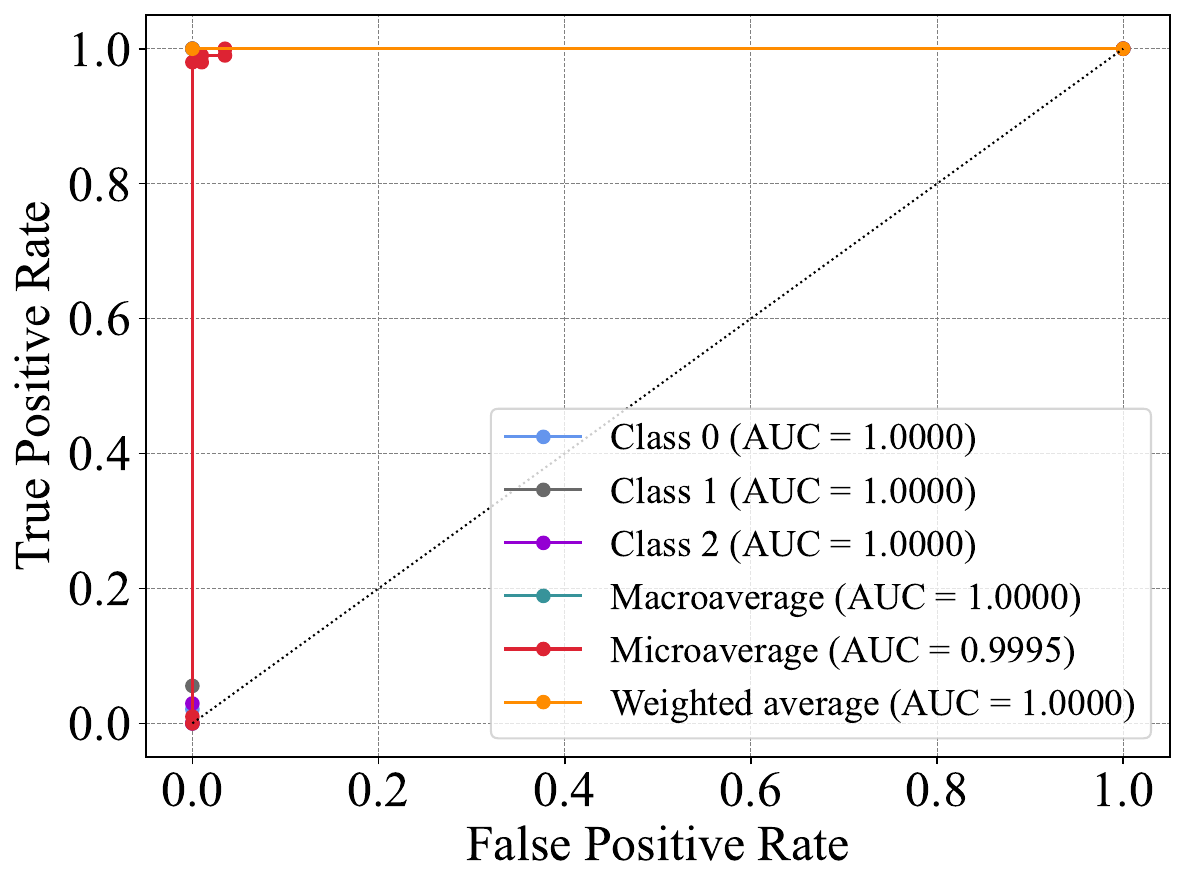}
		\label{figure_16c}}
	\hfill
	\subfloat[]{\includegraphics[width=2.321in]{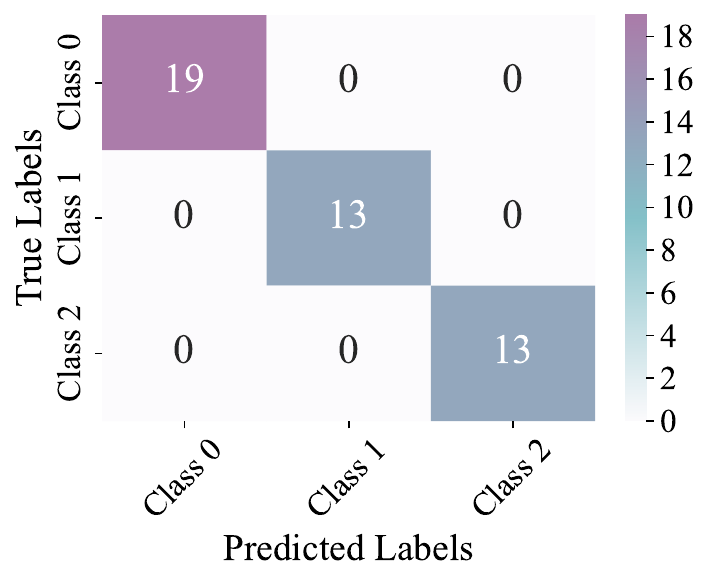}
		\label{figure_16d}}
	\hfill
	\subfloat[]{\includegraphics[width=2.321in]{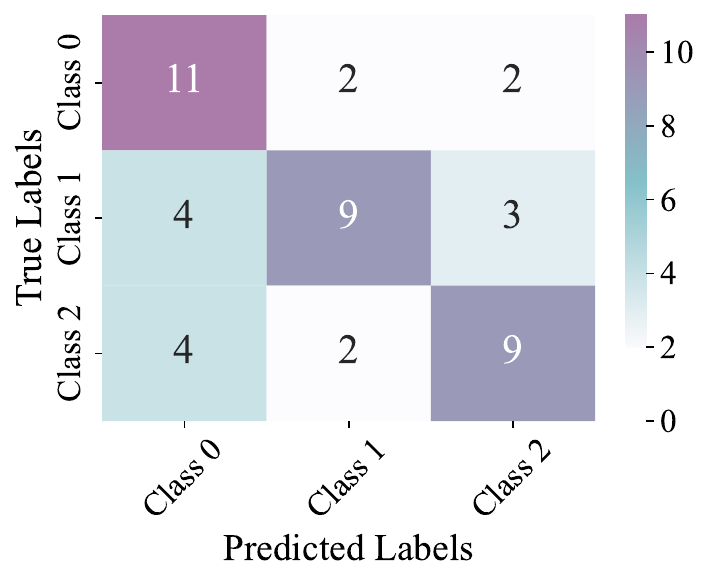}
		\label{figure_16e}}
	\hfill
	\subfloat[]{\includegraphics[width=2.321in]{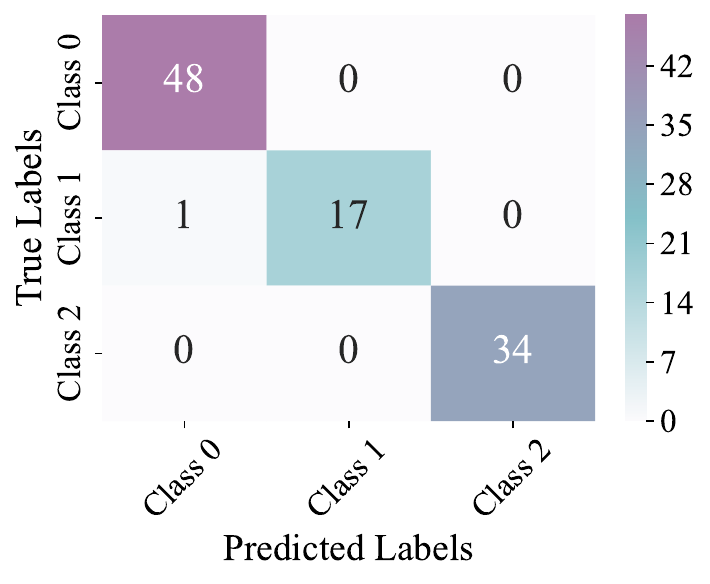}
		\label{figure_16f}}
	\caption{ROC curve and confusion matrix for each dataset, evaluated under depolarizing noise: (a) and (d) for the Iris dataset, (b) and (e) for the Tae dataset, and (c) and (f) for the Penguin dataset.}
	\label{figure_16}
\end{figure*}
To assess the effect of exponential concentration on generalization performance, we employ a relative testing loss $\vartheta$ relative to its initial values ($\mathrm{N}_{\text{init}} = 10$). As stated in Ref.~\cite{thanasilp2024exponential}, if $\mathrm{N}_{\text{init}} > 10$ and $\vartheta < 1$, it signifies superior generalization with larger training size. Fig.~\ref{figure_13} depicts the effects of exponential concentration on $4$-qubit to $10$-qubit simulations, with the solid red line representing the $40$-qubit simulation that achieves exponential concentration as detailed in Ref.~\cite{thanasilp2024exponential}. It is found from Fig.~\ref{figure_13} that for simulations ranging from $4$ to $10$ qubits, the condition $\mathrm{N}_{\text{init}} > 10$ and $\vartheta < 1$ holds, with generalization performance progressively improving as the training size increases. In addition, zero training loss is attained in all simulations. As shown in Table~\ref{tab:table2}, this paper utilizes six real-world datasets, with feature dimensions ranging from $4$ to $10$. As each feature is mapped to a qubit, the number of qubits in our simulations also ranges from $4$ to $10$. Therefore, such exponential concentration does not occur in the six real-world datasets used in this paper. To rigorously preclude the potential for exponential concentration, we further evaluate generalization performance using learning curves derived from six real-world datasets. As illustrated in Fig.~\ref{figure_14}, the testing accuracy on all six real-world datasets improves with the increasing training size, indicating a progressive improvement in generalization performance. Hence, in this paper, we do not observe the effects of exponential concentration.
\begin{figure}[t]
	\centering
	\subfloat[]{\includegraphics[width=0.9\linewidth]{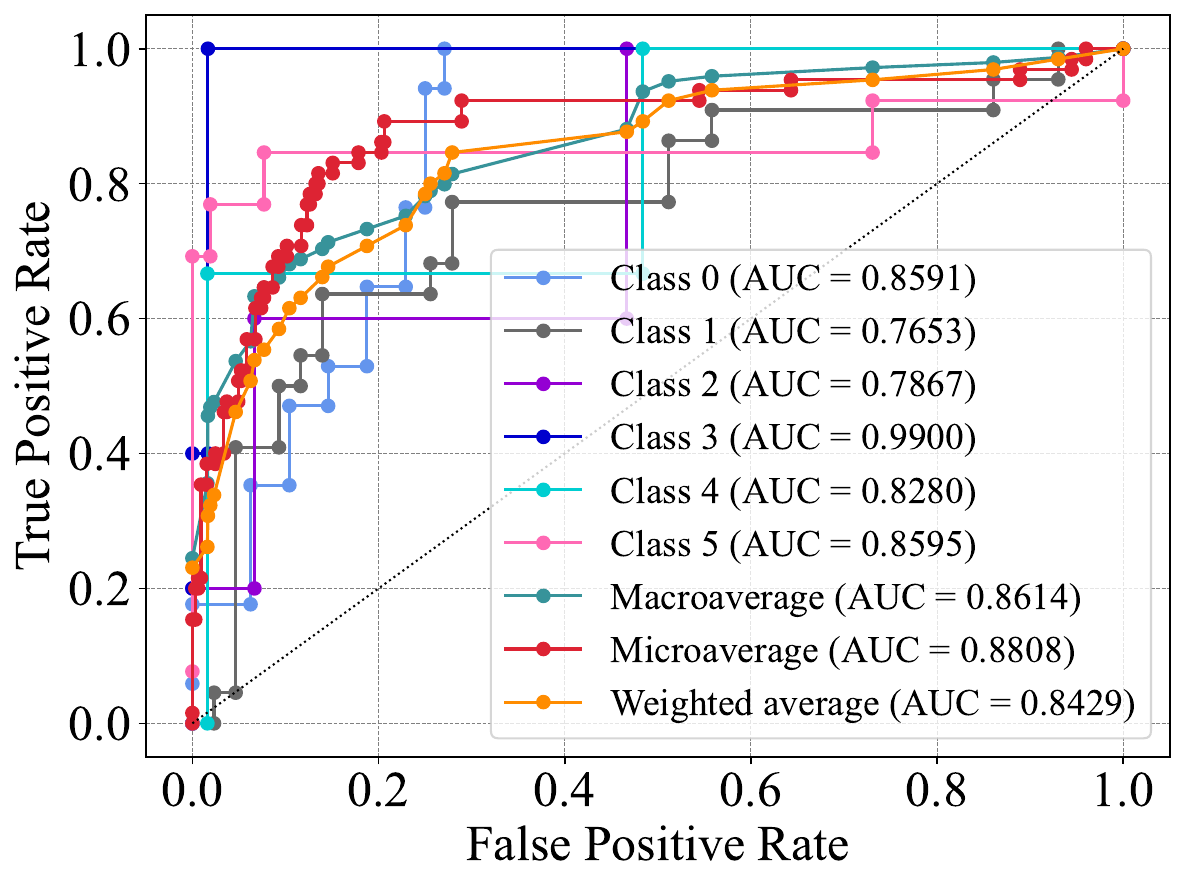}
		\label{figure_17a}}
	\hfill
	\subfloat[]{\includegraphics[width=0.8\linewidth]{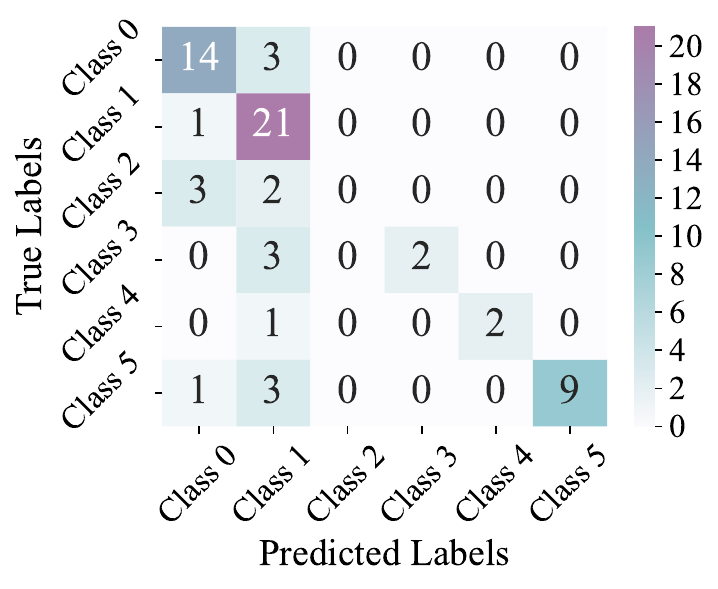}
		\label{figure_17b}}
	\caption{ROC curve (a) and confusion matrix (b) for the Glass dataset under depolarizing noise.}
	\label{figure_17}
\end{figure}
\begin{figure}[t]
	\centering
	\subfloat[]{\includegraphics[width=0.9\linewidth]{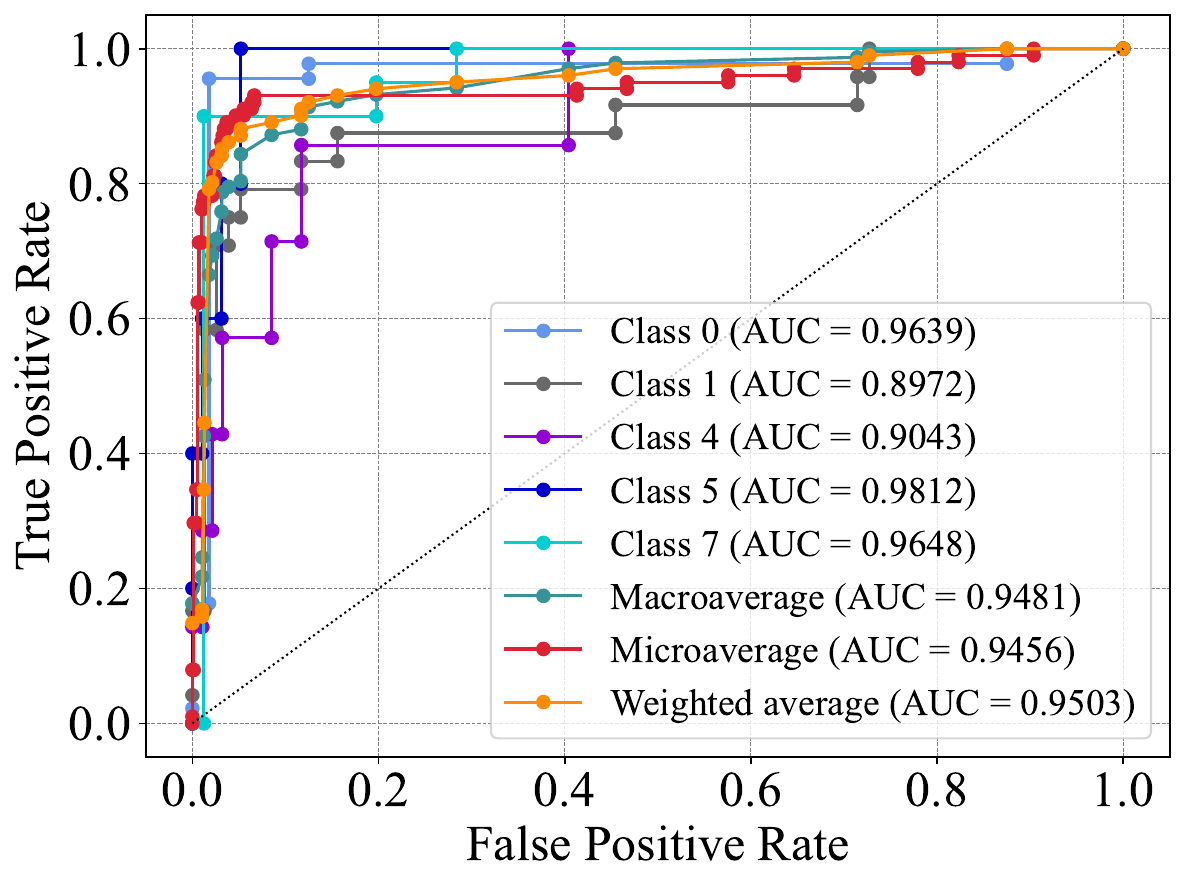}
		\label{figure_18a}}
	\hfill
	\subfloat[]{\includegraphics[width=0.8\linewidth]{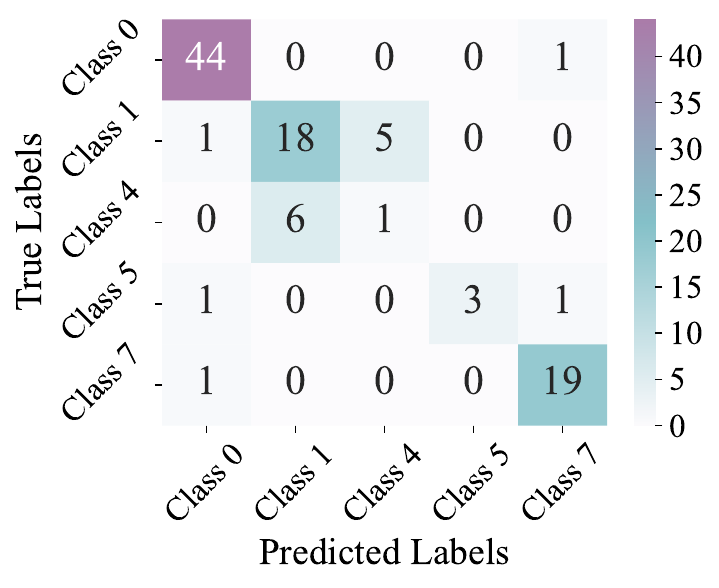}
		\label{figure_18b}}
	\caption{ROC curve (a) and confusion matrix (b) for the Ecoli dataset under depolarizing noise.}
	\label{figure_18}
\end{figure}

\begin{figure}[t]
	\centering
	\includegraphics[width=\linewidth]{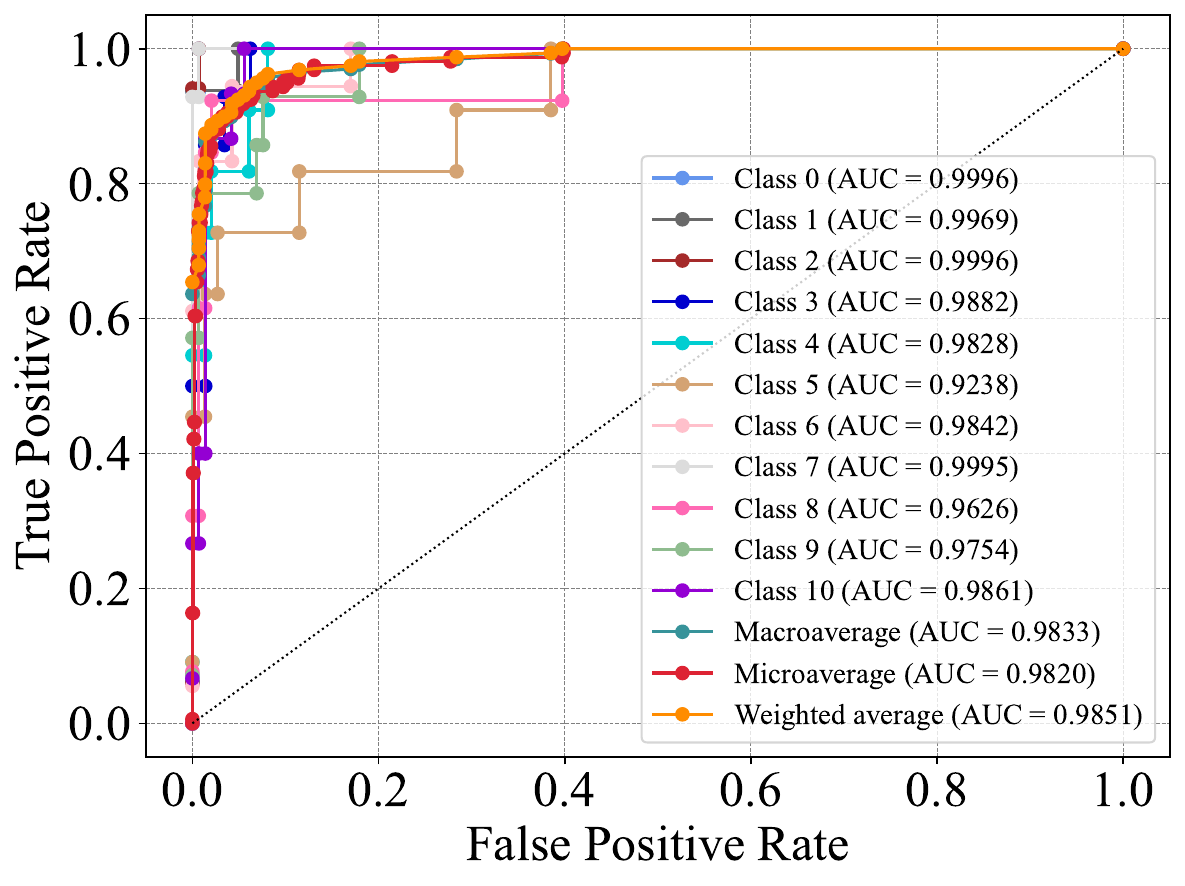}
	\caption{\label{figure_19}ROC curve for the Vowel dataset under depolarizing noise. The AUCs also all exceed $0.9000$.}
\end{figure}

\subsection{Effect of hardware noise}
Given the constraints of NISQ devices, hardware noise may disrupt quantum kernels constructed with parameterized quantum circuits, reducing the accuracy of similarity metrics. That is, hardware noise may bias the kernel values away from their true values, potentially affecting the classification accuracy of the proposed quantum algorithm. In this paper, we focus on depolarizing noise and conduct an analytical study of its effect on the quantum algorithm. To investigate the effect of depolarizing noise, we employ the noise model described in Refs.~\cite{thanasilp2024exponential,wang2021noise}. Suppose each layer of an ideal parameterized quantum circuit is followed by a depolarizing noise channel with a probability $\bar{p}$. The model for the depolarizing noise channel is denoted by $\mathcal{N}_{\bar{p}}=\mathcal{N}_{\bar{p}}^1 {\otimes} \cdots {\otimes} \mathcal{N}_{\bar{p}}^\mathrm{N}$. For any quantum state $\eta$, we have
\begin{equation}\label{noise}
	\mathcal{N}_{\bar{p}}^I(\eta)=(1-\bar{p}) \eta+\frac{\bar{p}}{3}\left(\sigma_x \eta \sigma_x +\sigma_y \eta \sigma_y + \sigma_z \eta \sigma_z\right)
\end{equation}
with the corresponding Kraus matrices $M_0 = \sqrt{1-\bar{p}}\sigma_0$, $M_1 = \sqrt{\bar{p}/3}\sigma_x$, $M_2 = \sqrt{\bar{p}/3}\sigma_y$, and $M_3 = \sqrt{\bar{p}/3}\sigma_z$. As shown in Fig.~\ref{figure_15}, the quantum kernel with depolarizing noise is expressed as
\begin{equation}\label{noise_kernel}
\!\! \kappa^{\prime}(\vec{x}_i, \vec{x}_j)=\operatorname{Tr}\left[\Pi \left(\mathcal{N}_{\bar{p}} \circ \mathcal{S}^{\dag}\left(\vec{x}_j\right) \circ \mathcal{N}_{\bar{p}} \circ \mathcal{S}\left(\vec{x}_i\right)\right)\Pi \right],
\end{equation}
where $\Pi = \left(|0\rangle\langle 0|\right)^{\otimes \mathrm{N}}$. From Eqs.~\myparencite{noise} and \myparencite{noise_kernel}, it is evident that the channel is noise-free only when the condition $\bar{p} = 0$ is met. Therefore, hardware noise can inevitably introduce bias, causing kernel values to shift from their true values.

\begin{figure}[t]
	\centering
	\includegraphics[width=0.95\linewidth]{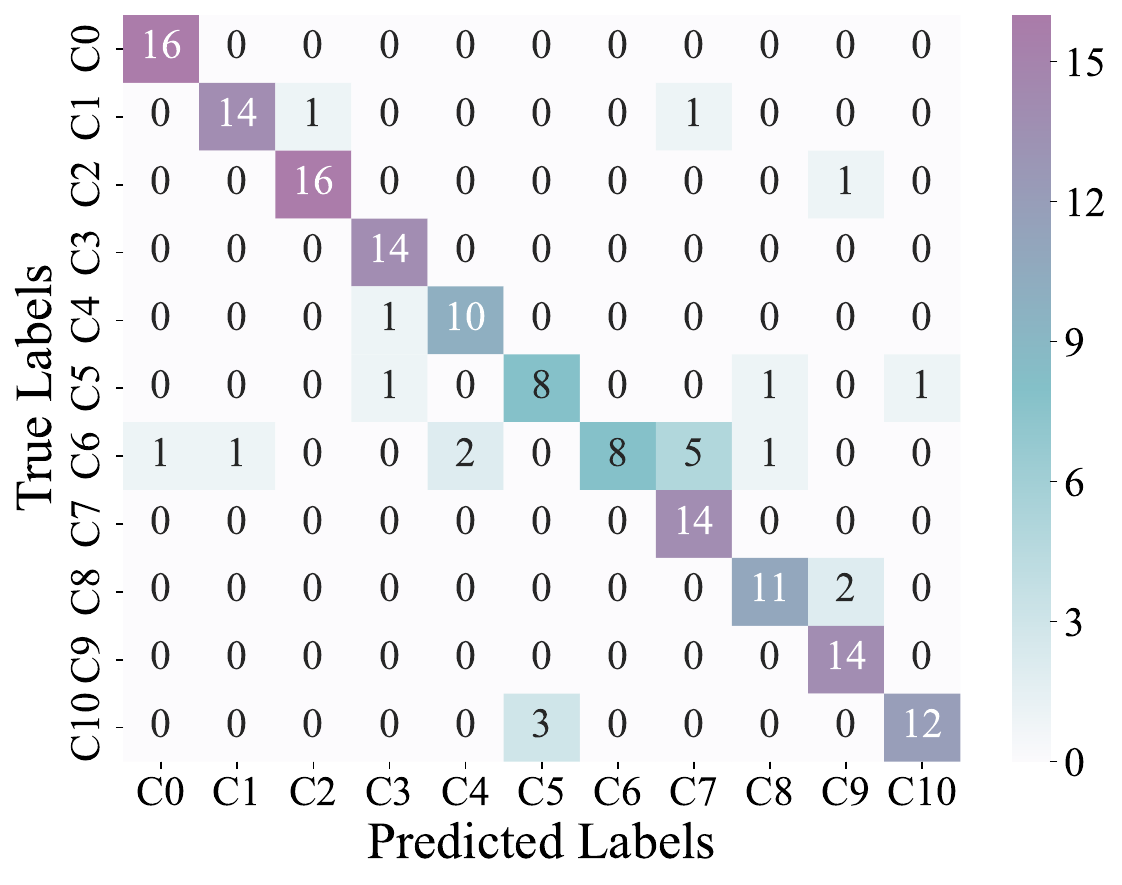}
	\caption{Confusion matrix for the Vowel dataset under depolarizing noise. C0--C10, Class $0$ to $10$.}
	\label{figure_20}
\end{figure}

To evaluate the proposed quantum algorithm under depolarizing noise, we construct six distinct noisy quantum kernels, detailed in Table~\ref{tab:table5}. Then, we employ confusion matrices to conduct an evaluation of the classification performance of the quantum algorithm using noisy quantum kernels. As shown in Table~\ref{tab:table6}, the noisy Pauli-X, Pauli-Y, and Pauli-Z quantum kernels exhibit identical performance, with their equivalence proven in the Appendix~\ref{appendixc}. From the comparative analysis in Tables~\ref{tab:table4} and \ref{tab:table6}, we further obtain the following results: for the Iris and Penguin datasets, the presence of depolarizing noise does not affect the classification performance of the quantum algorithm. Interestingly, for the Tae datasets, depolarizing noise appears to improve the quantum algorithm's classification accuracy. In contrast, for the Glass, Ecoli, and Vowel datasets, depolarizing noise negatively affects the quantum algorithm's classification performance. This negative effect can be further analyzed in detail through Figs.~\ref{figure_9b}, \ref{figure_10b}, \ref{figure_12}, \ref{figure_17b}, \ref{figure_18b}, and \ref{figure_20}. It is evident that depolarizing noise mainly affects the classification of Class $3$ in the Glass dataset, Class $1$ in the Ecoli dataset, and Classes $1$, $2$, $4$, $5$, $6$, and $10$ in the Vowel dataset. To analyze the effects of depolarizing noise on overall performance, we also adopt ROC curves. It is found from Figs.~\ref{figure_9a}, \ref{figure_10a}, \ref{figure_11}, \ref{figure_17a}, \ref{figure_18a}, and \ref{figure_19} that the overall performance of the quantum algorithm is affected to varying extents under depolarizing noise.

Building on the analysis above, hardware noise introduces bias to kernel values, leading to varied outcomes: it may enhance category separation and boost classification performance, blur boundaries and diminish accuracy, or have no effect at all. Therefore, future work should focus on exploring non-trivial quantum kernels, particularly those with entangled structures, to further improve generalization and classification performance. We also aim to investigate the trade-off between kernel complexity and hardware noise, with a focus on real-world applications constrained by limited resources.

\section{Conclusion}\label{conclusion}
In the current NISQ era, it is crucial to develop quantum machine learning algorithms tailored for effective operation on NISQ devices. This paper proposed a quantum machine learning algorithm for multiclass classification problems on NISQ devices, termed quantum-enhanced multiclass SVMs. The results from quantum simulations demonstrated that the proposed quantum algorithm surpasses its classical counterparts in handling six real-world multiclass classification problems. The quantum simulations further illuminated the effects of quantum kernel methods on generalization and classification performance, demonstrating their potential in advancing quantum machine learning. Furthermore, this paper successfully addressed the future work proposed by Havl\'\i\v{c}ek $et \ al.$ in Ref.~\cite{havlivcek2019supervised} and further expanded upon their research.

\begin{acknowledgments}
This work is supported by Singapore Quantum Engineering Program (NRF2021-QEP2-01-P01, NRF2021-QEP2-01-P02, NRF2021-QEP2-03-P01, NRF2021-QEP2-03-P09, NRF2021-QEP2-03-P10, NRF2021-QEP2-03-P11, NRF2022-QEP2-02-P13); ASTAR (M21K2c0116, M24M8b0004); Singapore National Research Foundation (NRF-CRP22-2019-0004, NRF-CRP30-2023-0003, NRF2023-ITC004-001, and NRF-MSG-2023-0002), and Singapore Ministry of Education Tier 2 Grant (MOE-T2EP50221-0005, MOE-T2EP50222-0018). This work is also supported by National Natural Science Foundation of China (Grants No.~61703151 and 62076091) and Natural Science Foundation of Hunan Province (Grant No.~2023JJ30167).
\end{acknowledgments}

\appendix
\renewcommand{\appendixname}{APPENDIX~}
\section{PERFORMANCE METRICS}\label{appendixa}
\setcounter{equation}{0}
\renewcommand\theequation{A\arabic{equation}}
\begin{figure}[b]
	\centering
	\includegraphics[width=\linewidth]{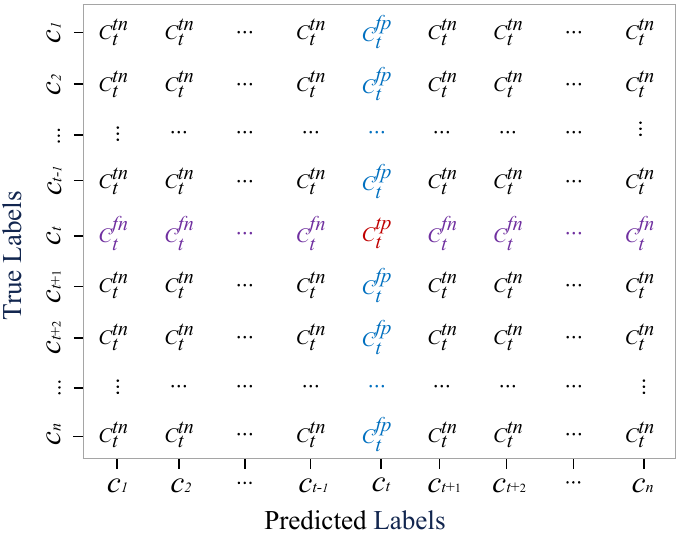}
	\caption{\label{figure_21}Schematic diagram of the confusion matrix for Class $t$ ($\mathrm{c}_{t}$).}
	\vspace{-1ex}
\end{figure}

In this appendix, we provide a comprehensive set of performance metrics \cite{luque2019impact,grandini2020metrics} for evaluating the effectiveness of the proposed quantum-enhanced multiclass SVMs. These metrics are meticulously designed to precisely evaluate the predictive capability, accuracy, and other key aspects of the quantum algorithm in multiclass classification tasks. As illustrated in Fig.~\ref{figure_21}, the precision $M^\phi_t$, recall $M^\theta_t$, and F1 score $M^\tau_t$ for Class $t$ can be denoted as follows:
\begin{gather}
	M^\phi_t=\frac{C^{\mathrm{\textit{tp}}}_t}{C^{\mathrm{\textit{tp}}}_t+C^{\mathrm{\textit{fp}}}_t},	\\
	M^\theta_t=\frac{C^{\mathrm{\textit{tp}}}_t}{C^{\mathrm{\textit{tp}}}_t+C^{\mathrm{\textit{fn}}}_t},\\
	M^\tau_t=\frac{2C^{\mathrm{\textit{tp}}}_t}{2C^{\mathrm{\textit{tp}}}_t+C^{\mathrm{\textit{fn}}}_t+C^{\mathrm{\textit{fp}}}_t}.
\end{gather}
Note that in Fig.~\ref{figure_21}, ${C^{\mathrm{\textit{tp}}}_t}$ represents testing instances correctly predicted as Class $t$; ${C^{\mathrm{\textit{fn}}}_t}$ denotes testing instances incorrectly predicted as not Class $t$ despite belonging to it; ${C^{\mathrm{\textit{fp}}}_t}$ is testing instances incorrectly predicted as Class $t$ despite not belonging to it; ${C^{\mathrm{\textit{tn}}}_t}$ describes testing instances correctly predicted as not Class $t$. Given a classification task with $l$ classes, calculating the performance metrics for each class is essential. To achieve a thorough evaluation of algorithm performance within such tasks, we integrate macroaverage, microaverage, and weighted average approaches. In the macroaverage approach, the computational expressions for macroaverage precision $M^\phi_{\mathrm{\textit{mac}}}$, macroaverage recall $M^\theta_{\mathrm{\textit{mac}}}$, and macroaverage F1 score $M^\tau_{\mathrm{\textit{mac}}}$ are delineated as follows:
\begin{align}
	M^\phi_{\mathrm{\textit{mac}}} &= \frac{1}{l}\sum_{i=1}^{l}M^{\phi}_i,\\
	M^\theta_{\mathrm{\textit{mac}}} &= \frac{1}{l}\sum_{i=1}^{l}M^{\theta}_i, \\
	M^\tau_{\mathrm{\textit{mac}}} &=\frac{1}{l}\sum_{i=1}^{l}M^{\tau}_i.
\end{align}
In the weighted average approach, the computational expressions for weighted average precision $M^\phi_{\mathrm{\textit{wgt}}}$, weighted average recall $M^\theta_{\mathrm{\textit{wgt}}}$, and weighted average F1 score $M^\tau_{\mathrm{\textit{wgt}}}$ are expressed as follows:
\begin{align}
	M^\phi_{\mathrm{\textit{wgt}}} &=  \sum_{i=1}^{l}\gamma_i M^{\phi}_i,\\
	M^\theta_{\mathrm{\textit{wgt}}} &= \sum_{i=1}^{l}\gamma_i M^{\theta}_i, \\
	M^\tau_{\mathrm{\textit{wgt}}} &= \sum_{i=1}^{l}\gamma_i M^{\tau}_i, 
\end{align}
where $\gamma_{i}$ is the weight of Class $i$. In the microaverage approach, the computational expressions for microaverage precision $M^\phi_{\mathrm{\textit{mic}}}$, microaverage recall $M^\theta_{\mathrm{\textit{mic}}}$, and microaverage F1 score $M^\tau_{\mathrm{\textit{mic}}}$ are described as follows:
\begin{gather}
	M^\phi_{\mathrm{\textit{mic}}} =   \frac{\sum_{i=1}^{l} C^{\mathrm{\textit{tp}}}_i}{\sum_{i=1}^{l} \left(C^{\mathrm{\textit{tp}}}_i + C^{\mathrm{\textit{fp}}}_i\right) }, \\
	M^\theta_{\mathrm{\textit{mic}}} = \frac{\sum_{i=1}^{l} C^{\mathrm{\textit{tp}}}_i}{\sum_{i=1}^{l} \left(C^{\mathrm{\textit{tp}}}_i + C^{\mathrm{\textit{fn}}}_i\right) },\\
	M^\tau_{\mathrm{\textit{mic}}} = \frac{\sum_{i=1}^{l} 2C^{\mathrm{\textit{tp}}}_i}{\sum_{i=1}^{l} \left(2C^{\mathrm{\textit{tp}}}_i+ C^{\mathrm{\textit{fn}}}_i+ C^{\mathrm{\textit{fp}}}_i\right)},
\end{gather}
where $M^\tau_{\mathrm{\textit{mic}}}$ is the harmonic mean of $M^\phi_{\mathrm{\textit{mic}}}$ and $M^\theta_{\mathrm{\textit{mic}}}$. Furthermore, overall accuracy is regarded as a performance metric, expressed as
\begin{equation}
	M^\prime = \frac{1}{C^{\prime}}\sum_{i=1}^{l} C^{\mathrm{\textit{tp}}}_i,
\end{equation}
where $C^\prime$ denotes all testing instances. According to the performance metrics discussed above, a higher metric value signifies a stronger classification performance of the proposed quantum algorithm, whereas a lower value indicates inferior classification performance.


\section{PROOF OF EQUIVALENCE AMONG PAULI-X, PAULI-Y, AND PAULI-Z QUANTUM KERNELS}\label{appendixb}
\setcounter{equation}{0}
\renewcommand\theequation{B\arabic{equation}}
In Sec.~\ref{experiments}, Pauli-X, Pauli-Y, and Pauli-Z quantum kernels achieve identical mean accuracy in multiclass classification tasks on various real-world datasets. Although these quantum kernels exhibit different physical properties and mechanisms, our experimental results reveal that their classification performance is remarkably consistent on various multiclass classification tasks. This consistency has prompted further investigation into the potential theoretical equivalence of these quantum kernels. 

\subsection{Mathematical analysis of Pauli-X quantum kernels}
For any input classical state $\vec{x}_i = (2x_0, 2x_1, \ldots, 2x_{\mathrm{N}-1})^T \in \mathbb{R}^{\mathrm{N}}$, the Pauli-X quantum kernel has the form
\begin{equation}\label{appc:kernel_xi}
	\kappa_x(\vec{x}_i,\vec{x}^{\prime}_i) = |\langle \phi_x(\vec{x}_i) | \phi_x(\vec{x}^{\prime}_i) \rangle|^2.
\end{equation}
Let us first consider the process of encoding $\vec{x}_i$ as a quantum state, which we denote by 
\begin{align}\label{appc:state_xi}
	| \phi_x(\vec{x}_i)\rangle = \left[\begin{array}{c}
		\cos x_0 \\
		-i\sin x_0
	\end{array}\right]{\otimes} \cdots {\otimes}	\left[\begin{array}{c}
		\cos x_{\mathrm{N}-1} \\
		-i\sin x_{\mathrm{N}-1}
	\end{array}\right].
\end{align}
Substituting Eq.~\myparencite{appc:state_xi} into Eq.~\myparencite{appc:kernel_xi}, we have
\begin{equation}\label{appc:Pauli_x}
	\begin{split}
		\kappa_x(\vec{x}_i,\vec{x}^{\prime}_i) = \prod_{i=0}^{\mathrm{N}-1}\left|\cos \left(x_i-x_i^{\prime}\right)\right|^2.
	\end{split}
\end{equation}
Here we utilize $\langle 0| 0\rangle = \langle 1| 1\rangle = 1$ and $\langle 0| 1\rangle = \langle 1| 0\rangle = 0$.

\subsection{Mathematical analysis of Pauli-Y quantum kernels}
Similar to the Pauli-X quantum kernel, the Pauli-Y quantum kernel is denoted as
\begin{equation}\label{appc:kernel_yi}
	\kappa_y(\vec{x}_i,\vec{x}^{\prime}_i)=|\langle \phi_y(\vec{x}_i) | \phi_y(\vec{x}^{\prime}_i) \rangle|^2.
\end{equation}
Then we encode $\vec{x}_i$ as a quantum state, given by
\begin{align}\label{appc:state_yi}
	|\phi_y(\vec{x}_i)\rangle = \left[\begin{array}{c}
		\cos x_0 \\
		\sin x_0
	\end{array}\right]{\otimes} \cdots {\otimes}	\left[\begin{array}{c}
		\cos x_{\mathrm{N}-1} \\ 
		\sin x_{\mathrm{N}-1}
	\end{array}\right].
\end{align}
Substituting Eq.~\myparencite{appc:state_yi} into Eq.~\myparencite{appc:kernel_yi}, we have
\begin{equation}\label{appc:Pauli_y}
	\begin{split}
		\kappa_y(\vec{x}_i,\vec{x}^{\prime}_i)= \prod_{i=0}^{\mathrm{N}-1}\left|\cos \left(x_i-x_i^{\prime}\right)\right|^2.
	\end{split}
\end{equation}

\begin{table}[t]
	\centering
	\caption{\label{appd:tab:01}Generalization analysis of various kernel functions.}
	\begin{ruledtabular}
		\begin{tabular}{lllcc}
			Dataset& Kernel function & $\ \ \|K\|_F$ & $\mathfrak{\hat{R}}_\mathrm{E}(\mathcal{F})$ & Upper bound \\ 
			\hline \rule{0pt}{10pt}Iris  & LK & 303.5553 & 2.5593 & 2.8910 \\
				& PK & 527.1340  & 4.8091 & 5.0203  \\
				& SK & 76.3759 & 0.6851 & 0.7274 \\
				& GK & 50.4150 & 0.4601 & 0.4801  \\
			 & FQK & 39.9393 & 0.3589 & 0.3804 \\
				& LQK & 41.9019 & 0.3814 & 0.3991 \\
				& CQK & 40.4679 & 0.3683 & 0.3854 \\
				& XQK, YQK, ZQK & 49.4697 & 0.4519 & 0.4711 \\
			Tae  & LK & 266.1292 & 2.4115& 2.5346 \\
				& PK & 445.2876  & 4.1964& 4.2408  \\
				& SK & 76.4926 & 0.6603 & 0.7285 \\
				& GK & 37.7070 & 0.3385& 0.3591  \\
			 & FQK & 26.2685 & 0.2370& 0.2502  \\
				& LQK & 28.5439 & 0.2559 & 0.2718 \\
				& CQK & 27.5773 & 0.2491 & 0.2626 \\
				& XQK, YQK, ZQK & 32.2922 & 0.2928 & 0.3075 \\
			Penguin  & LK & 778.3569 & 2.6868 & 3.3406 \\
				& PK & 1074.1190  & 4.0841 & 4.6100 \\
				& SK & 168.8427 & 0.6911 & 0.7246 \\
				& GK & 92.4048 & 0.3862& 0.3966  \\
				& FQK & 54.5493 & 0.2353 & 0.2341 \\
				& LQK & 62.1466 & 0.2649&  0.2667 \\
				& CQK & 58.5630 & 0.2501 & 0.2513 \\
				& XQK, YQK, ZQK & 81.9560 & 0.3430 & 0.3517 \\
			Glass  & LK & 573.0572 & 3.3799 &3.8460 \\
				& PK & 10762.2241  & 69.6747 &72.2297  \\
				& SK & 109.0790 & 0.6680 &0.7321  \\
				& GK & 66.5629 & 0.4125 & 0.4467 \\
				& FQK & 49.7011 & 0.3135&  0.3336 \\
				& LQK & 54.1024 & 0.3412&  0.3631 \\
				& CQK & 53.1684 & 0.3361&  0.3568 \\
				& XQK, YQK, ZQK & 60.8262 & 0.3812& 0.4082  \\
			Ecoli  & LK & 795.9544 & 3.1618& 3.3870 \\
				& PK & 424827.4131  & 1807.6520  &1807.7762 \\
				& SK & 169.9102 & 0.7020& 0.7230  \\
				& GK & 97.1009 & 0.4020  &0.4132 \\
				& FQK & 62.1455 & 0.2549 & 0.2644 \\
				& LQK & 70.9959 & 0.2872  &0.3021 \\
				& CQK & 69.5493 & 0.2822 & 0.2960\\
				& XQK, YQK, ZQK & 86.7654 & 0.3574& 0.3692  \\
			Vowel  & LK & 1443.9961 & 3.3813 & 3.9133 \\
				& PK & 2169.7484  &  5.3771& 5.8801 \\
				& SK & 268.2608 &  0.6948 &0.7270 \\
				& GK & 87.5059 &  0.2337& 0.2371  \\
				& FQK & 40.6547 & 0.1107 & 0.1102 \\
				& LQK & 47.2345 &  0.1259& 0.1280  \\
				& CQK & 43.8518 & 0.1176&  0.1188 \\
				& XQK, YQK, ZQK & 63.6651 & 0.1700& 0.1725 \\
		\end{tabular}
	\end{ruledtabular}
\end{table}
\begin{table}[t]
	\renewcommand{\tablename}{ALGORITHM}
	\renewcommand{\thetable}{1}
	\caption{Quantum-enhanced multiclass SVMs}
	\centering
	\begin{ruledtabular}
		\begin{tabular}{c}
			\begin{minipage}{.48\textwidth}
				\begin{algorithmic}[1]
					\Require $Training$ $dataset$ $\mathcal{X}=\{\left(\vec{x}_1, y_1\right), \ldots,\left(\vec{x}_m, y_m\right)\}$, $where$ $\vec{x}_i = (x_1, x_2, \ldots, x_{\mathrm{N}})^T$ $is$ $the$ $feature$ $vector$ $of$ $the$ $ith$ $data$ $point$ $and$ $y_i$ $is$ $the$ $corresponding$ $label$; $Label$ $matrix$ $L$; $Testing$ $instance$ $\vec{x}_t$	
					\Ensure $Predicted$ $class$ $label$ $\tilde{y}$
					\State \textbf{Initialization:}
					\State $Initialize$ $parameters:$ $\vec{\alpha}$, $\vec{\theta}$, $b$
					\State $Set$ $C$ $as$ $the$ $penalty$ $parameter$
					\State $Set$ $M$ $as$ $the$ $resulting$ $bit$ $strings$ 
					\State $Set$ ${Z}$ $as$ $the$ $number$ $of$ $measurements$
					\State $Prepare$ $quantum$ $registers$ $for$ $initial$ $states$ $\ket{0}^{\otimes \mathrm{N}}$
					\State $Calibrate$ $the$ $NISQ$ $device$ $to$ $construct$ $the$ $quantum$ $circuit$ 
					\State \textbf{Quantum kernel estimation:}
					\For{$each$ $pair$ $of$ $data$ $points$ $(\vec{x}_i, \vec{x}_j)$}
					\State $Set$ $the$ $counter$ $R = \mathrm{r}(0,\ldots,0)=0$
					\For {$z \gets 1$ \textbf{to} $Z$}
					\State $Apply$ $\mathcal{S}\left(\vec{x}_i\right)$ $to$ $prepare$ $quantum$ $state$ $\ket{\phi(\vec{x}_i)}$
					\State $Apply$ $\mathcal{S}^{\dag}(\vec{x}_j)$ $to$ $the$ $prepared$ $quantum$ $state$ $\ket{\phi(\vec{x}_i)}$
					\State $Measure$ $the$ $resulting$ $state$ $\mathcal{S}^\dag(\vec{x}_j) \mathcal{S}(\vec{x}_i)|0\rangle^{\otimes \mathrm{N}}$
					\If{$M = 0^\mathrm{N}$}
					\State ${R}={R} + 1$
					\Else
					\State ${R}={R}$
					\EndIf
					\EndFor
					\State $Use$ $the$ $frequency$ ${R}Z^{-1}$ $to$ $estimate$ $K_{ij}$
					\State $Store$ $K_{ij}$ $in$ $the$ $quantum$ $kernel$ $matrix$ $K$
					\EndFor
					\State \textbf{Return:} $Quantum$ $kernel$ $matrix$ $K$
					\State \textbf{Training:}
					\For{$s \gets 1$ \textbf{to} $l$}
					\State $Construct$ $the$ $binary$ $label$ $vector$ $L_s$
					\For{$i \gets 1$ \textbf{to} $m$}
					\If{$L[i] = s$}
					\State $L_s[i] = 1$
					\Else
					\State $L_s[i] = -1$
					\EndIf
					\EndFor
					\State \textbf{Update to optimal parameters:}	
					\State $\vec{\alpha}^{s}, \vec{\theta}_s, b_s \gets $ SMO($K, L_s, C$)
					\EndFor
					\State \textbf{Return:} $Trained$ $parameters$ $(\vec{\alpha}, \vec{\theta}, b)$
					\State \textbf{Prediction:}
					\For{$s \gets 1$ \textbf{to} $l$}
					\State $Calculate$ $the$ $sth$ $decision$ $function$ $using$ $the$ $trained$ $parameters$ $(\vec{\alpha}, \vec{\theta}, b)$ $and$ $the$ $testing$ $instance$ $\vec{x}_t:$ $\tilde{f}_s(\vec{x}_t)=\sum_{i \in \Omega} \alpha_i^{s} y_i K_{it}+b_s$
					\EndFor
					\State \textbf{Return:} $\tilde{y} = \arg \max_{s=1, \ldots, l} \tilde{f}_s(\vec{x}_t)$
				\end{algorithmic}
				\label{app:alg1}
			\end{minipage}
		\end{tabular}
	\end{ruledtabular}
\end{table}

\subsection{Mathematical analysis of Pauli-Z quantum kernels}
Following the approach for deriving the Pauli-X and Pauli-Y quantum kernels, the Pauli-Z quantum kernel is expressed as
\begin{equation}\label{appc:kernel_z}
	\kappa_z(\vec{x}_i,\vec{x}^{\prime}_i)=|\langle \phi_z(\vec{x}_i) | \phi_z(\vec{x}^{\prime}_i) \rangle|^2.
\end{equation}
Then we encode $\vec{x}_i$ as a quantum state, denoted by
\begin{align}\label{appc:state_zi}
	| \phi_z(\vec{x}_i)\rangle &= \left(\frac{\sqrt{2}}{2}\right)^{\mathrm{N}} \left[\begin{array}{c}
		\mathcal{M}_{(x_0)} \\
		\mathcal{N}_{(x_0)}
	\end{array}\right]{\otimes} \cdots {\otimes}	\left[\begin{array}{c}
		\mathcal{M}_{(x_\mathrm{N-1})} \\ 
		\mathcal{N}_{(x_\mathrm{N-1})}
	\end{array}\right]
\end{align}
with $\mathcal{M}_{(x)} = \cos x-i\sin x$, $\mathcal{N}_{(x)} = \cos x+i\sin x$. The complex conjugate transpose of $| \phi_z(\vec{x}_i)\rangle$ is thus given by
\begin{align}\label{appc:state_zzi}
	\langle\phi_z(\vec{x}^{\prime}_i)| = \left(\frac{\sqrt{2}}{2}\right)^{\mathrm{N}} \left[\begin{array}{c}
		\mathcal{N}_{(x^{\prime}_0)}  \\ 
		\mathcal{M}_{(x^{\prime}_0)}
	\end{array}\right]^T {\otimes} \cdots {\otimes}	\left[\begin{array}{c}
		\mathcal{N}_{(x^{\prime}_\mathrm{N-1})} \\
		\mathcal{M}_{(x^{\prime}_\mathrm{N-1})}
	\end{array}\right]^T.
\end{align}
Substituting Eq.~\myparencite{appc:state_zi} and Eq.~\myparencite{appc:state_zzi} into Eq.~\myparencite{appc:kernel_z}, we have
\begin{align}\label{appc:Pauli_z}
	\kappa_z(\vec{x}_i,\vec{x}^{\prime}_i) &= \left(\frac{1}{4}\right)^{\mathrm{N}} \prod_{i=0}^{\mathrm{N-1}}\left|\left(\mathcal{N}_{\left(x_i\right)} \mathcal{M}_{\left(x_i^{\prime}\right)}+\mathcal{M}_{\left(x_i\right)} \mathcal{N}_{\left(x_i^{\prime}\right)}\right)\right|^2  \\ \nonumber
	&=\prod_{i=0}^{\mathrm{N-1}}\left|\cos \left(x_i-x_i^{\prime}\right)\right|^2. 
\end{align}
From Eqs.~\myparencite{appc:Pauli_x}, \myparencite{appc:Pauli_y}, and \myparencite{appc:Pauli_z}, we can conclude that Pauli-X, Pauli-Y, and Pauli-Z quantum kernels are equivalent.

\section{PROOF OF EQUIVALENCE AMONG NOISY PAULI-X, PAULI-Y, AND PAULI-Z QUANTUM KERNELS}\label{appendixc}
\setcounter{equation}{0}
\renewcommand\theequation{C\arabic{equation}}
In Sec.~\ref{limitation}, the noisy Pauli-X, Pauli-Y, and Pauli-Z quantum kernels all attain the same accuracy in multiclass classification tasks across different real-world datasets. Therefore, in this appendix, we further explore the theoretical equivalence of these noisy quantum kernels. As indicated in Eq.~\myparencite{noise} of Sec.~\ref{limitation}, the depolarizing noise channel for a single qubit can be equivalently expressed as
\begin{equation}
	\mathcal{N}^I_{\bar{p}}(\eta)=\left(1-\frac{4\bar{p}}{3}\right) \eta+ \frac{2\bar{p}}{3}\mathbb{I}.
\end{equation}

\begin{widetext}
In noisy Pauli-X and Pauli-Y quantum kernels, we first consider the case of a single-qubit system, where
\begin{equation}\label{appd:xy}
	\Gamma_k = \left(1-\frac{4\bar{p}}{3}\right)^2 {\mathsf{R}}^{\dag}(\frac{{x}^{\prime}_k}{2}) {\mathsf{R}}(\frac{{x}_k}{2}) |0\rangle\langle 0| {\mathsf{R}}^{\dag}(\frac{{x}_k}{2}) {\mathsf{R}}(\frac{{x}^{\prime}_k}{2})+ \left(1-\frac{4\bar{p}}{3}\right)\frac{2\bar{p}}{3}\mathbbm{I} +\frac{2\bar{p}}{3}\mathbbm{I}
\end{equation}
with $\mathsf{R} \in \{R_x, R_y\}$. Similarly, in noisy Pauli-Z quantum kernels, we also first explore the case of a single-qubit system, where
\begin{equation}\label{appd:z}
	\Gamma_k = \left(1-\frac{4\bar{p}}{3}\right)^2 H{R_z}^{\dag}(\frac{{x}^{\prime}_k}{2})  {R_z}(\frac{{x}_k}{2}) H |0\rangle\langle 0| H {R_z}^{\dag}(\frac{{x}_k}{2})  {R_z}(\frac{{x}^{\prime}_k}{2}) H+ \left(1-\frac{4\bar{p}}{3}\right)\frac{2\bar{p}}{3}\mathbbm{I} +\frac{2\bar{p}}{3}\mathbbm{I}.
\end{equation}
Let $\Gamma$ represent the density matrix of an $\mathrm{N}$-qubit system. In this case, $\Gamma = \Gamma_0 \otimes \Gamma_1 \otimes \cdots \otimes \Gamma_\mathrm{N-1}$. Given that the projector $\Pi = \left(|0\rangle\langle 0|\right)^{\otimes \mathrm{N}}$, the noisy quantum kernel is given by
\begin{equation}\label{appd:noise}
	\kappa^{\prime}(\vec{x}_i, \vec{x}^{\prime}_i) = \operatorname{Tr}\left[\Pi \left(\Gamma_0 \otimes \Gamma_1 \otimes \cdots \otimes \Gamma_{\mathrm{N}-1}\right)\right]=\prod_{k=0}^{\mathrm{N}-1}\langle 0| \Gamma_k|0\rangle.
\end{equation}
The following mathematical analysis studies the noisy Pauli-X, Pauli-Y, and Pauli-Z quantum kernels individually.

\subsection{Mathematical analysis of noisy Pauli-X quantum kernels}
For any input classical state $\vec{x}_i = (2x_0, 2x_1, \ldots, 2x_{\mathrm{N}-1})^T \in \mathbb{R}^{\mathrm{N}}$, the noisy Pauli-X quantum kernel is given by
\begin{equation}\label{appd:nxqk}
	\kappa_x^{\prime}(\vec{x}_i, \vec{x}^{\prime}_i)=\operatorname{Tr}\left[\Pi \left(\mathcal{N}_{\bar{p}} \circ {U_x}^{\dag}\left(\vec{x}^{\prime}_i\right) \circ \mathcal{N}_{\bar{p}} \circ {U_x}\left(\vec{x}_i\right)\right)\Pi \right].
\end{equation}
Combining Eqs.~\myparencite{appd:xy} and \myparencite{appd:noise}, Eq.~\myparencite{appd:nxqk} can be expanded as:
\begin{equation}
	\kappa_x^{\prime}(\vec{x}_i, \vec{x}^{\prime}_i)=  \prod_{k=0}^{\mathrm{N}-1}\left[\left(1-\frac{4\bar{p}}{3}\right)^{2}| \langle 0| {R_x}^{\dag}({x}^{\prime}_k)  {R_x}\left({x}_k\right)|0\rangle |^2 + \left(2-\frac{4\bar{p}}{3}\right)\frac{2\bar{p}}{3}\right].
\end{equation}
Substituting Eq.~\myparencite{rx} leads to a simplification, yielding
\begin{equation}\label{appd:xx}
	\kappa_x^{\prime}(\vec{x}_i, \vec{x}^{\prime}_i)=  \prod_{k=0}^{\mathrm{N}-1}\left[ \left(1-\frac{4\bar{p}}{3}\right)^{2}| \cos \left(x_k - x_k^{\prime}\right)|^2 + \left(2-\frac{4\bar{p}}{3}\right)\frac{2\bar{p}}{3}\right].
\end{equation}

\subsection{Mathematical analysis of noisy Pauli-Y quantum kernels}
Similar to the noisy Pauli-X quantum kernel, the noisy Pauli-Y quantum kernel is expressed as
\begin{equation}\label{appd:nyqk}
	\kappa_y^{\prime}(\vec{x}_i, \vec{x}^{\prime}_i)=\operatorname{Tr}\left[\Pi \left(\mathcal{N}_{\bar{p}} \circ {U_y}^{\dag}\left(\vec{x}^{\prime}_i\right) \circ \mathcal{N}_{\bar{p}} \circ {U_y}\left(\vec{x}_i\right)\right)\Pi \right].
\end{equation}
Combining Eqs.~\myparencite{appd:xy} and \myparencite{appd:noise}, Eq.~\myparencite{appd:nyqk} can be expanded as:
\begin{equation}
	\kappa_y^{\prime}(\vec{x}_i, \vec{x}^{\prime}_i)=  \prod_{k=0}^{\mathrm{N}-1}\left[\left(1-\frac{4\bar{p}}{3}\right)^{2}| \langle 0| {R_y}^{\dag}({x}^{\prime}_k)  {R_y}\left({x}_k\right)|0\rangle |^2 + \left(2-\frac{4\bar{p}}{3}\right)\frac{2\bar{p}}{3}\right].
\end{equation}
Substituting Eq.~\myparencite{ry} results in a simplification, yielding
\begin{equation}\label{appd:yy}
	\kappa_y^{\prime}(\vec{x}_i, \vec{x}^{\prime}_i)=  \prod_{k=0}^{\mathrm{N}-1}\left[ \left(1-\frac{4\bar{p}}{3}\right)^{2}| \cos \left(x_k - x_k^{\prime}\right)|^2 + \left(2-\frac{4\bar{p}}{3}\right)\frac{2\bar{p}}{3}\right].
\end{equation}

\subsection{Mathematical analysis of noisy Pauli-Z quantum kernels}
Building on the approach used to derive the noisy Pauli-X and Pauli-Y quantum kernels, the noisy Pauli-Z quantum kernel is described as
\begin{equation}\label{appd:nzqk}
	\kappa_z^{\prime}(\vec{x}_i, \vec{x}^{\prime}_i)=\operatorname{Tr}\left[\Pi \left(\mathcal{N}_{\bar{p}} \circ {U_z}^{\dag}\left(\vec{x}^{\prime}_i\right) \circ \mathcal{N}_{\bar{p}} \circ {U_z}\left(\vec{x}_i\right)\right)\Pi \right].
\end{equation}
Combining Eqs.~\myparencite{appd:z} and \myparencite{appd:noise}, Eq.~\myparencite{appd:nzqk} can be expanded as:
\begin{equation}
	\kappa_z^{\prime}(\vec{x}_i, \vec{x}^{\prime}_i)=  \prod_{k=0}^{\mathrm{N}-1}\left[\left(1-\frac{4\bar{p}}{3}\right)^{2}| \langle 0| H{R_z}^{\dag}({x}^{\prime}_k) {R_z}\left({x}_k\right)H|0\rangle |^2 + \left(2-\frac{4\bar{p}}{3}\right)\frac{2\bar{p}}{3}\right].
\end{equation}
Substituting Eq.~\myparencite{rotation_z} simplifies to
\begin{equation}\label{appd:zz}
	\kappa_z^{\prime}(\vec{x}_i, \vec{x}^{\prime}_i)=  \prod_{k=0}^{\mathrm{N}-1}\left[ \left(1-\frac{4\bar{p}}{3}\right)^{2}| \cos \left(x_k - x_k^{\prime}\right)|^2 + \left(2-\frac{4\bar{p}}{3}\right)\frac{2\bar{p}}{3}\right].
\end{equation}
From Eqs.~\myparencite{appd:xx}, \myparencite{appd:yy}, and \myparencite{appd:zz}, we can conclude that noisy Pauli-X, Pauli-Y, and Pauli-Z quantum kernels are also equivalent.
\end{widetext}

\section{THEORETICAL AND NUMERICAL GENERALIZATION ANALYSIS OF QUANTUM KERNELS}\label{appendixd}
\setcounter{equation}{0}
\renewcommand\theequation{D\arabic{equation}}
In this appendix, we focus on the generalization error bound of quantum kernels, providing a theoretical and numerical analysis of their generalization capabilities. As noted in Ref.~\cite{huangPowerDataQuantum2021}, we can derive an upper bound on the empirical Rademacher complexity to obtain a generalization error bound for quantum kernels. In other words, investigating the empirical Rademacher complexity $\mathfrak{\hat{R}}_\mathrm{E}(\mathcal{F})$ of quantum kernels provides insights into its ability to generalize.

Given a sample $\mathrm{E}=\left\{\vec{x}_1, \vec{x}_2, \ldots, \vec{x}_\mathrm{D}\right\}$, where $\vec{x}_i \in \mathbb{R}^{\mathrm{N}}$. Let $\kappa: \mathbb{R}^{\mathrm{N}} \times \mathbb{R}^{\mathrm{N}} \rightarrow \mathbb{R}$ denote a quantum kernel function, and let $K_{i j}=\kappa \left(\vec{x}_i, \vec{x}_j\right)$ be the corresponding quantum kernel matrix. Furthermore, a class of functions can be denoted by
\begin{equation}
	\mathcal{F}=\left\{f(x)=\sum_{i=1}^\mathrm{D} O_i \kappa\left(\vec{x}, \vec{x}_i\right) \mid\|\vec{O}\|_2 \leqslant \Delta \right\}
\end{equation}
with $\vec{O} = \left(O_1, \ldots, O_\mathrm{D}\right)^{T} \in \mathbb{R}^\mathrm{D}$. From the description in Ref.~\cite{xiao2022adversarial}, the empirical Rademacher complexity of $\mathcal{F}$ is defined as
\begin{equation}\label{appd:radc}
	\mathfrak{\hat{R}}_\mathrm{E}(\mathcal{F})=\mathbb{E}_\varpi \left[\sup _{\|\vec{O}\|_2 \leqslant \Delta} \frac{1}{\mathrm{D}} \sum_{i=1}^\mathrm{D} \varpi_i \sum_{j=1}^\mathrm{D} O_j \kappa\left(\vec{x}_i, \vec{x}_j\right)\right],
\end{equation}
where $\varpi_1, \varpi_2, \ldots, \varpi_\mathrm{D}$ are i.i.d. Rademacher random variables uniformly drawn from $\{-1,1\}$. Reordering the summation, Eq.\myparencite{appd:radc} becomes:
\begin{equation}\label{appd:radd}
	\mathfrak{\hat{R}}_\mathrm{E}(\mathcal{F})=\mathbb{E}_\varpi \left[\sup _{\|\vec{O}\|_2 \leqslant \Delta} \frac{1}{\mathrm{D}} \sum_{j=1}^\mathrm{D} O_j \vec{\upsilon}_j \right],
\end{equation}
where 
\begin{equation}
	\vec{\upsilon}_j := \sum_{i=1}^\mathrm{D} \varpi_i  \kappa\left(\vec{x}_i, \vec{x}_j\right).
\end{equation}
Given that $\|\vec{O}\|_2 \leqslant \Delta$, we employ the Cauchy–Schwarz inequality to deduce
\begin{equation}\label{appd:detal}
	\sum_{j=1}^\mathrm{D} O_j \vec{\upsilon}_j \leqslant \|\vec{O}\| \|\vec{\upsilon}\| \leqslant \Delta\|\vec{\upsilon}\|.
\end{equation}
Substituting Eq.~\myparencite{appd:detal} results in a simplification of Eq.~\myparencite{appd:radd}, yielding
\begin{equation}
	\mathfrak{\hat{R}}_\mathrm{E}(\mathcal{F})=\frac{\Delta}{\mathrm{D}} \mathbb{E}_\varpi[\|\vec{\upsilon}\|].
\end{equation}
By the Cauchy–Schwarz inequality, we have
\begin{equation}
	\left(\mathbb{E}_\varpi[\|\vec{\upsilon}\|]\right)^2 \leqslant \mathbb{E}_\varpi\left[\|\vec{\upsilon}\|^2\right].
\end{equation}
Therefore, $\frac{\Delta}{\mathrm{D}} \sqrt{\mathbb{E}_\varpi\left[\|\vec{\upsilon}\|^2\right]}$ can denote the upper bound of $\mathfrak{\hat{R}}_\mathrm{E}(\mathcal{F})$, i.e.,
\begin{equation}\label{appd:inequality}
	\mathfrak{\hat{R}}_\mathrm{E}(\mathcal{F}) \leqslant \frac{\Delta}{\mathrm{D}} \sqrt{\mathbb{E}_\varpi\left[\|\vec{\upsilon}\|^2\right]}.
\end{equation}
To determine the upper bound of $\mathfrak{\hat{R}}_\mathrm{E}(\mathcal{F})$, we first analyze $\|\vec{\upsilon}\|^2$, where
\begin{align}
	\|\vec{\upsilon}\|^2 &=\sum_{j=1}^\mathrm{D} \vec{\upsilon}_j^2 \\
	 &=\sum_{j=1}^\mathrm{D}\left(\sum_{i=1}^\mathrm{D} \varpi_i \kappa\left(\vec{x}_i, \vec{x}_j\right)\right)^2 \\
	 &=\sum_{j=1}^\mathrm{D} \sum_{i=1}^\mathrm{D} \sum_{i^{\prime}=1}^\mathrm{D} \varpi_i \varpi_{i^{\prime}} \kappa\left(\vec{x}_i, \vec{x}_j\right) \kappa\left(\vec{x}_{i^{\prime}}, \vec{x}_j\right).
\end{align}
Given that the Rademacher random variables satisfy $\mathbb{E}_\varpi\left[\varpi_i \varpi_{i^{\prime}}\right]=\left\{\begin{array}{ll}1, & i=i^{\prime} \\ 0, & i \neq i^{\prime}\end{array}\right.$, we have
\begin{equation}\label{appd:vv}
	\mathbb{E}_\varpi\left[\|\vec{\upsilon}\|^2\right]=\sum_{i, j}^\mathrm{D} \kappa\left(\vec{x}_i, \vec{x}_j\right)^2=\|K\|_F^2,
\end{equation}
where $\|K\|_F$ is the Frobenius norm of the quantum kernel matrix $K$. By substituting Eq.~\myparencite{appd:vv} into Eq.~\myparencite{appd:inequality}, we conclude that $\mathfrak{\hat{R}}_\mathrm{E}(\mathcal{F})$ satisfies the condition
\begin{equation}
	\mathfrak{\hat{R}}_\mathrm{E}(\mathcal{F}) \leqslant \frac{\Delta}{\mathrm{D}} \|K\|_F.
\end{equation}
Hence, the upper bound of the empirical Rademacher complexity is proportional to the Frobenius norm of the quantum kernel matrix. Furthermore, as shown in Table~\ref{appd:tab:01}, the quantum kernels employed in this paper exhibit a lower Frobenius norm and empirical Rademacher complexity across all six real-world datasets, outperforming classical kernels. These numerical results also demonstrate that quantum kernels achieve superior generalization performance over classical kernels on these datasets.

\section{PSEUDOCODE DESCRIPTION}\label{appendixe}
\setcounter{equation}{0}
\renewcommand\theequation{E\arabic{equation}}
In this appendix, we provide the pseudocode for the proposed quantum-enhanced multiclass SVMs, as detailed in ALGORITHM~\ref{app:alg1}.


%

\end{document}